%% file: main.tex
\pdfoutput=1
\documentclass[12pt,a4paper]{article}
\usepackage{ifthen}
\newboolean{pdflatex}
\setboolean{pdflatex}{true}

\newboolean{articletitles}
\setboolean{articletitles}{true}

\newboolean{uprightparticles}
\setboolean{uprightparticles}{false}

\def\paperauthors{LHCb collaboration}

\def\papertitle{Observation of structure\\ in the $\jpsi$-pair mass spectrum}

\def\papercopyright{2020\ CERN for the benefit of the LHCb collaboration}
\def\paperlicence{CC BY 4.0 licence}
\def\paperlicenceurl{https://creativecommons.org/licenses/by/4.0/}

\usepackage{subfig}
\usepackage{floatrow}
\input{preamble}
\input{mydefine}

\begin{document}

%%%%%%%%%%%%%%%%%%%%%%%%%
%%%%% Title     %%%%%%%%%
%%%%%%%%%%%%%%%%%%%%%%%%%
\renewcommand{\thefootnote}{\fnsymbol{footnote}}
\setcounter{footnote}{1}

\input{title-LHCb-PAPER}

\renewcommand{\thefootnote}{\arabic{footnote}}
\setcounter{footnote}{0}

\cleardoublepage

%%%%%%%%%%%%%%%%%%%%%%%%%
%%%%% Main text %%%%%%%%%
%%%%%%%%%%%%%%%%%%%%%%%%%

\pagestyle{plain}
\setcounter{page}{1}
\pagenumbering{arabic}

\input{body}

\input{acknowledgements}

\input{app}

\addcontentsline{toc}{section}{References}
\bibliographystyle{LHCb}
\bibliography{local}
 
\newpage
\input{LHCb_Authorship_28-Apr-2020}

\end{document}

%% file: preamble.tex
\usepackage{microtype}
\usepackage{lineno}  % for line numbering during review
\usepackage{xspace} % To avoid problems with missing or double spaces after
\usepackage{caption} %these three command get the figure and table captions automatically small
\renewcommand{\captionfont}{\small}
\renewcommand{\captionlabelfont}{\small}

\usepackage{graphicx}  % to include figures (can also use other packages)
\usepackage{color}
\usepackage{colortbl}
\graphicspath{{./figs/}} % Make Latex search fig subdir for figures

\usepackage{amsmath} % Adds a large collection of math symbols
\usepackage{amssymb}
\usepackage{amsfonts}
\usepackage{upgreek} % Adds in support for greek letters in roman typeset

\newcommand*\patchAmsMathEnvironmentForLineno[1]{%
\expandafter\let\csname old#1\expandafter\endcsname\csname #1\endcsname
\expandafter\let\csname oldend#1\expandafter\endcsname\csname
end#1\endcsname
 \renewenvironment{#1}%
   {\linenomath\csname old#1\endcsname}%
   {\csname oldend#1\endcsname\endlinenomath}%
}
\newcommand*\patchBothAmsMathEnvironmentsForLineno[1]{%
  \patchAmsMathEnvironmentForLineno{#1}%
  \patchAmsMathEnvironmentForLineno{#1*}%
}
\AtBeginDocument{%
\patchBothAmsMathEnvironmentsForLineno{equation}%
\patchBothAmsMathEnvironmentsForLineno{align}%
\patchBothAmsMathEnvironmentsForLineno{flalign}%
\patchBothAmsMathEnvironmentsForLineno{alignat}%
\patchBothAmsMathEnvironmentsForLineno{gather}%
\patchBothAmsMathEnvironmentsForLineno{multline}%
\patchBothAmsMathEnvironmentsForLineno{eqnarray}%
}

\usepackage{hyperref}    % Hyperlinks in references
\usepackage[all]{hypcap} % Internal hyperlinks to floats.

\input{lhcb-symbols-def} % Add in the predefined LHCb symbols

\usepackage{cite} % Allows for ranges in citations
\usepackage{mciteplus}

%% file: lhcb-symbols-def.tex
\usepackage{xspace} 
\usepackage{upgreek}

\def\lhcb   {\mbox{LHCb}\xspace}

\def\cms    {\mbox{CMS}\xspace}

\def\lhc    {\mbox{LHC}\xspace}

\def\tevatron {Tevatron\xspace}

\def\MagUp {\mbox{\em Mag\kern -0.05em Up}\xspace}

\ifthenelse{\boolean{uprightparticles}}%
{

 \def\Pmu         {\ensuremath{\upmu}\xspace}

 \def\Pchi        {\ensuremath{\upchi}\xspace}                 
 \def\Ppsi        {\ensuremath{\uppsi}\xspace}

 \def\PDelta      {\ensuremath{\Delta}\xspace}                 
 \def\PXi         {\ensuremath{\Xi}\xspace}                 
 \def\PLambda     {\ensuremath{\Lambda}\xspace}                 
 \def\PSigma      {\ensuremath{\Sigma}\xspace}                 
 \def\POmega      {\ensuremath{\Omega}\xspace}                 
 \def\PUpsilon    {\ensuremath{\Upsilon}\xspace}

 \def\PB      {\ensuremath{\mathrm{B}}\xspace}                 
                  
 \def\PD      {\ensuremath{\mathrm{D}}\xspace}

 \def\PJ      {\ensuremath{\mathrm{J}}\xspace}                 
 \def\PK      {\ensuremath{\mathrm{K}}\xspace}

 \def\Pb      {\ensuremath{\mathrm{b}}\xspace}                 
 \def\Pc      {\ensuremath{\mathrm{c}}\xspace}

 \def\Pi      {\ensuremath{\mathrm{i}}\xspace}

 \def\Pq      {\ensuremath{\mathrm{q}}\xspace}                 
                  
 \def\Ps      {\ensuremath{\mathrm{s}}\xspace}

 \def\thebaroffset{0.0em}
}
{

 \def\Pmu         {\ensuremath{\mu}\xspace}

 \def\Pchi        {\ensuremath{\chi}\xspace}                 
 \def\Ppsi        {\ensuremath{\psi}\xspace}                 
                  
 \mathchardef\PDelta="7101
 \mathchardef\PXi="7104
 \mathchardef\PLambda="7103
 \mathchardef\PSigma="7106
 \mathchardef\POmega="710A
 \mathchardef\PUpsilon="7107
                  
 \def\PB      {\ensuremath{B}\xspace}                 
                  
 \def\PD      {\ensuremath{D}\xspace}

 \def\PJ      {\ensuremath{J}\xspace}                 
 \def\PK      {\ensuremath{K}\xspace}

 \def\Pb      {\ensuremath{b}\xspace}                 
 \def\Pc      {\ensuremath{c}\xspace}

 \def\Pi      {\ensuremath{i}\xspace}

 \def\Pq      {\ensuremath{q}\xspace}                 
                  
 \def\Ps      {\ensuremath{s}\xspace}

 \def\thebaroffset{0.18em}
}
\newcommand{\offsetoverline}[2][\thebaroffset]{\kern #1\overline{\kern -#1 #2}}%

\makeatletter
\ifcase \@ptsize \relax% 10pt
  \newcommand{\miniscule}{\@setfontsize\miniscule{4}{5}}% \tiny: 5/6
\or% 11pt
  \newcommand{\miniscule}{\@setfontsize\miniscule{5}{6}}% \tiny: 6/7
\or% 12pt
  \newcommand{\miniscule}{\@setfontsize\miniscule{5}{6}}% \tiny: 6/7
\fi
\makeatother

\DeclareRobustCommand{\optbar}[1]{\shortstack{{\miniscule (\rule[.5ex]{1.25em}{.18mm})}
  \\ [-.7ex] $#1$}}

\def\mup        {{\ensuremath{\Pmu^+}}\xspace}
\def\mun        {{\ensuremath{\Pmu^-}}\xspace} % muon negative (\mum is taken)

\def\mumu       {{\ensuremath{\Pmu^+\Pmu^-}}\xspace}

\def\quark     {{\ensuremath{\Pq}}\xspace}
\def\quarkbar  {{\ensuremath{\overline \quark}}\xspace}

\def\squark    {{\ensuremath{\Ps}}\xspace}

\def\cquark    {{\ensuremath{\Pc}}\xspace}
\def\cquarkbar {{\ensuremath{\overline \cquark}}\xspace}

\def\bquark    {{\ensuremath{\Pb}}\xspace}
\def\bquarkbar {{\ensuremath{\overline \bquark}}\xspace}

\def\KorKbar {\kern \thebaroffset\optbar{\kern -\thebaroffset \PK}{}\xspace}

\def\D       {{\ensuremath{\PD}}\xspace}

\def\DorDbar {\kern \thebaroffset\optbar{\kern -\thebaroffset \PD}\xspace}

\def\Dp      {{\ensuremath{\D^+}}\xspace}
\def\Dm      {{\ensuremath{\D^-}}\xspace}

\def\DpDm    {\ensuremath{\Dp {\kern -0.16em \Dm}}\xspace}

\def\B       {{\ensuremath{\PB}}\xspace}

\def\BorBbar {\kern \thebaroffset\optbar{\kern -\thebaroffset \PB}\xspace}

\def\Bd      {{\ensuremath{\B^0}}\xspace}

\def\BdorBdbar {\kern \thebaroffset\optbar{\kern -\thebaroffset \Bd}\xspace}

\def\Bs      {{\ensuremath{\B^0_\squark}}\xspace}

\def\BsorBsbar {\kern \thebaroffset\optbar{\kern -\thebaroffset \Bs}\xspace}

\def\jpsi     {{\ensuremath{{\PJ\mskip -3mu/\mskip -2mu\Ppsi}}}\xspace}

\def\chic     {{\ensuremath{\Pchi_\cquark}}\xspace}
\def\chiczero {{\ensuremath{\Pchi_{\cquark 0}}}\xspace}
\def\chicone  {{\ensuremath{\Pchi_{\cquark 1}}}\xspace}

\def\Upsilonres  {{\ensuremath{\PUpsilon}}\xspace}
\def\Y#1S{\ensuremath{\PUpsilon{(#1S)}}\xspace}

\def\LorLbar     {\kern \thebaroffset\optbar{\kern -\thebaroffset \PLambda}\xspace}

\def\Xires       {{\ensuremath{\PXi}}\xspace}

\def\Xiccpp      {{\ensuremath{\Xires^{++}_{\cquark\cquark}}}\xspace}

\def\BF         {{\ensuremath{\mathcal{B}}}\xspace}

\def\to                 {\ensuremath{\rightarrow}\xspace}

\def\AT#1     {\ensuremath{A_{\mathrm{T}}^{#1}}\xspace}           % 2

\def\C#1      {\ensuremath{\mathcal{C}_{#1}}\xspace}                       % 9
\def\Cp#1     {\ensuremath{\mathcal{C}_{#1}^{'}}\xspace}                    % 7
\def\Ceff#1   {\ensuremath{\mathcal{C}_{#1}^{\mathrm{(eff)}}}\xspace}        % 9  
\def\Cpeff#1  {\ensuremath{\mathcal{C}_{#1}^{'\mathrm{(eff)}}}\xspace}       % 7
\def\Ope#1    {\ensuremath{\mathcal{O}_{#1}}\xspace}                       % 2
\def\Opep#1   {\ensuremath{\mathcal{O}_{#1}^{'}}\xspace}                    % 7

\newcommand{\aunit}[1]{\ensuremath{\text{\,#1}}}       
\newcommand{\tev}{\aunit{Te\kern -0.1em V}\xspace}
\newcommand{\gev}{\aunit{Ge\kern -0.1em V}\xspace}
\newcommand{\mev}{\aunit{Me\kern -0.1em V}\xspace}
\newcommand{\kev}{\aunit{ke\kern -0.1em V}\xspace}
\newcommand{\ev}{\aunit{e\kern -0.1em V}\xspace}
 
\newcommand{\mevc}{\ensuremath{\aunit{Me\kern -0.1em V\!/}c}\xspace}
\newcommand{\gevc}{\ensuremath{\aunit{Ge\kern -0.1em V\!/}c}\xspace}
\newcommand{\mevcc}{\ensuremath{\aunit{Me\kern -0.1em V\!/}c^2}\xspace}
\newcommand{\gevcc}{\ensuremath{\aunit{Ge\kern -0.1em V\!/}c^2}\xspace}
\def\fb   {\ensuremath{\aunit{fb}}\xspace}
\def\invfb   {\ensuremath{\fb^{-1}}\xspace}

\newcommand{\stat}{\aunit{(stat)}\xspace}
\newcommand{\syst}{\aunit{(syst)}\xspace}

\newcommand{\chisq}{\ensuremath{\chi^2}\xspace}
\newcommand{\chisqndf}{\ensuremath{\chi^2/\mathrm{ndf}}\xspace}

\def\gsim{{~\raise.15em\hbox{$>$}\kern-.85em
          \lower.35em\hbox{$\sim$}~}\xspace}
\def\lsim{{~\raise.15em\hbox{$<$}\kern-.85em
          \lower.35em\hbox{$\sim$}~}\xspace}

\def\sPlot{\mbox{\em sPlot}\xspace}

\def\sqs   {\ensuremath{\protect\sqrt{s}}\xspace}

\def\pt         {\ensuremath{p_{\mathrm{T}}}\xspace}

\def\evtgen     {\mbox{\textsc{EvtGen}}\xspace}

\def\geant      {\mbox{\textsc{Geant4}}\xspace}

\def\photos     {\mbox{\textsc{Photos}}\xspace}

\def\pythia     {\mbox{\textsc{Pythia}}\xspace}

\def\tell1  {TELL1\xspace}
\def\ukl1   {UKL1\xspace}

\newcommand{\ie}{\mbox{\itshape i.e.}\xspace}

%% file: mydefine.tex
\newcommand{\xx}{\ensuremath{\kern 0.5em }}

\def\jpsione  {{\ensuremath{{\PJ\mskip -3mu/\mskip -2mu\Ppsi_1\mskip 2mu}}}\xspace}
\def\jpsitwo  {{\ensuremath{{\PJ\mskip -3mu/\mskip -2mu\Ppsi_2\mskip 2mu}}}\xspace}
\def\diM      {{\ensuremath{M_{{\rm di}{\text -}\PJ\mskip -3mu/\mskip -2mu\Ppsi}}}\xspace}
\def\dipt      {{\ensuremath{p^{{\rm di}{\text -}\PJ\mskip -3mu/\mskip -2mu\Ppsi}_{\rm T}}}\xspace}

\def\bbbb  {{\ensuremath{T_{\bquark\bquark\bquarkbar\bquarkbar}}}\xspace} 
\def\cccc  {{\ensuremath{T_{\cquark\cquark\cquarkbar\cquarkbar}}}\xspace}

\usepackage{multirow} % for complicated table
\usepackage{booktabs} % for complicated table
\usepackage{rotating}

%% file: title-LHCb-PAPER.tex
%%%%%%%%%%%%%%%%%%%%%%%%%
%%%%%  TITLE PAGE  %%%%%%
%%%%%%%%%%%%%%%%%%%%%%%%%
\begin{titlepage}
\pagenumbering{roman}

% Header ---------------------------------------------------
\vspace*{-1.5cm}
\centerline{\large EUROPEAN ORGANIZATION FOR NUCLEAR RESEARCH (CERN)}
\vspace*{1.5cm}
\noindent
\begin{tabular*}{\linewidth}{lc@{\extracolsep{\fill}}r@{\extracolsep{0pt}}}
\ifthenelse{\boolean{pdflatex}}% Logo format choice
{\vspace*{-1.5cm}\mbox{\!\!\!\includegraphics[width=.14\textwidth]{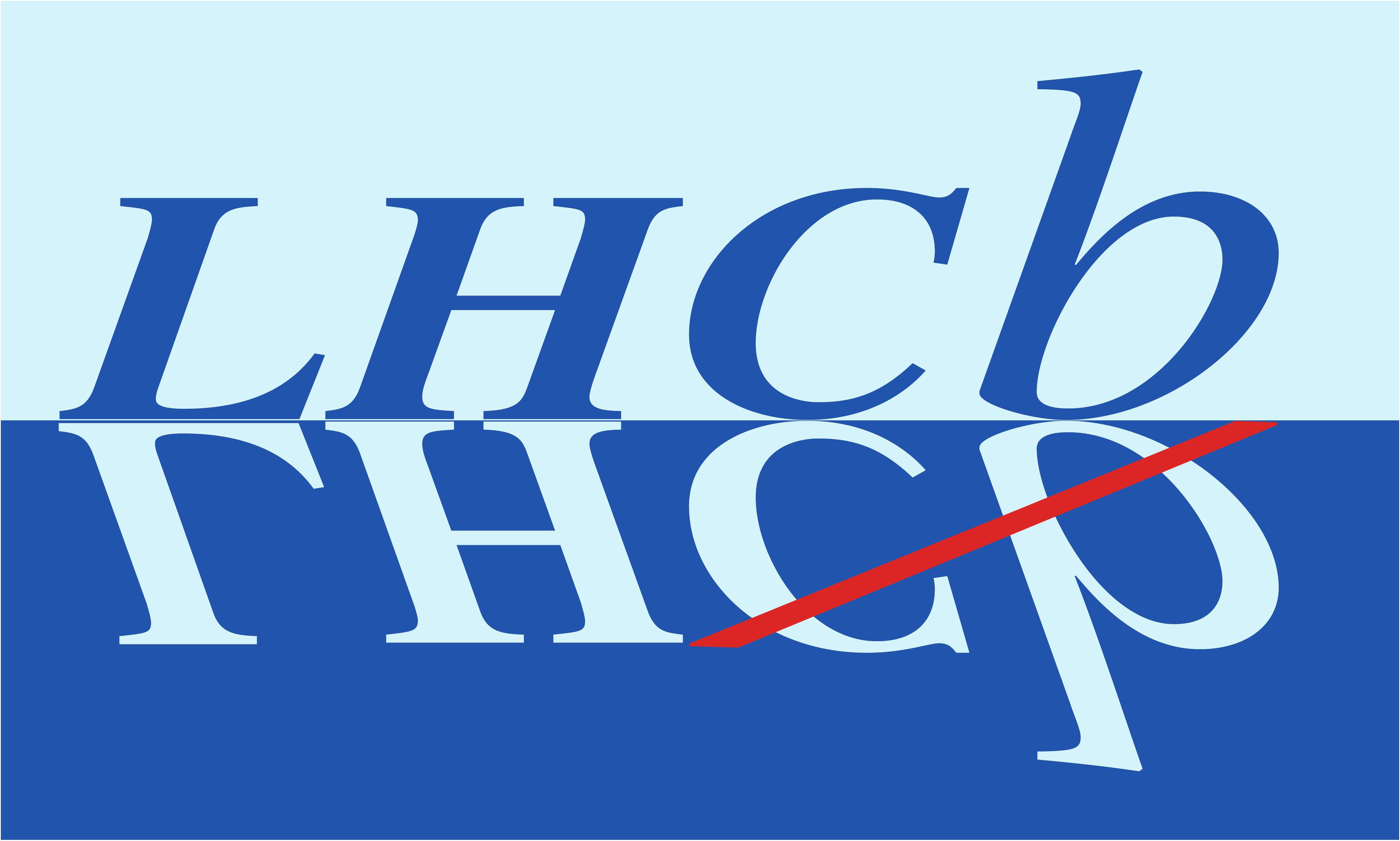}} & &}%
{\vspace*{-1.2cm}\mbox{\!\!\!\includegraphics[width=.12\textwidth]{figs/lhcb-logo.eps}} & &}%
\\
 & & CERN-EP-2020-115 \\  % ID 
 & & LHCb-PAPER-2020-011 \\  % ID 
 & & November 10, 2020 \\ % Date - Can also hardwire e.g.: 23 March 2010
 & & \\
% not in paper \hline
\end{tabular*}

\vspace*{4.0cm}

% Title --------------------------------------------------
{\normalfont\bfseries\boldmath\huge
\begin{center}
  \papertitle 
\end{center}
}

\vspace*{2.0cm}

% Authors -------------------------------------------------
\begin{center}
\paperauthors\footnote{Authors are listed at the end of this paper.}
\end{center}

\vspace{\fill}

% Abstract -----------------------------------------------
\begin{abstract}
  \noindent
   Using proton-proton collision data at centre-of-mass energies of $\sqs = 7$, $8$ and $13\tev$
   recorded by the LHCb experiment at the Large Hadron Collider,
   corresponding to an integrated luminosity of $9\invfb$,
   the invariant mass spectrum of $\jpsi$ pairs is studied.
   A narrow structure around $6.9\gevcc$ matching the lineshape of a resonance
   and a broad structure just above twice the $\jpsi$ mass are observed.
   The deviation of the data from nonresonant \jpsi-pair production 
   is above five standard deviations in the mass region between $6.2$ and $7.4\gevcc$, covering predicted masses of states composed of four charm quarks.
   The mass and natural width of the narrow $X(6900)$ structure are measured
   assuming a Breit--Wigner lineshape.

\end{abstract}

\vspace*{0.5cm}


\vspace*{1.0cm}

\begin{center}
   Published in Science Bulletin 2020 65(23)1983-1993
\end{center}

\vspace{\fill}

{\footnotesize 
\centerline{\copyright~\papercopyright. \href{\paperlicenceurl}{\paperlicence}.}}
\vspace*{2mm}

\end{titlepage}

%%%%%%%%%%%%%%%%%%%%%%%%%%%%%%%%
%%%%%  EOD OF TITLE PAGE  %%%%%%
%%%%%%%%%%%%%%%%%%%%%%%%%%%%%%%%

%  empty page follows the title page ----
\newpage
\setcounter{page}{2}
\mbox{~}
%\newpage
%

%% file: body.tex
\section{Introduction}
%%%%% Introduction
The strong interaction is one of the fundamental forces of nature and it governs the dynamics of quarks and gluons. According to quantum chromodynamics (QCD), the theory describing the strong interaction, quarks are confined into hadrons, in agreement with experimental observations.
The quark model~\cite{GellMann:1964nj,Zweig:352337} classifies hadrons into conventional mesons~($\quark\quarkbar$) and baryons~($\quark\quark\quark$ or $\quarkbar\quarkbar\quarkbar$), and also allows for the existence of
exotic hadrons such as tetraquarks~($\quark\quark\quarkbar\quarkbar$) and pentaquarks~($\quark\quark\quark\quarkbar\quarkbar$). Exotic states provide a unique environment to study the strong interaction and the confinement mechanism~\cite{Brambilla:2014jmp}. 
The first experimental evidence for an exotic hadron candidate was the $\chi_{c1}(3872)$ state observed in 2003 by the Belle collaboration~\cite{X3872}. Since then
a series of novel states consistent with 
containing four quarks have been
discovered~\cite{Olsen:2017bmm}. 
Recently, the \lhcb collaboration observed resonances interpreted to be 
pentaquark states~\cite{LHCb-PAPER-2015-029,LHCb-PAPER-2016-009,LHCb-PAPER-2016-015,LHCb-PAPER-2019-014}. 
All hadrons observed to date, including those of exotic nature, 
contain at most two heavy charm~($\cquark$) or bottom~($\bquark$) quarks, whereas many QCD-motivated phenomenological models also predict
the existence of states consisting of four heavy quarks,
\ie $T_{Q_1 Q_2 \overline{Q}_3 \overline{Q}_4}$, where $Q_i$ is a \cquark or a \bquark quark~\cite{Iwasaki:1976cn,Chao:1980dv,PhysRevD.25.2370,PhysRevD.29.426,Badalian:1985es,
Berezhnoy:2011xn,Wu:2016vtq,Karliner:2016zzc,
Barnea:2006sd,Debastiani:2017msn,Liu:2019zuc,
Chen:2016jxd,
Wang:2019rdo,Bedolla:2019zwg,Lloyd:2003yc,
Chen:2020lgj,Wang:2018poa,
Anwar:2017toa,Liu:2019zoy,Esposito:2018cwh,
Becchi:2020mjz,Bai:2016int,Richard:2017vry,Vega-Morales:2017pmm,Chen:2019vrj,Berezhnoy:2012tu}.
Theoretically, the interpretation of the internal structure of such states usually assumes the formation of a $Q_1 Q_2$ diquark and a $\overline{Q}_3\overline{Q}_4$ antidiquark attracting each other.
Application of this diquark model successfully predicts
the mass of the doubly charmed 
baryon $\Xiccpp$ \cite{Karliner:2014gca, LHCb-PAPER-2017-018} and helps to explain the relative rates of bottom baryon decays~\cite{LHCb-PAPER-2019-039}.

Tetraquark states comprising only bottom quarks, $\bbbb$, have been searched for by the \lhcb and \cms collaborations
in the $\PUpsilon\mumu$ decay~\cite{LHCb-PAPER-2018-027,Sirunyan:2020txn}, with the $\PUpsilon$ state consisting of a $\bquark \bquarkbar$ pair. However, the 
four-charm states, $\cccc$, have not yet been
studied in detail experimentally.
A $\cccc$ state could disintegrate into a pair of charmonium states such as $\jpsi$ mesons, with each consisting of a $\cquark \cquarkbar$ pair. Decays to
a \jpsi meson plus a heavier charmonium state, or two heavier charmonium states, with the heavier states decaying subsequently into a \jpsi meson and accompanying particles, are also possible.
Predictions for the masses of $\cccc$ states vary from $5.8$ to $7.4\gevcc$
~\cite{Iwasaki:1976cn,Chao:1980dv,PhysRevD.25.2370,PhysRevD.29.426,Badalian:1985es,Berezhnoy:2011xn,Wu:2016vtq,Karliner:2016zzc,Barnea:2006sd,Debastiani:2017msn,Liu:2019zuc,Chen:2016jxd,Wang:2019rdo,Bedolla:2019zwg,Lloyd:2003yc,Chen:2020lgj,Wang:2018poa},
which are above the masses of 
 known charmonia and charmonium-like exotic states and below those of bottomonium hadrons.~\footnote{Energy units $\mev=10^{6}\ev$, $\gev=10^{9}\ev$ and $\tev=10^{12}\ev$ are used in this paper.}
This mass range guarantees a clean experimental environment to identify possible $\cccc$ states
in the $\jpsi$-pair (also referred to as di-$\jpsi$) invariant mass~($\diM$) spectrum.

In proton-proton ($pp$) collisions, a pair of $\jpsi$ mesons can be produced in two separate interactions of gluons or quarks, named double-parton scattering\,(DPS)~\cite{Calucci:1997ii,Calucci:1999yz,DelFabbro:2000ds}, or in a single interaction, named single-parton scattering\,(SPS)~\cite{Sun:2014gca,Likhoded:2016zmk,Shao:2012iz,Shao:2015vga,Baranov:2011zz,Lansberg:2013qka,Lansberg:2014swa,Lansberg:2015lva}. The SPS process
includes both resonant production via intermediate states, which could be $\cccc$ tetraquarks, and nonresonant production.
Within the DPS process,
the two $\jpsi$ mesons are usually assumed to be produced independently,
thus the distribution of any di-$\jpsi$ observable
can be constructed using the kinematics from single $\jpsi$ production.
Evidence of DPS has been found in  
studies at the Large Hadron Collider (\lhc) experiments~\cite{ 
W2jets,CMSDJ,DUpsilon,JpsiW,JpsiZ,Aaboud:2016dea,ATLASDJ,LHCb-PAPER-2012-003,LHCb-PAPER-2013-062,LHCb-PAPER-2015-046} and the AFS experiment~\cite{4jets1} in $pp$ collisions,
and at the \tevatron experiments~\cite{Abe:1993rv,Gamma3jets1,Gamma3jets2,Abazov:2014fha,Abazov:2015nnn} and the UA2 experiment~\cite{4jets2} in proton-antiproton collisions.
The \lhcb experiment has measured
the di-$\jpsi$ production in $pp$ collisions 
at centre-of-mass energies of $\sqs=7$~\cite{LHCb-PAPER-2011-013} and $13\tev$~\cite{LHCb-PAPER-2016-057}.
The DPS contribution is found to dominate the high
$\diM$ region, in agreement with expectation.

In this paper, fully charmed
tetraquark states $\cccc$ are searched for in the di-$\jpsi$ invariant mass spectrum,
using $pp$ collision data collected by \lhcb
at $\sqs=7, 8$
and $13\tev$,
corresponding to an integrated luminosity of $9\invfb$.
The two $\jpsi$ candidates in a pair are reconstructed through the $\jpsi\to\mup\mun$ decay, and are labelled randomly as either $\jpsione$ or $\jpsitwo$.

\section{Detector and data set}
%%%%% Detector
The LHCb detector is designed to study particles containing \bquark\ or \cquark\ quarks
at the \lhc.
It is a single-arm forward spectrometer covering the pseudorapidity range $2 < \eta < 5$, 
described in detail in Refs.~\cite{LHCb-DP-2008-001,LHCb-DP-2014-002}.
The online event selection is performed by a trigger, 
which consists of a hardware stage, based on information from the calorimeter and muon
systems, followed by a software stage, which applies a full event
reconstruction.
At the hardware stage, 
events are required to have at least one muon with high momentum transverse to the beamline, $\pt$. 
At the software stage, 
two oppositely charged muon candidates
are required to have high \pt and to form a common vertex.
Events are retained if there is at least one $\jpsi$ candidate passing both the hardware and software trigger requirements.
Imperfections in the description of the
magnetic field and misalignment of subdetectors lead to a bias in the momentum
measurement of charged particles, which is calibrated using 
reconstructed $\jpsi$ and $B^+$ mesons~\cite{LHCb-PAPER-2011-035},
with well-known masses.

Simulated $\jpsi\to\mup\mun$ decays are used 
to study the signal properties, including the invariant mass resolution and the reconstruction efficiency. 
In the simulation, $pp$ collisions are generated using
\pythia~\cite{Sjostrand:2007gs} 
 with a specific \lhcb
configuration~\cite{LHCb-PROC-2010-056}.  
Decays of unstable particles
are described by \evtgen~\cite{Lange:2001uf}, in which final-state
radiation is generated using \photos~\cite{Golonka:2005pn}. 
The interaction of the generated particles with the detector and its response
are implemented using the \geant
toolkit~\cite{Allison:2006ve}, as described in
Ref.~\cite{LHCb-PROC-2011-006}.

\section{Candidate selection}
%%%%% Selection
In the offline selection, 
two pairs of oppositely charged muon candidate tracks
are reconstructed,
with each pair forming a vertex of a $\jpsi$ candidate.
Each muon track must have \mbox{$\pt>0.65\gevc$} and momentum \mbox{$p>6\gevc$}.
The $\jpsi$ candidates are required to have a dimuon invariant mass in the range $3.0<M_{\mu\mu}<3.2\gevcc$.
A kinematic fit is performed for each \jpsi candidate constraining its vertex to coincide with a primary  $pp$ collision vertex (PV)~\cite{Hulsbergen:2005pu}.
The requirement of a good kinematic fit quality strongly suppresses
the contamination of di-$\jpsi$ candidates stemming from feed-down of $\bquark$-hadrons, 
which decay at displaced vertices.
The four muon tracks in  a $\jpsi$-pair candidate are required to originate from the same PV, reducing to a negligible level the number of pile-up candidates with the two $\jpsi$ candidates produced in 
separated $pp$ collisions. 
Fake di-$\jpsi$ candidates, comprising two muon-track candidates reconstructed from the same real particle, are rejected by requiring muons of the same charge to have trajectories separated by an angle inconsistent with zero.
For events with more than one reconstructed di-$\jpsi$ candidate, accounting for about 0.8\% of the total sample,
only one pair is randomly chosen.

\begin{figure}[t]
\begin{center}
\includegraphics[width=1.0\linewidth]{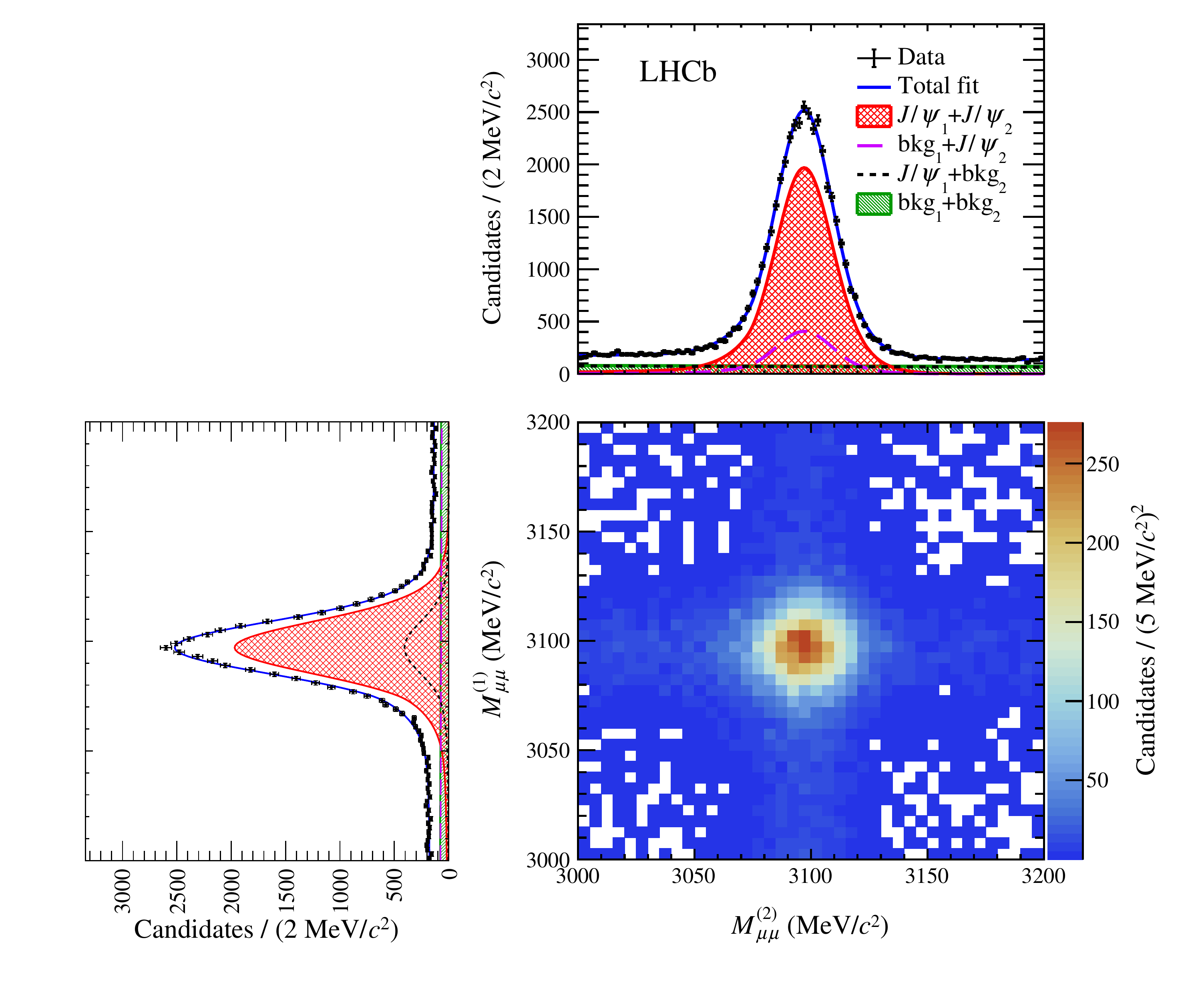}
\vspace*{-1.cm}
\end{center}
   \caption{\small
	(Bottom right) Two-dimensional \mbox{$(M^{(1)}_{\mu\mu},M^{(2)}_{\mu\mu})$} distribution
   of di-$\jpsi$ candidates
   and its projections on (bottom left)~$M^{(1)}_{\mu\mu}$ and (top)~$M^{(2)}_{\mu\mu}$. Four components are present as each projection consists of signal and background $\jpsi$ candidates. The labels $\jpsi_{1,2}$ and bkg${}_{1,2}$ represent the signal and background contributions, respectively, in the $M^{(1),(2)}_{\mu\mu}$ distribution.
   }\label{fig:2DM}
\end{figure}

The di-$\jpsi$ signal yield is extracted by
performing an extended unbinned maximum-likelihood fit to
the two\nobreakdash-dimensional distribution of $\jpsione$ and $\jpsitwo$ invariant masses, \mbox{$(M^{(1)}_{\mu\mu},M^{(2)}_{\mu\mu})$},
as displayed in Fig.~\ref{fig:2DM},
where projections of the data and the fit result are shown.
For both $\jpsi$ candidates,
the signal mass shape is modelled by a Gaussian kernel with power-law tails~\cite{Skwarnicki:1986xj}.
Each component of combinatorial background,
consisting of random combinations of muon tracks, is described by an exponential function.
The total di-$\jpsi$ signal yield is measured to be \mbox{$(33.57\pm0.23)\times10^3$}, where the uncertainty is statistical.

The di-$\jpsi$ transverse momentum ($\dipt$) in SPS production is expected to be, on average, higher than that in DPS~\cite{Lansberg:2014swa}. 
The high-$\dipt$ region is thus exploited to select a data sample with enhanced SPS production, which could include contributions from $\cccc$ states. Two different approaches are applied.
In the first approach~(denoted as $\dipt$-threshold), $\jpsi$-pair candidates are selected with the requirement  $\dipt>5.2\gevc$,
which maximises the statistical significance of the SPS signal yield, $N_{\rm SPS}/\sqrt{N_{\rm total}}$.
$N_{\rm SPS}$ and $N_{\rm total}$ are yields of the SPS component and total di-$\jpsi$ candidates in the $\diM$ range between $6.2$ and $7.4\gevcc$, respectively. This mass region covers the predicted masses of  $\cccc$ states decaying into a $\jpsi$ pair.
In the second approach~(denoted as $\dipt$-binned), di-$\jpsi$ candidates are categorised  into six $\dipt$ intervals with boundaries $\left\{ 0, 5, 6, 8, 9.5, 12, 50 \right\}\gevc$,
defined to obtain equally populated bins of SPS signal events in the $6.2<\diM<7.4\gevcc$ range. 
For both scenarios, the DPS yield in the $\cccc$ signal region is extrapolated from the high-$\diM$ region using the wide-range distribution
constructed from available double-differential \jpsi cross-sections~\cite{LHCb-PAPER-2011-003,LHCb-PAPER-2013-016,LHCb-PAPER-2015-037}
as performed in~\cite{LHCb-PAPER-2016-057}.
The high-$\diM$ region is chosen such that the SPS yield is negligible compared to DPS.
The SPS yield is obtained by subtracting the DPS contribution from the total number of $\jpsi$-pair signals.

\begin{figure}[htb!]
\begin{center}
\includegraphics[width=0.6\linewidth]{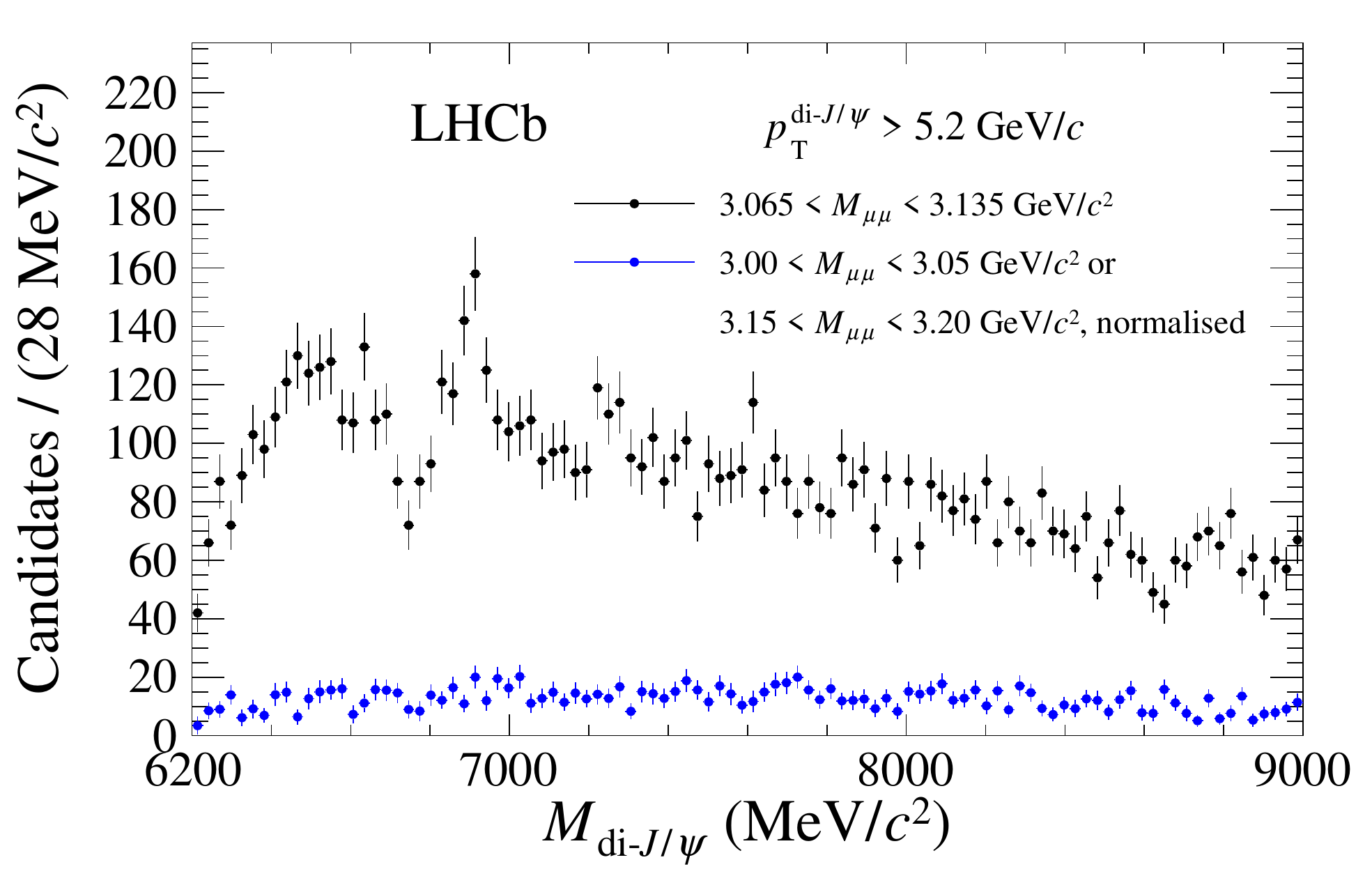}
\vspace*{-0.1cm}
\end{center}
   \caption{\small Invariant mass spectrum of $\jpsi$-pair  candidates passing the $\dipt>5.2\gevc$ requirement with reconstructed $\jpsi$ masses in the  (black)~signal and (blue)~background regions,
   respectively.
   }\label{fig:Xmass}
\end{figure}

The $\diM$ distribution for candidates with \mbox{$\dipt>5.2\gevc$} 
and \mbox{$3.065< M_{\mu\mu}^{(1),(2)}<3.135\gevcc$} is shown in
Fig.~\ref{fig:Xmass}. The di-$\jpsi$ mass is calculated by constraining the reconstructed mass of each $\jpsi$ candidate to its known value~\cite{PDG2019}.
The spectrum shows a broad structure just above twice the $\jpsi$ mass threshold 
ranging from $6.2$ to $6.8\gevcc$ (dubbed threshold enhancement in the following)
and a narrower structure at about $6.9\gevcc$, referred to hereafter as $X(6900)$.
There is also a hint of another structure around $7.2\gevcc$, 
whereas there are no evident structures at higher invariant mass.
Several cross-checks are performed to investigate the origin of these structures and to exclude that they are experimental artifacts.
The threshold enhancement and the $X(6900)$ structure become more pronounced in higher $\dipt$ intervals, and they are present
in subsamples split according to different beam or detector configurations for data collection.
The structures are not caused by the experimental  efficiency, since the efficiency variation across the whole $\diM$ range is found to be marginal. 
Residual background, in which a muon track is reused  or at least one \jpsi candidate is produced from a $\bquark$-hadron decay, is observed to have no structure.
The possible contribution of $\jpsi$ pairs from $\Upsilonres$ decays 
is estimated to be negligible and distributed uniformly in the $\diM$ distribution. In Fig.~\ref{fig:Xmass}, the $\diM$ distribution for background pairs with 
$M_{\mu\mu}^{(1),(2)}$ in the range $3.00-3.05\gevcc$ or $3.15-3.20\gevcc$
is also shown, with the yield
normalised by interpolating the background into the $\jpsi$ signal region,
which accounts for around 15\% of the total candidates.
There is no evidence of structures in the $\diM$ distribution of background candidates.

%%%%%%%%%%%%%%%%%%%%%%%%%%%%%%%%%%%
%%%%%%%%%%%%%%%%%%%%%%%%%%%%%%%%%%%

\section{Investigation of the $\jpsi$-pair invariant mass spectrum}
\begin{figure}[htb!]
\begin{center}
   \sidesubfloat[]{\includegraphics[width=0.48\linewidth]{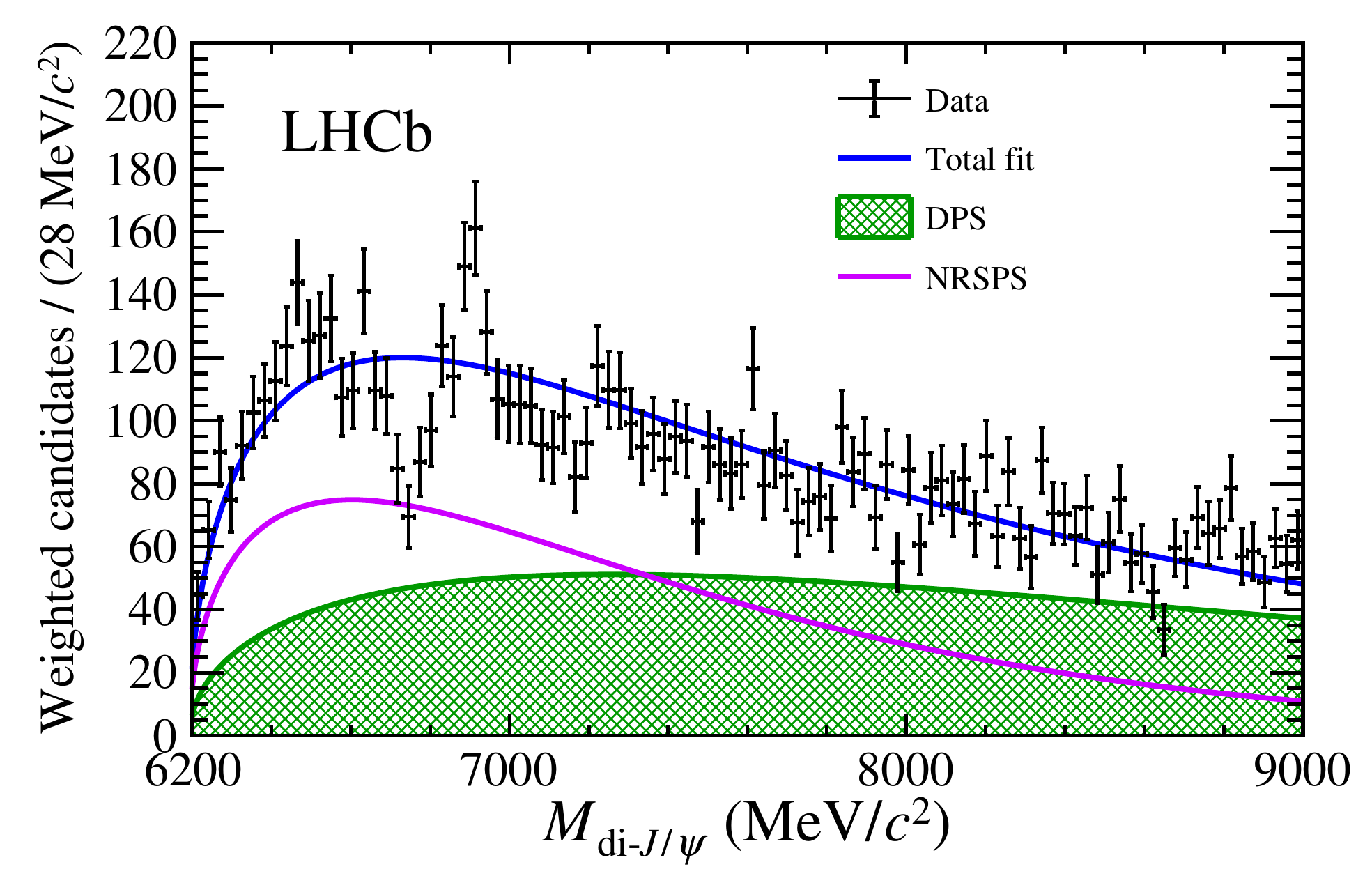}}\\
   \sidesubfloat[]{\includegraphics[width=0.48\linewidth]{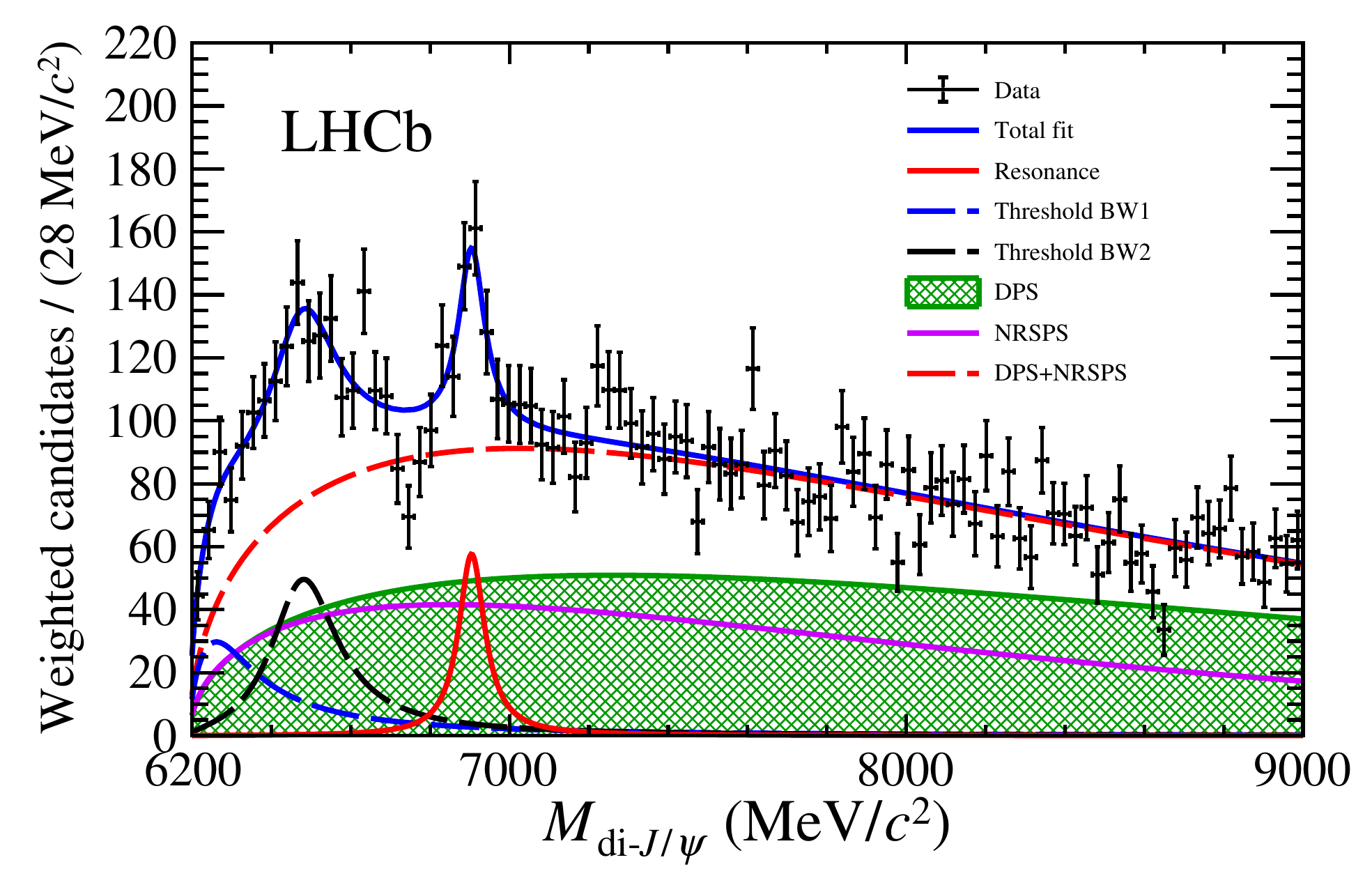}}
   \sidesubfloat[]{\includegraphics[width=0.48\linewidth]{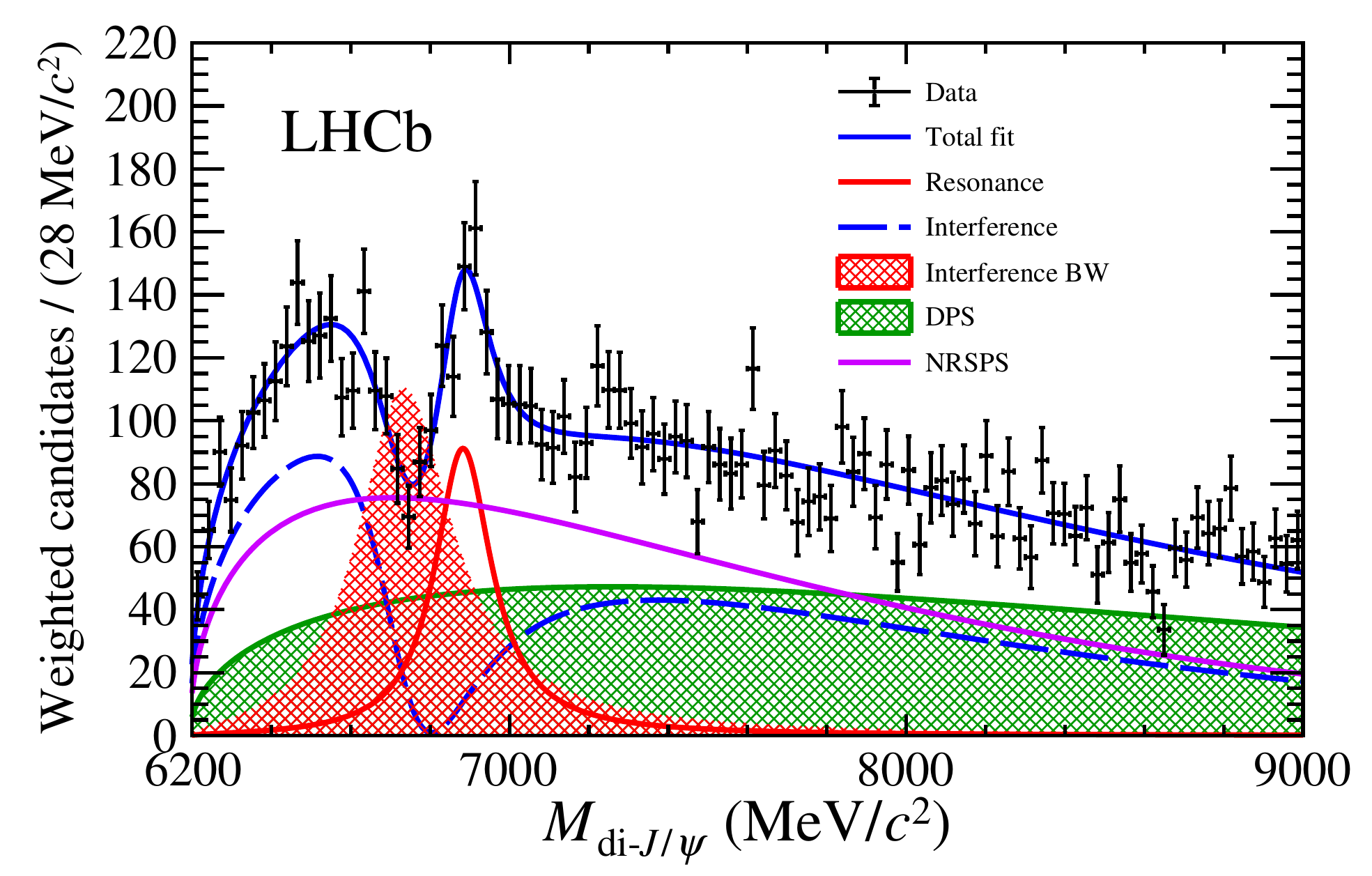}}
\vspace*{-0.1cm}
\end{center}
   \caption{\small Invariant mass spectra of weighted di-$\jpsi$ candidates with $\dipt>5.2\gevc$ and overlaid projections of the $\dipt$-threshold fit using (a)~the NRSPS plus DPS model, (b) model I,
   and (c) model II.
   }\label{fig:fitXcut}
\end{figure}

\begin{figure}[!tb]
\begin{center}
   \sidesubfloat[]{\includegraphics[width=0.45\linewidth]{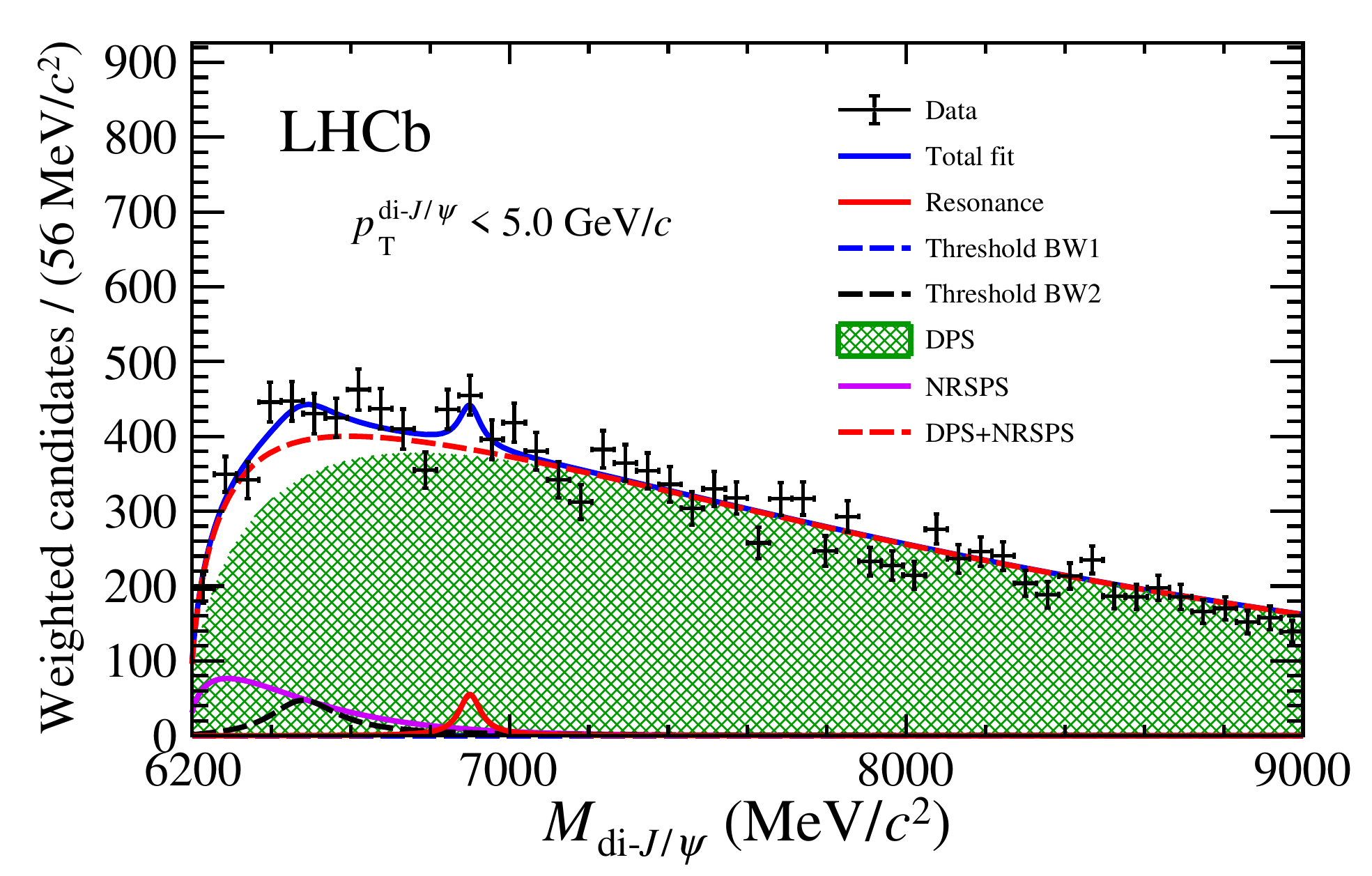}}
   \sidesubfloat[]{\includegraphics[width=0.45\linewidth]{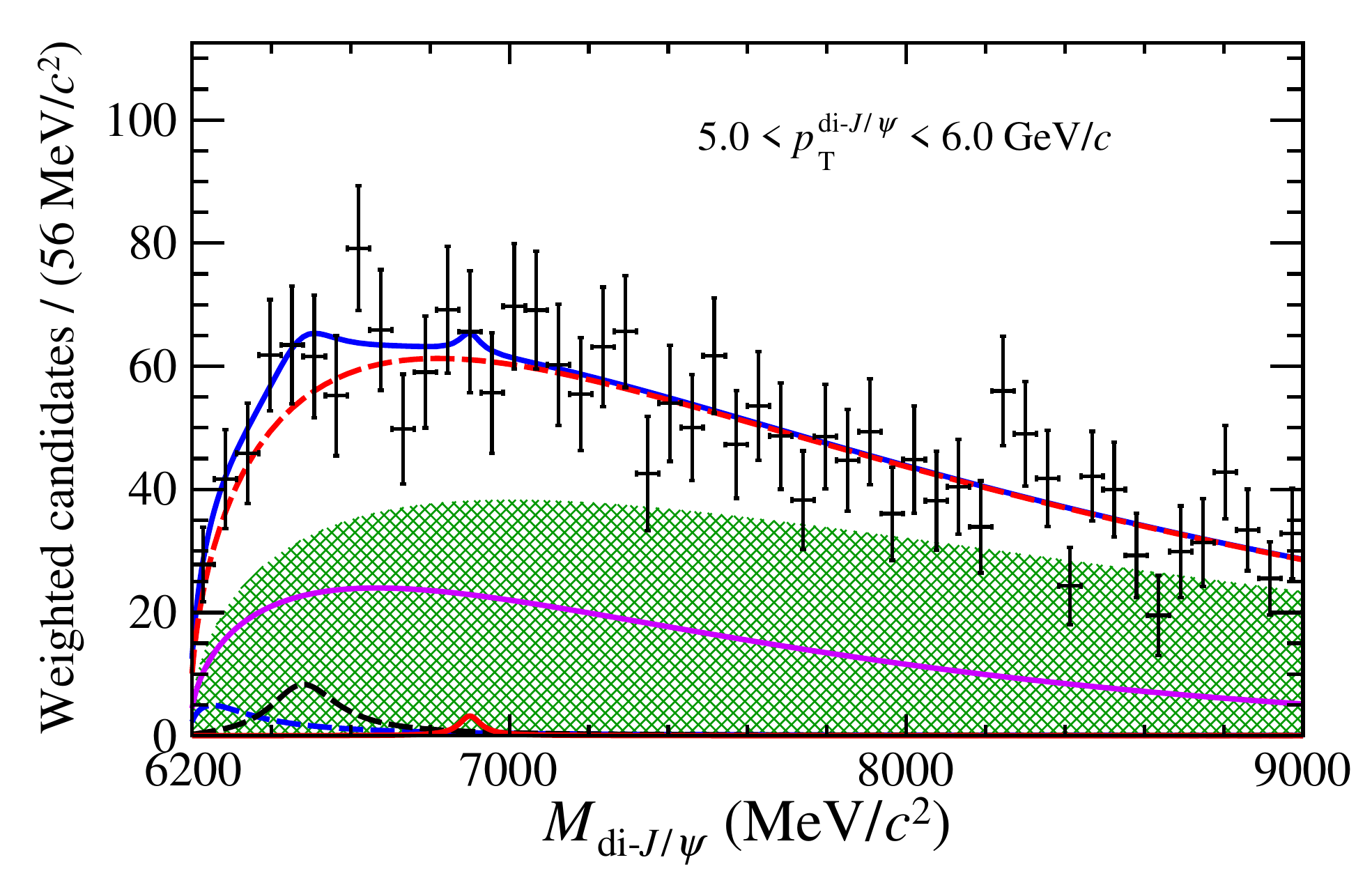}}\\
   \sidesubfloat[]{\includegraphics[width=0.45\linewidth]{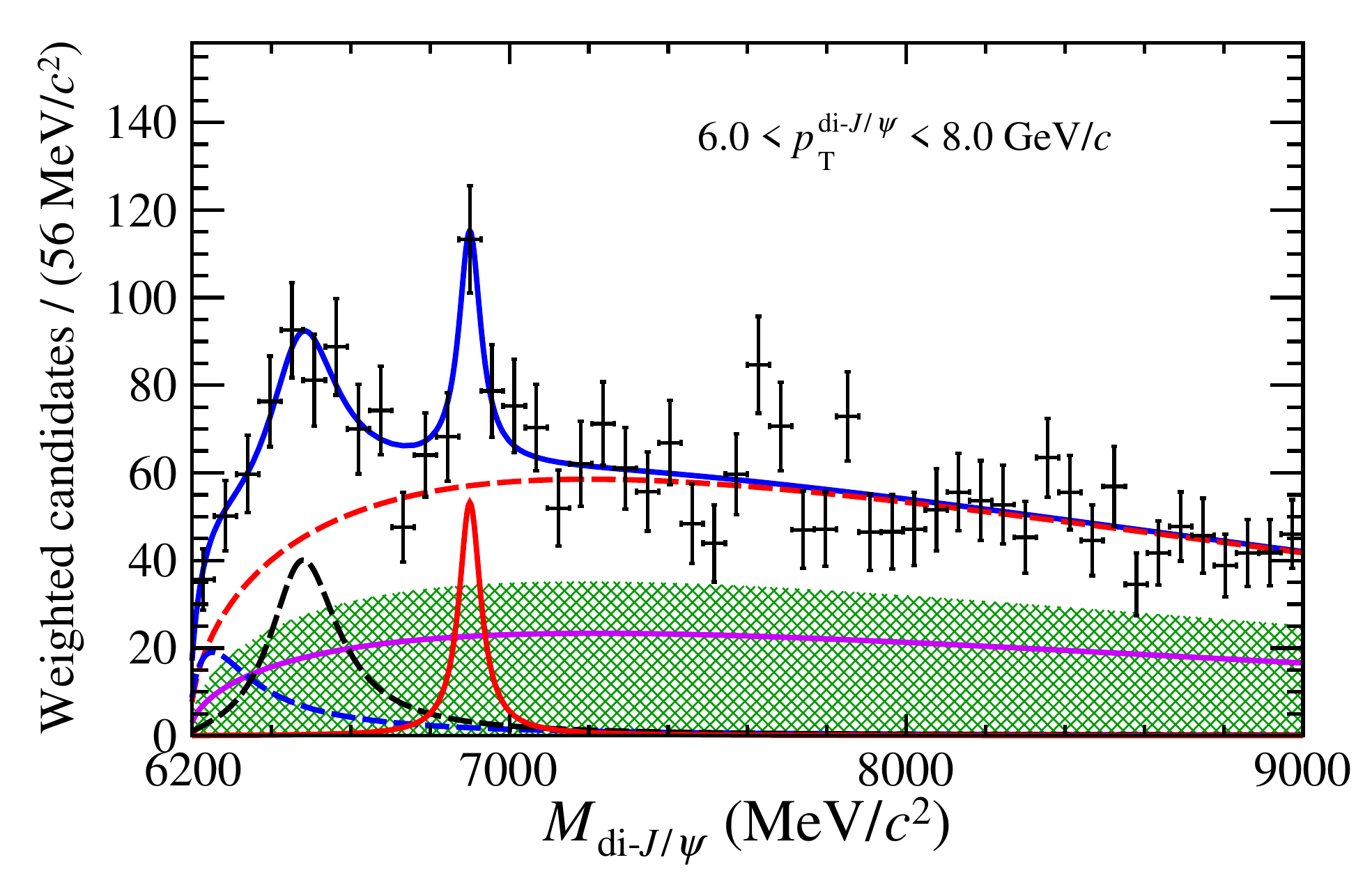}}
   \sidesubfloat[]{\includegraphics[width=0.45\linewidth]{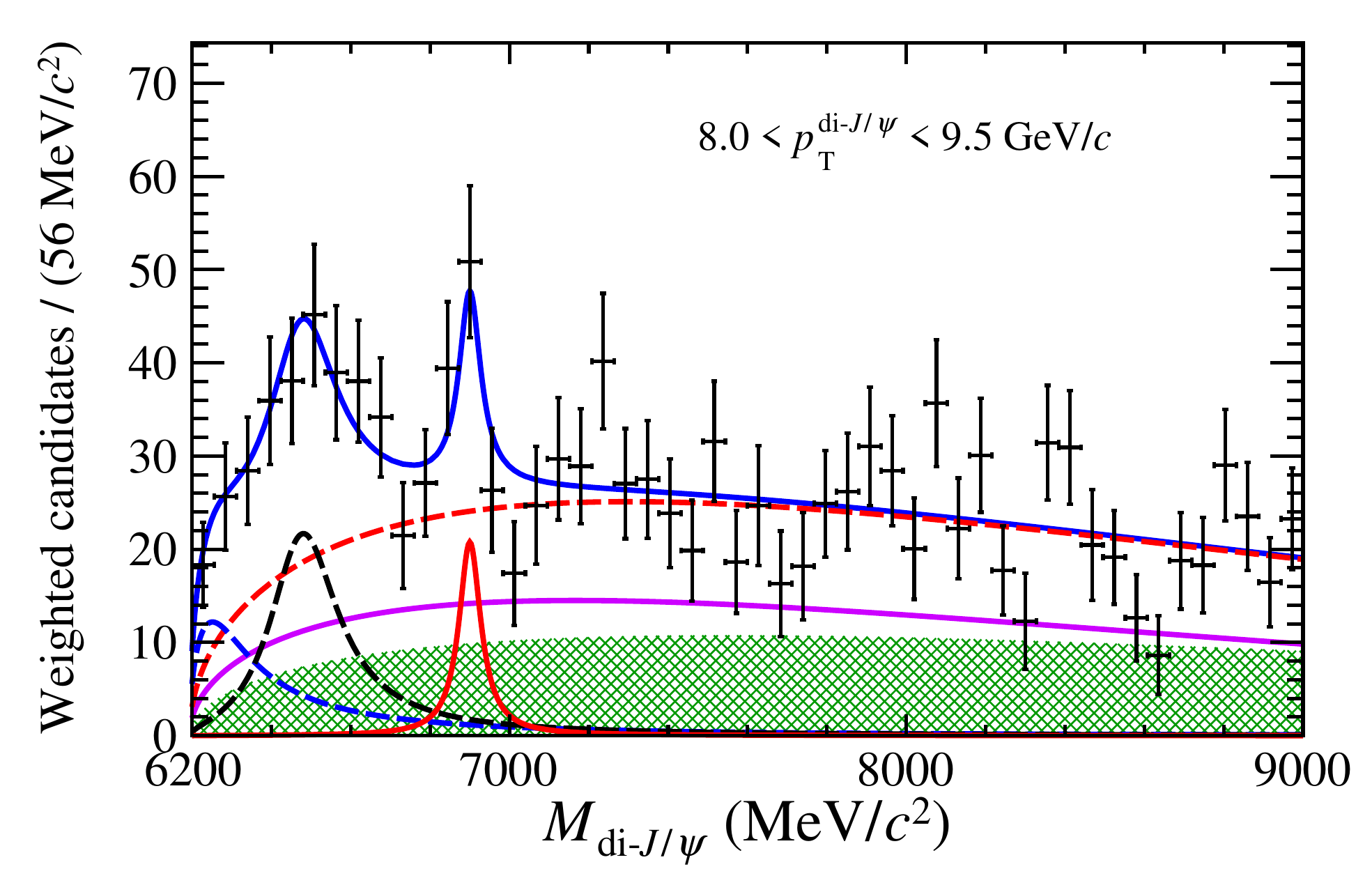}}\\
   \sidesubfloat[]{\includegraphics[width=0.45\linewidth]{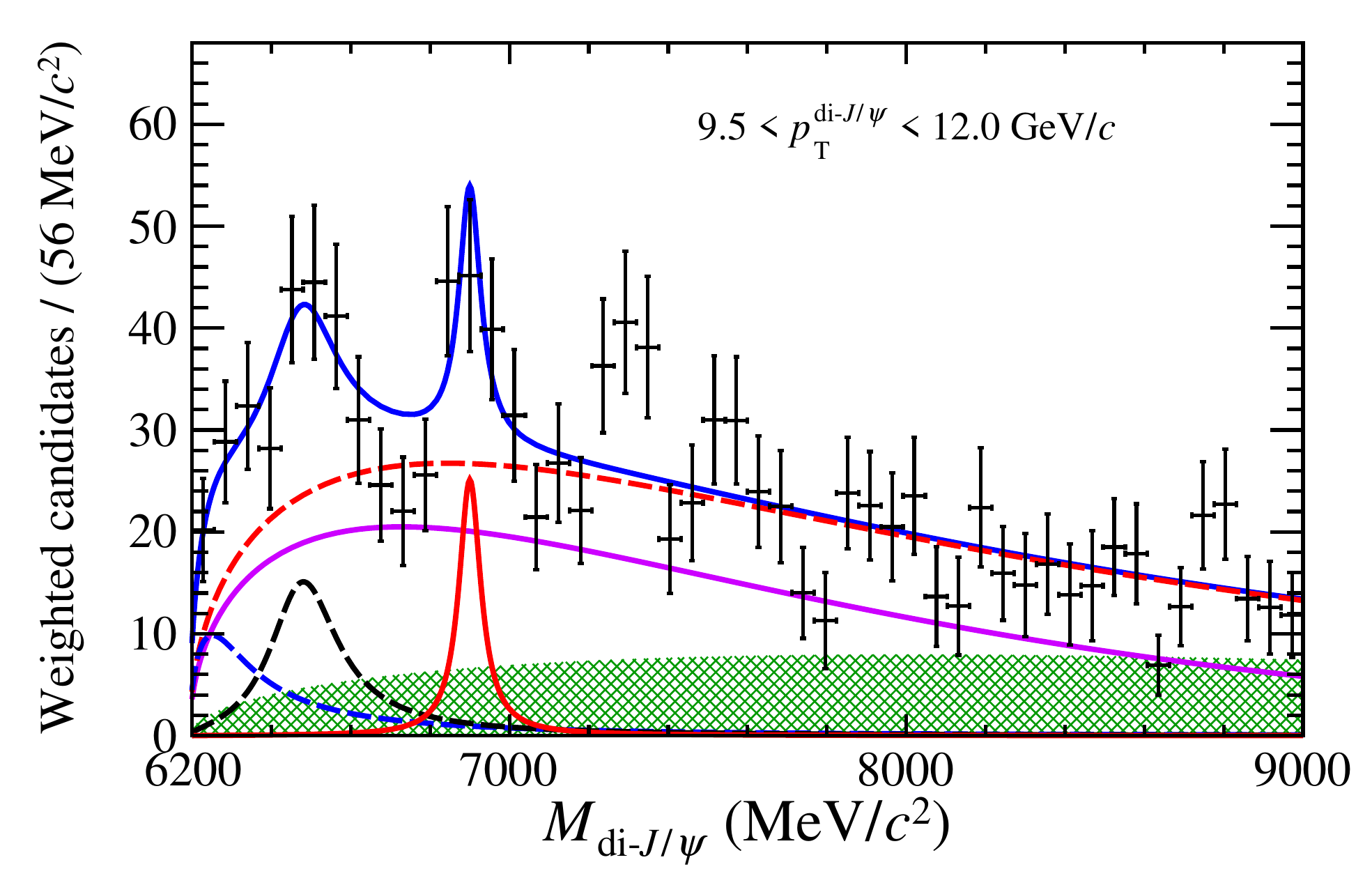}}
   \sidesubfloat[]{\includegraphics[width=0.45\linewidth]{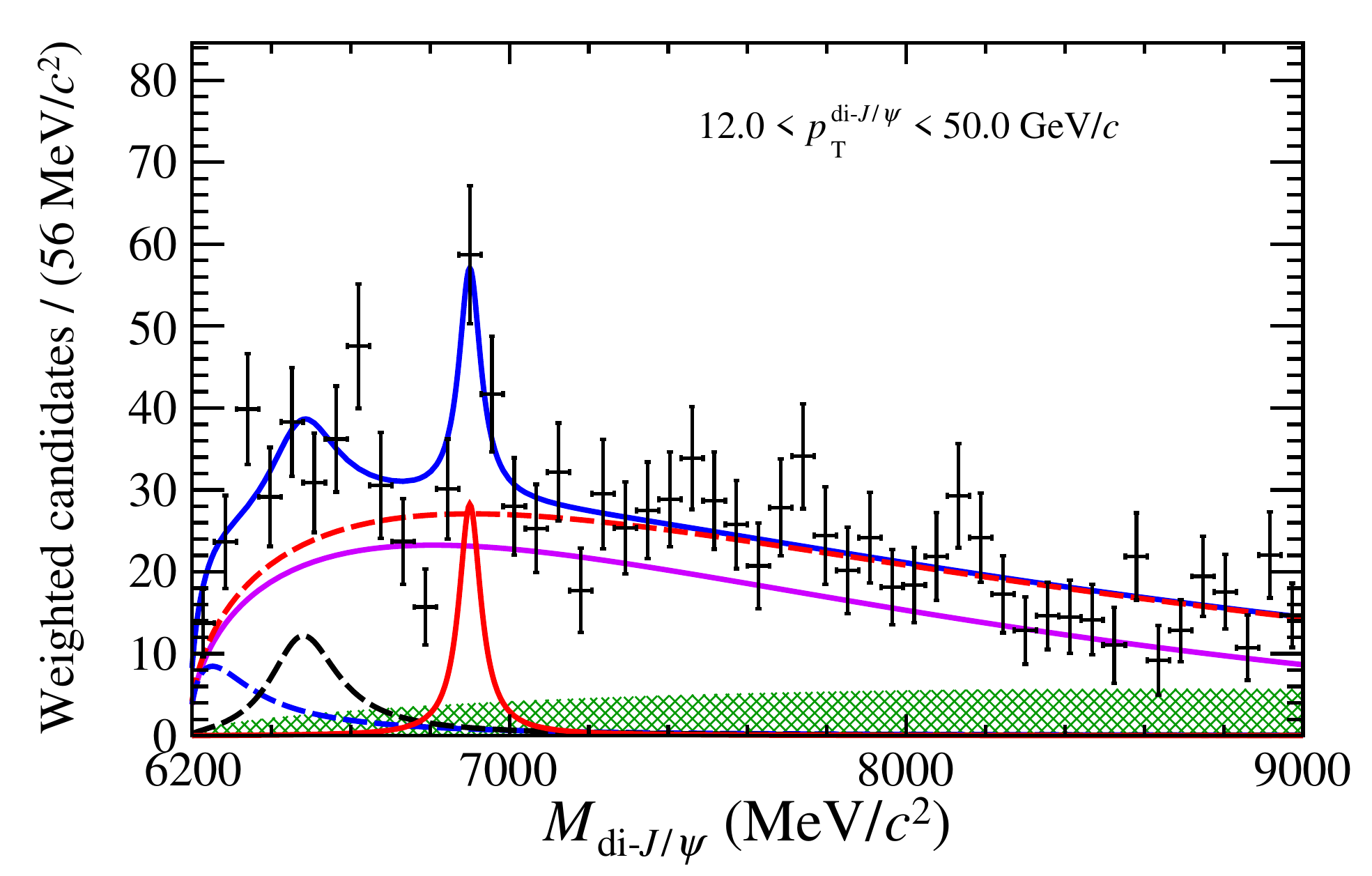}}\\ 
\vspace*{-0.1cm}
\end{center}
   \caption{\small Invariant mass spectra of weighted di-$\jpsi$ candidates in bins of $\dipt$ and overlaid projections of the $\dipt$-binned fit with 
  model I.
   }\label{fig:fitXbinned_2peak}
\end{figure}

%%%%% Fit
%%%%%%%%%%% model %%%%%%%%%%%%%%
To remove background pairs that have at least one background \jpsi candidate, the \sPlot weighting method~\cite{Pivk:2004ty} is applied, where the weights are calculated from the fit to the two\nobreakdash-dimensional  \mbox{$(M^{(1)}_{\mu\mu},M^{(2)}_{\mu\mu})$} distribution. 
The background-subtracted di-$\jpsi$ spectra in the range \mbox{$6.2<\diM<9.0\gevcc$} are shown in Fig.~\ref{fig:fitXcut} for candidates with $\dipt>5.2\gevc$ and~Fig.~\ref{fig:fitXbinned_2peak} for candidates in the six $\dipt$ intervals,
which are investigated by weighted unbinned maximum-likelihood fits~\cite{Xie:2009rka}.
The $\diM$ distribution of signal events is expected to be dominated by the sum of the nonresonant SPS~(NRSPS) and DPS production, which have smooth shapes~(referred to as continuum in the following).
The DPS continuum is described by
a two-body phase-space function multiplied by the product of an exponential function and a second order polynomial function, whose parameters are fixed according to the $\diM$ distribution constructed from $\jpsi$ differential cross-sections. Its yield is determined by extrapolation from the $\diM>12\gevcc$ region, which is dominated and well described by the DPS distribution.
The continuum NRSPS is modelled by
a two-body phase-space distribution multiplied by an exponential function 
determined from the data.
The combination of continuum NRSPS and DPS does not provide a good description of the data, as is illustrated in Fig.~\ref{fig:fitXcut}(a).
The $\diM$ spectrum in the data is tested against the hypothesis that only  NRSPS and DPS components are present
in the range $6.2<\diM<7.4\gevcc$ (null hypothesis) using a $\chisq$ test statistic.
Pseudoexperiments are generated and fitted according to the null hypothesis, and the fraction of these fits with a $\chisq$ value exceeding that in the data is converted into a significance.
Considering the sample in the $\dipt>5.2\gevc$ region,
the null hypothesis is inconsistent with the data at $3.4$ standard deviations ($\sigma$).
A test performed simultaneously in the aforementioned six $\dipt$ regions  yields a discrepancy  of $6.0\,\sigma$ with the null hypothesis.
A higher value is obtained in the latter case attributed to the presence of the structure at the same $\diM$ location in different $\dipt$ intervals.
A cross-check is performed by dividing the data into five or seven $\dipt$ regions instead,
which results in significance values consistent with the nominal $6.0\,\sigma$.
The significance values are summarised in Table~\ref{tab:signif}~(Any structure beyond NRSPS plus DPS).

The structures in the $\diM$ distribution can have various interpretations. There may be one or more resonant states $\cccc$ decaying directly into a pair of $\jpsi$ mesons, or $\cccc$ states decaying into a pair of $\jpsi$ mesons through feed-down of heavier quarkonia, for example $\cccc\to\chic(\to\jpsi\gamma)\jpsi$ where the photon escapes detection. 
In the latter case, such a state would be expected to peak at a lower $\diM$ position, close to the di-$\jpsi$ mass threshold, and its structure would be broader compared to that from a direct decay.
This feed-down is unlikely an explanation for the narrow $X(6900)$ structure.
Rescattering of two charmonium states produced by SPS close to their mass threshold may also generate a narrow structure ~\cite{Braaten:2020iye,Liu:2015fea,Xie:2016lvs,Guo:2016bjq}. 
The two thresholds close to the $X(6900)$ structure could be formed by $\chiczero\chiczero$ pairs at $6829.4\mevcc$ and $\chicone\chiczero$ pairs at $6925.4\mevcc$, respectively.
Whereas a resonance is often described by a relativistic Breit--Wigner (BW) function~\cite{PDG2019}, the lineshape of a structure with rescattering effects taken into account is more complex. In principle, resonant production can interfere with NRSPS of the same spin-parity quantum numbers~($J^\mathrm{PC}$), resulting in a coherent sum of the two components and thus a modification of the total $\diM$ distribution.

Two different models of the structure lineshape providing a reasonable description of the data are investigated.
The $X(6900)$ lineshape parameters and yields are derived from fits to the $\dipt$-threshold sample.
Simultaneous $\dipt$-binned fits
are also performed as a cross-check and the variation of lineshape parameters is considered as a source of systematic uncertainties. 
Due to its low significance, the structure around $7.2\gevcc$ has been neglected.

In model I,
the $X(6900)$ structure is considered as a resonance, whereas the threshold enhancement is described through
a superposition of two resonances.
The lineshapes of these resonances are described by $S$-wave relativistic BW functions
multiplied by a two-body phase-space distribution.
The experimental resolution on $\diM$ is below $5\mevcc$ over the full mass range and negligible compared to the widths of the structures.
The projections of the $\dipt$-threshold fit using this model
are shown in
Fig.~\ref{fig:fitXcut}(b).
The mass, natural width and yield 
are determined to be
\mbox{$m[X(6900)]=6905\pm11\mevcc$}, \mbox{$\Gamma[X(6900)]=80\pm19\mev$} and \mbox{$N_{\rm sig}=252\pm63$}, where biases on the statistical uncertainties have been corrected using a bootstrap method~\cite{Langenbruch:2019nwe}.
The goodness of fit is studied using a $\chisq$ test statistic and found to be $\chisq/{\rm ndof}=112.7/89$,
corresponding to a probability of $4.6\%$.
The fit is also performed assuming the threshold enhancement as due to  a single wide resonance (see Supplementary Material); the fit quality is found  significantly poorer and thus this model is not further investigated.

A comparison between the best fit result of model I
and the data reveals a tension around $6.75\gevcc$, where the data shows a dip.
In an attempt to describe the dip, %the second model 
model II
allows for interference between the NRSPS component and a resonance for the threshold enhancement. The coherent sum of the two components is defined as
\begin{equation}
   \label{eq:itf}
   \left| A e^{i\phi} \sqrt{f_{\rm nr}(\diM)} + {\rm BW}(\diM) \right|^2,
\end{equation}
where $A$ and $\phi$ are the magnitude and phase of the nonresonant component, relative to the BW lineshape  for the  resonance, assumed to be independent of $\diM$,
and $f_{\rm nr}(\diM)$ is an exponential function. 
The interference term in Eq.~(\ref{eq:itf}) is then added incoherently to the BW function describing the X(6900) structure and the DPS description. 
The fit to the $\dipt$-threshold sample with this model has a probability of $15.5\%$~($\chisqndf=104.7/91$), and 
its projections are illustrated in Fig.~\ref{fig:fitXcut}(c).
In this case, the mass, natural width and yield 
are determined to be
\mbox{$m[X(6900)]=6886\pm11\mevcc$}, \mbox{$\Gamma[X(6900)]=168\pm33\mev$} and \mbox{$N_{\rm sig}=784\pm148$}.
A larger $X(6900)$ width and yield are preferred in comparison to 
model I.
Here it is assumed that the whole NRSPS production is involved in the interference with the lower-mass resonance despite that there may be several  components with different quantum numbers in the NRSPS and more than one resonance in the threshold enhancement.

Fits to the $\diM$ distributions in the six individual $\dipt$ bins are shown in Fig.~\ref{fig:fitXbinned_2peak} for 
%the first model, 
model I,
while those for 
%the second model 
model II
are given in the Supplementary Material.
An additional model describing the dip and the $X(6900)$ structure simultaneously by using the interference between the NRSPS and a BW resonance around $6.9\gevcc$ is also considered, however the fit quality is clearly poorer, as illustrated in the Supplementary Material.
Alternative lineshapes, other than the BW, may also be possible to describe these structures and will be the subject of future studies.

%%%%%%%%%%%%%%%%%%%%%%%%
%%%%%%%%%%%%%%%%%%%%%%%%

%%%%%%%%%%% significance %%%%%%%%%%%%%%
%the $6.9\gevcc$ structure and the threshold enhancement
The increase of the likelihood between the fits with or without considering the $X(6900)$ and the threshold enhancement structures on top of the continuum NRSPS plus DPS model is taken as the test statistic to calculate the combined global significance of the two structures~\cite{Cowan:2010js} 
in the \mbox{$6.2<\diM<7.4\gevcc$} region,
where pseudoexperiments are also generated to evaluate the significance. Only model I is studied, where the interference between the NRSPS 
and the threshold enhancement
is not included.
Similarly, the significance for either the threshold enhancement or the $X(6900)$ structure is evaluated assuming the presence of the other along with the NRSPS and DPS continuum.
The significance is determined from 
both $\dipt$-threshold and $\dipt$-binned fits,
and summarised in Table~\ref{tab:signif}. The results are above $5\,\sigma$ for the two structures, with slightly higher significance for the $\dipt$-binned case.

%%%%%%%%%%%%%%%%%%%%%%
\begin{table}[!tb]
\caption{\small Global significance evaluated under the various assumptions described in the text.
   }\label{tab:signif}
\begin{center}
\begin{tabular}{c|c|c}
\toprule
   \multirow{2}{*}{Structure} & \multicolumn{2}{c}{Significance} \\
   & $\dipt$-threshold & $\dipt$-binned \\
\hline
   Any structure beyond NRSPS plus DPS & $3.4\,\sigma$ & $6.0\,\sigma$ \\
\hline
   Threshold enhancement plus $X(6900)$ & $6.4\,\sigma$ & $6.9\,\sigma$ \\ 
   Threshold enhancement & $6.0\,\sigma$ & $6.5\,\sigma$ \\
   $X(6900)$ & $5.1\,\sigma$ & $5.4\,\sigma$ \\ 
\bottomrule
\end{tabular}
\end{center}
\end{table}
%%%%%%%%%%%%%%%%%%%%%%

%%%%% Systematics
%%%%%%%%%%%%%%%%%%%%%%
\begin{table}[!tb]
   \caption{\small Systematic uncertainties on the mass~($m$) and natural width~($\Gamma$) of the $X(6900)$ structure.
   }\label{tab:sys}
\begin{center}
\begin{tabular}{c|cc|cc}
\toprule
   & \multicolumn{2}{c|}{Without interference} & \multicolumn{2}{c}{With interference} \\
   Component & $m$~[$\mevcc$] & $\Gamma$~[$\mev$] & $m$~[$\mevcc$] & $\Gamma$~[$\mev$] \\
\hline
\sPlot weights             & 0.8 & 10.3 & 4.4 & 36.9 \\
Experimental resolution   & 0.0 & 1.4  & 0.0 & 0.6 \\
NRSPS+DPS modelling       & 0.8 & 16.1 & 3.5 & 9.3 \\
$X(6900)$ shape              & 0.0 & 0.3  & 0.4 & 0.2 \\
Dependence on $\dipt$       & 4.6 & 13.5 & 6.2 & 56.7 \\
$\bquark$-hadron feed-down & 0.0 & 0.2  & 0.0 & 5.3 \\
Structure at $7.2\gevcc$          & 1.3 & 9.2  & 6.7 & 5.2 \\
Threshold structure shape & 5.2 & 20.5 & --  & -- \\
NRSPS phase               & --  & --   & 0.3 & 1.3 \\
\hline
Total & 7 & 33 & 11 & 69 \\
\bottomrule
\end{tabular}
\end{center}
\end{table}
%%%%%%%%%%%%%%%%%%%%%%

Systematic uncertainties on the measurements of the mass and natural width of the $X(6900)$ structure
are reported in Table~\ref{tab:sys}.
They include variations of the results obtained by:
including an explicit component in the $\diM$ fits  for the \jpsi combinatorial background rather than subtracting it using the weighting method (\sPlot weights in Table~\ref{tab:sys});
convolving the $\diM$ fit functions with a Gaussian function of $5\mevcc$ width to account for the invariant mass resolution (Experimental resolution);
using alternative functions to describe the NRSPS component  and varying the DPS yield (NRSPS plus DPS modelling);
using  an alternative $P$-wave BW function for the $X(6900)$ structure and varying the hadron radius in the BW function from $2$ to $5\gev^{-1}$ [$X(6900)$ shape];
obtaining results from a simultaneous fit to the $\diM$ distributions in the six $\dipt$ bins which covers the uncertainty due to variations of the NRSPS, DPS shapes and the NRSPS-resonance interference with respect to $\dipt$ (Dependence on $\dipt$);
including an explicit contribution for \jpsi mesons from $\bquark$-hadron feed-down ($b$-hadron feed-down) or adding a BW component for the $7.2\gevcc$ structure (Structure at $7.2\gevcc$);
modelling the threshold structure using an alternative Gaussian function with asymmetric power-law tails,
or fitting in a reduced $\diM$ range excluding the threshold structure (Threshold structure shape);
allowing the relative phase in the NRSPS component to vary linearly with $\diM$ (NRSPS phase).
The total uncertainties 
are determined to be
$7\mevcc$ and $33\mev$ for the mass and natural width, respectively, without considering any interference, and
$11\mevcc$ and $69\mev$ when the interference between NRSPS and the threshold structure is introduced.

%%%%% Production
For the scenario without interference, 
the production cross-section of the $X(6900)$ structure relative to that of all $\jpsi$ pairs~(inclusive), times the branching fraction  $\BF(X(6900)\to\jpsi\jpsi)$, $\mathcal{R}$,
is determined in the $pp$ collision data at $\sqs=13\tev$.
The measurement is obtained for both $\jpsi$ mesons in the fiducial region of
transverse momentum 
below $10\gevc$
and rapidity 
between $2.0$ and $4.5$.
An event-by-event efficiency correction is performed to obtain the signal yield at production.
The residual contamination from $\bquark$-hadron feed-down is subtracted from inclusive $\jpsi$-pair production following 
%the method in 
Ref.~\cite{LHCb-PAPER-2015-037}.
The systematic uncertainties on the $X(6900)$ yield
are estimated in a similar way to that for the mass and natural width,
while other systematic uncertainties mostly cancel in the ratio.
The production ratio is measured to be
\mbox{$\mathcal{R}=[1.1\pm0.4\stat\pm0.3\syst]\%$} without any $\dipt$ requirement and \mbox{$\mathcal{R}=[2.6\pm0.6\stat\pm0.8\syst]\%$} for $\dipt>5.2\gevc$.

\section{Summary}
%%%%% Summary
In conclusion,
using $pp$ collision data 
at centre-of-mass energies of $7$, $8$ and $13\tev$ collected with the \lhcb detector,
corresponding to an integrated luminosity of $9\invfb$,
the $\jpsi$-pair invariant mass spectrum is studied. 
The data in the mass range between $6.2$ and $7.4\gevcc$ are found to be inconsistent with the hypothesis of 
NRSPS plus DPS continuum.
A narrow structure, $X(6900)$, matching the lineshape of a resonance
and a broad structure next to the di-\jpsi mass threshold
are found.
The global significance of either the broad or the $X(6900)$ structure is determined to be larger than five standard deviations. 
Describing the $X(6900)$ structure with a Breit--Wigner lineshape,
its mass and natural width
are determined to be
\begin{equation*}
m[X(6900)]=6905\pm11\pm7\mevcc
\end{equation*}
and 
\begin{equation*}
\Gamma[X(6900)]=80\pm19\pm33\mev,
\end{equation*} 
assuming no interference with the NRSPS continuum is present,
where the first uncertainty is statistical and the second systematic.
When assuming the NRSPS continuum interferes with the broad structure close to the di-\jpsi mass threshold, they become
\begin{equation*}
m[X(6900)]=6886\pm11\pm11\mevcc
\end{equation*}
and 
\begin{equation*}
\Gamma[X(6900)]=168\pm33\pm69\mev.
\end{equation*}
The $X(6900)$ structure could originate from a hadron state consisting of four charm quarks, $\cccc$, predicted in various tetraquark models. The broad structure close to the di-\jpsi mass threshold could be due to a mixture of multiple four-charm quark states or have contributions from feed-down decays of four-charm states through heavier quarkonia.
Other interpretations cannot presently be ruled out, for example the rescattering of two charmonium states produced close to their mass threshold. More data along with additional measurements, including determination of the spin-parity quantum numbers and $\pt$ dependence of the production cross-section, 
are needed to provide further information about the nature of the observed structure.

%% file: acknowledgements.tex
\section*{Acknowledgements}
%
% These Acknowledgements valid from 3-May-2019
%
\noindent We express our gratitude to our colleagues in the CERN
accelerator departments for the excellent performance of the LHC. We
thank the technical and administrative staff at the LHCb
institutes.
We acknowledge support from CERN and from the national agencies:
CAPES, CNPq, FAPERJ and FINEP (Brazil); 
MOST and NSFC (China); 
CNRS/IN2P3 (France); 
BMBF, DFG and MPG (Germany); 
INFN (Italy); 
NWO (Netherlands); 
MNiSW and NCN (Poland); 
MEN/IFA (Romania); 
MSHE (Russia); 
MinECo (Spain); 
SNSF and SER (Switzerland); 
NASU (Ukraine); 
STFC (United Kingdom); 
DOE NP and NSF (USA).
We acknowledge the computing resources that are provided by CERN, IN2P3
(France), KIT and DESY (Germany), INFN (Italy), SURF (Netherlands),
PIC (Spain), GridPP (United Kingdom), RRCKI and Yandex
LLC (Russia), CSCS (Switzerland), IFIN-HH (Romania), CBPF (Brazil),
PL-GRID (Poland) and OSC (USA).
We are indebted to the communities behind the multiple open-source
software packages on which we depend.
Individual groups or members have received support from
AvH Foundation (Germany);
EPLANET, Marie Sk\l{}odowska-Curie Actions and ERC (European Union);
A*MIDEX, ANR, Labex P2IO and OCEVU, and R\'{e}gion Auvergne-Rh\^{o}ne-Alpes (France);
Key Research Program of Frontier Sciences of CAS, CAS PIFI, and the Thousand Talents Program (China);
RFBR, RSF and Yandex LLC (Russia);
GVA, XuntaGal and GENCAT (Spain);
the Royal Society
and the Leverhulme Trust (United Kingdom).

%% file: app.tex
\clearpage
\renewcommand{\thetable}{S\arabic{table}}  
\renewcommand{\thefigure}{S\arabic{figure}}
\renewcommand{\theequation}{S\arabic{equation}}
\setcounter{figure}{0}
\setcounter{table}{0}
\setcounter{equation}{0}
\setcounter{page}{1}

{\noindent\normalfont\bfseries\Large Supplementary Material}

\appendix

\vspace{0.5cm}
\noindent
In the Supplementary Material,
the $\jpsi$-pair mass distributions in bins of $\dipt$ are shown in Sec.~\ref{app:Mbinned},
the fits using several additional models to the $\jpsi$-pair mass spectrum are presented in Sec.~\ref{app:morefits},
and some supplemental information to the fit result of model II is given in Sec.~\ref{app:itffit}.

\boldmath
\section{$\jpsi$-pair mass distributions in bins of $\dipt$}
\unboldmath
\label{app:Mbinned}
\begin{figure}[htbp]
\begin{center}
\includegraphics[width=1.0\linewidth]{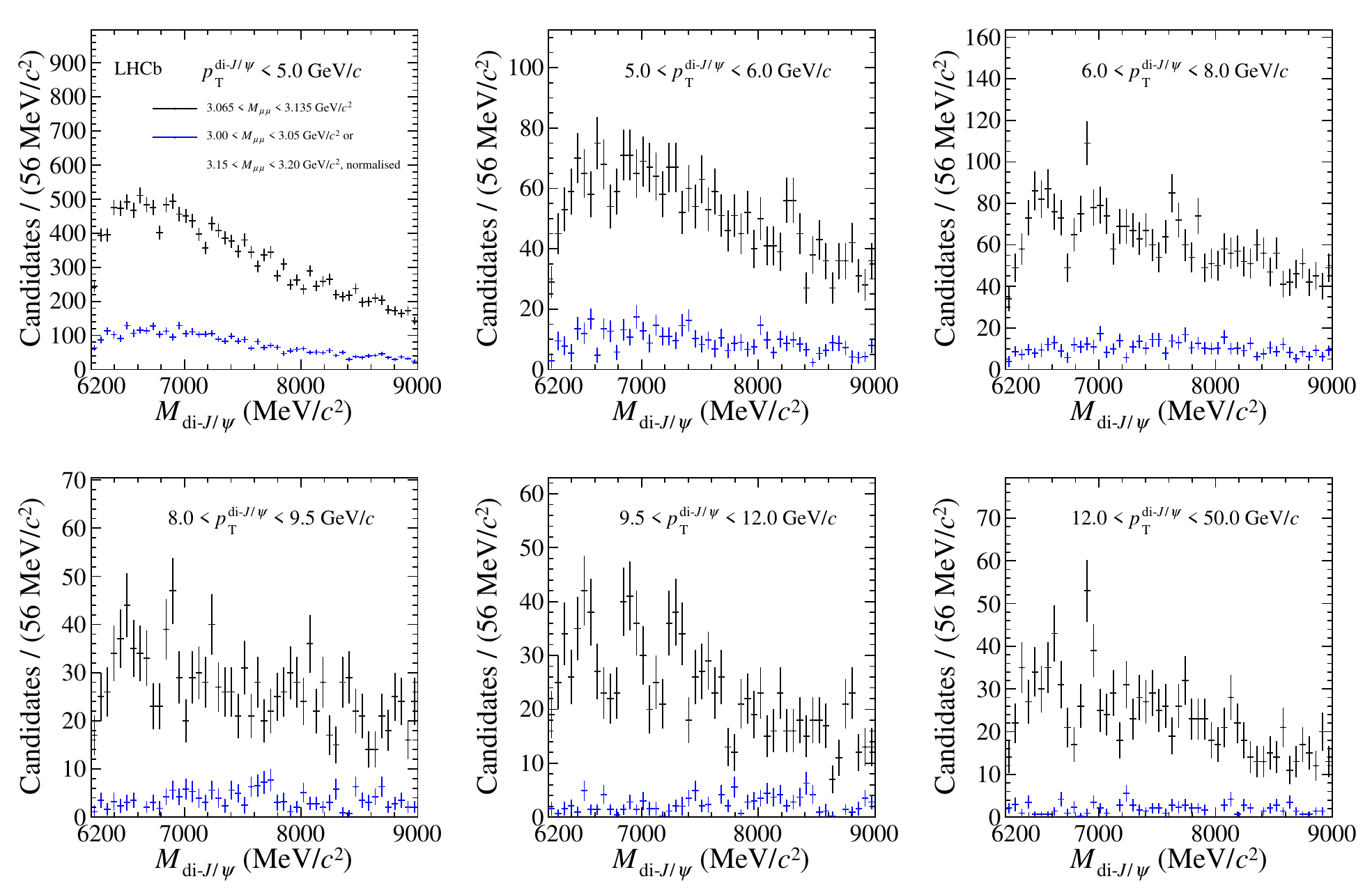}
\vspace*{-0.1cm}
\end{center}
   \caption{
   Invariant mass spectra of $\jpsi$-pair candidates in the six $\dipt$ regions with boundaries  
   $\left\{ 0, 5, 6, 8, 9.5, 12, 50 \right\}\gevc$
   with reconstructed $\jpsi$ masses in the  (black)~signal and (blue)~background regions,
   respectively. 
   }\label{fig:Xmassbinned}
\end{figure}

\boldmath
\section{Additional fits to the $\jpsi$-pair mass spectrum}
\begin{figure}[h]
\begin{center}
\sidesubfloat[]{\includegraphics[width=0.48\linewidth]{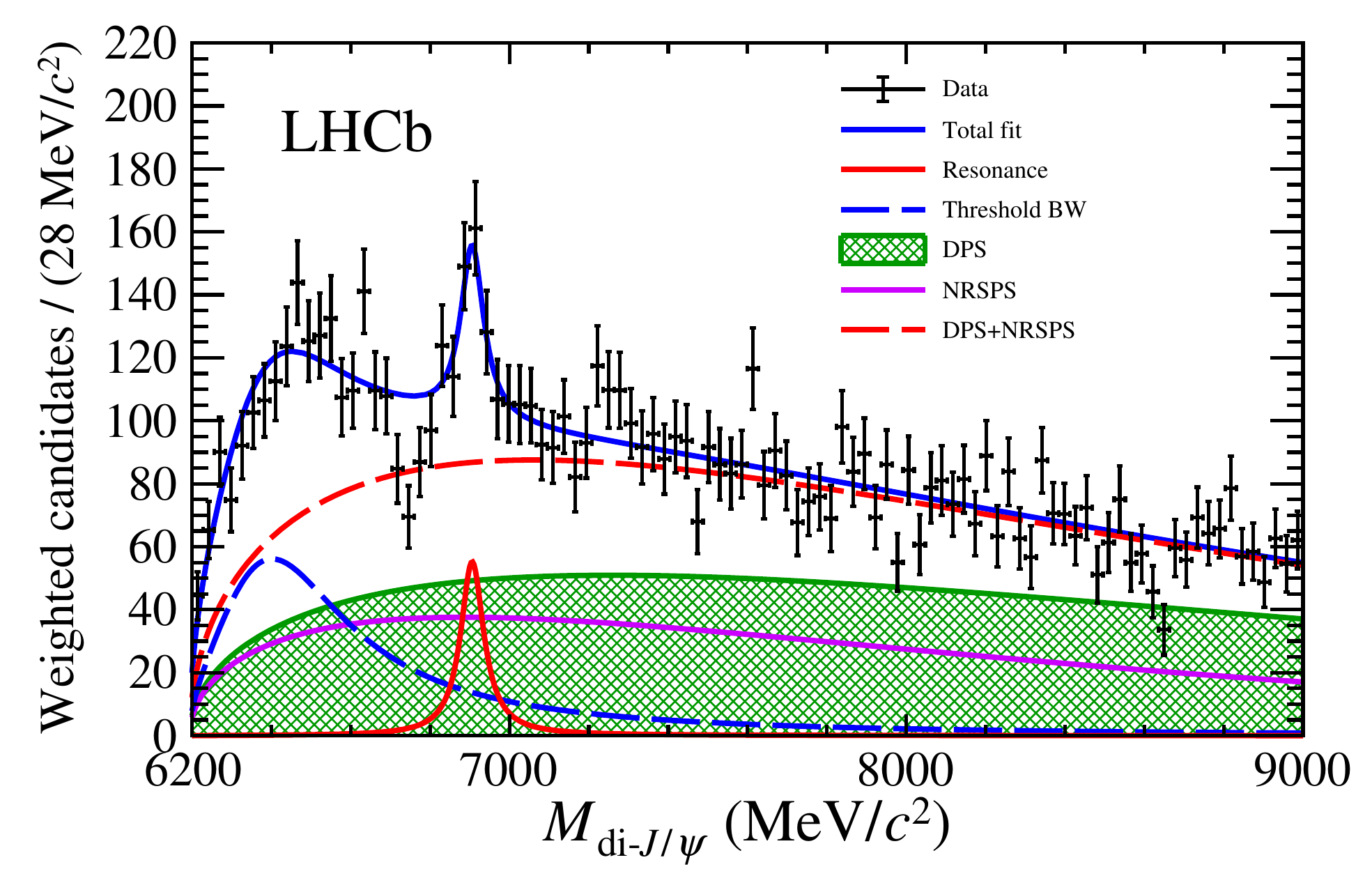}}
\sidesubfloat[]{\includegraphics[width=0.48\linewidth]{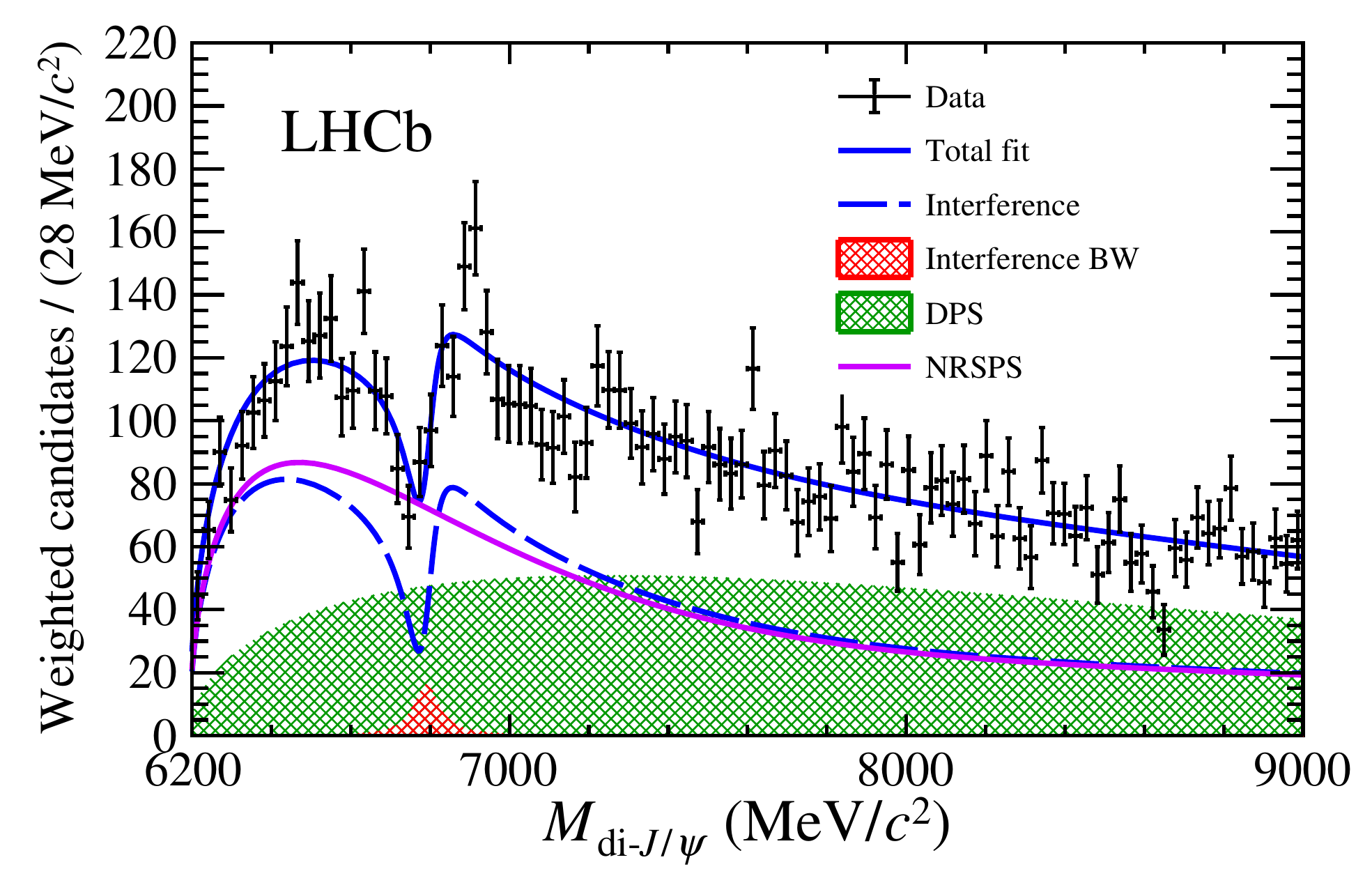}}\\
\vspace*{-0.1cm}
\end{center}
   \caption{Invariant mass spectra of weighted di-$\jpsi$ candidates with $\dipt>5.2\gevc$ and overlaid projections of the $\dipt$-threshold fit with (a)~the threshold structure described as a single BW function,
   and (b)~assuming a single BW 
   interfering with the SPS continuum.
   }\label{fig:fitXcut2}
\end{figure}

\begin{figure}[h]
\begin{center}
\includegraphics[width=0.48\linewidth]{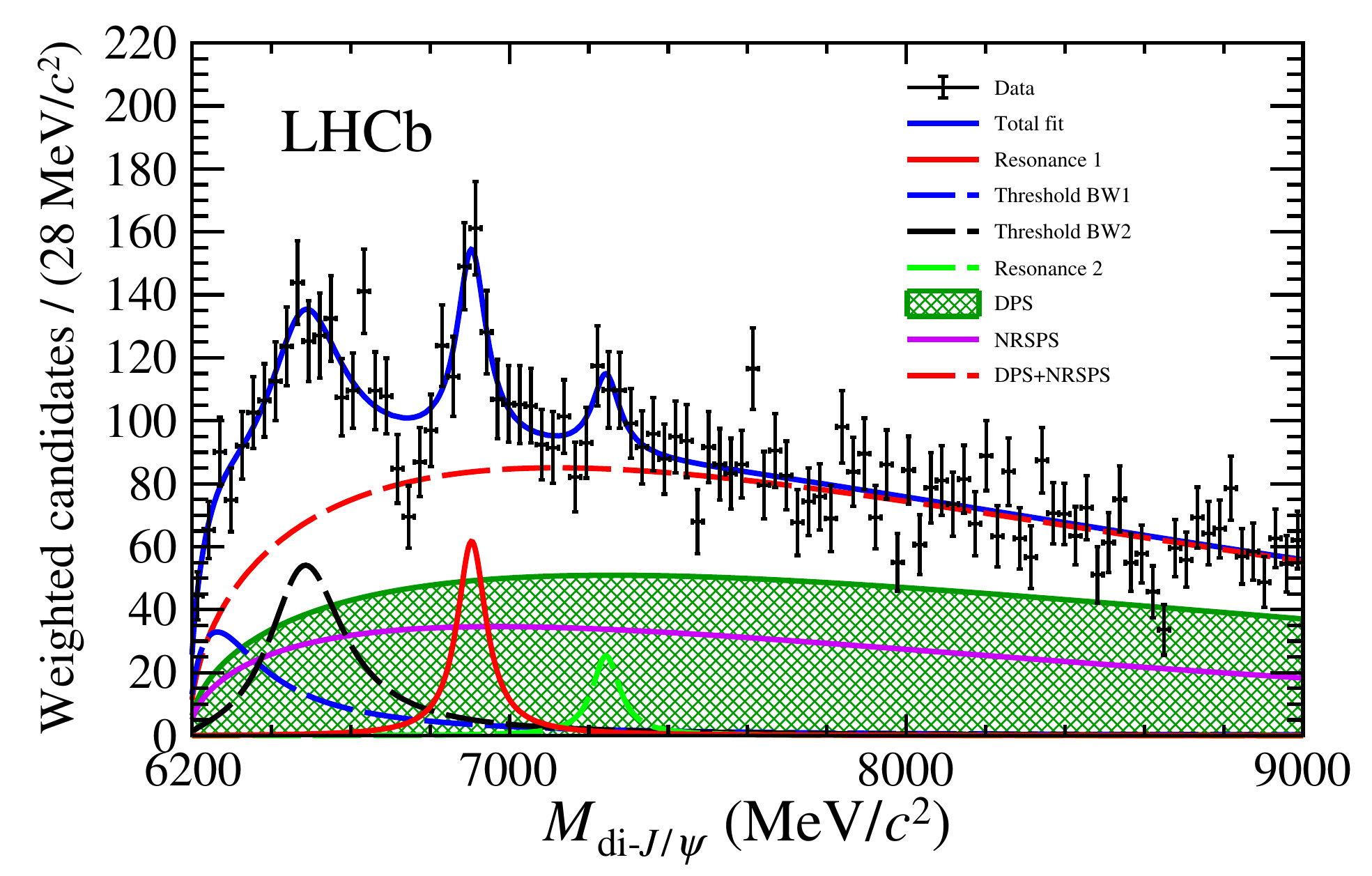}
\vspace*{-0.1cm}
\end{center}
   \caption{Invariant mass spectra of weighted di-$\jpsi$ candidates with $\dipt>5.2\gevc$ and overlaid projections of the $\dipt$-threshold fit with an additional BW function introduced to describe the $7.2\gevcc$ structure, based on the model that contains two BW lineshapes for the threshold structure and a BW shape for the $X(6900)$ structure on top of the NRSPS plus DPS continuum. 
   }\label{fig:fitXcut3}
\end{figure}

\unboldmath
\label{app:morefits}
Figure~\ref{fig:fitXcut2} shows the fits to the $\jpsi$-pair mass spectrum with (a) the threshold structure described by a single Breit--Wigner (BW) lineshape and (b) using a model that contains a single BW resonance interfering with the SPS continuum.
The $\chisq/{\rm ndof}$ of the two fits 
are $125.6/92$ 
and $118.6/91$, 
corresponding to a probability of $1.2\%$ and $2.8\%$, respectively.
Figure~\ref{fig:fitXcut3} shows the fit with an additional BW function introduced to describe the $7.2\gevcc$ structure, based on the model that contains two BW lineshapes for the threshold structure and a BW shape for the $X(6900)$ structure on top of the NRSPS plus DPS continuum.

\boldmath
\section{Supplement to fit result of model II}
\unboldmath
\label{app:itffit}
In model II that contains a BW lineshape for the threshold structure interfering with the NRSPS, a BW shape for the $X(6900)$ structure and the DPS continuum, 
the parameters of the lower-mass BW lineshape is determined to $M=6741\pm6\stat\mevcc$ and $\Gamma=288\pm16\stat\mev$.
The systematic uncertainties on the mass and natural width are not studied.
Due to the complex nature of the threshold structure, and the simple interference scenario considered,
this study is not considered to claim a state with the parameters reported here.

\begin{figure}[!htbp]
\begin{center}
   \sidesubfloat[]{\includegraphics[width=0.45\linewidth]{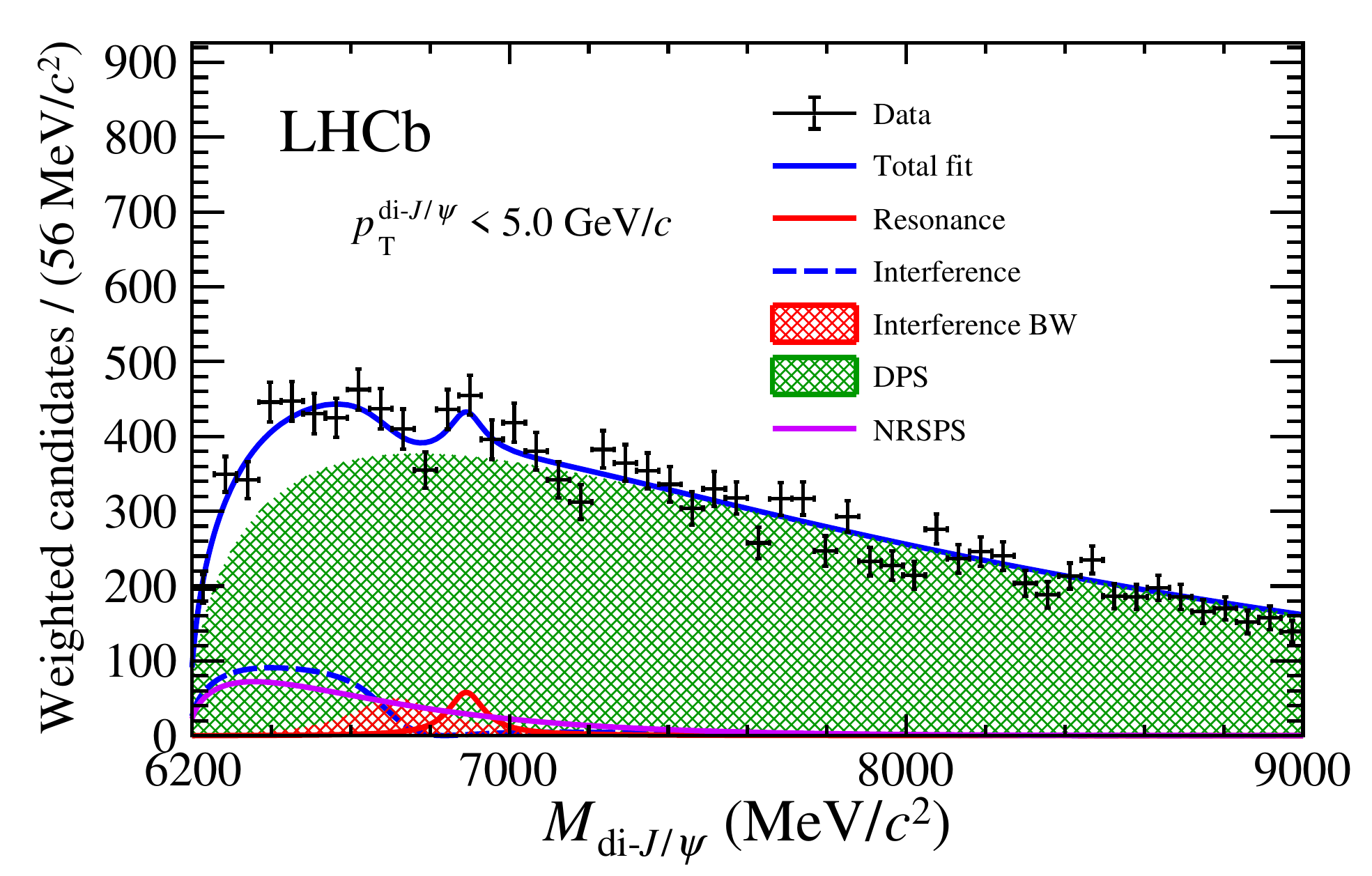}}
   \sidesubfloat[]{\includegraphics[width=0.45\linewidth]{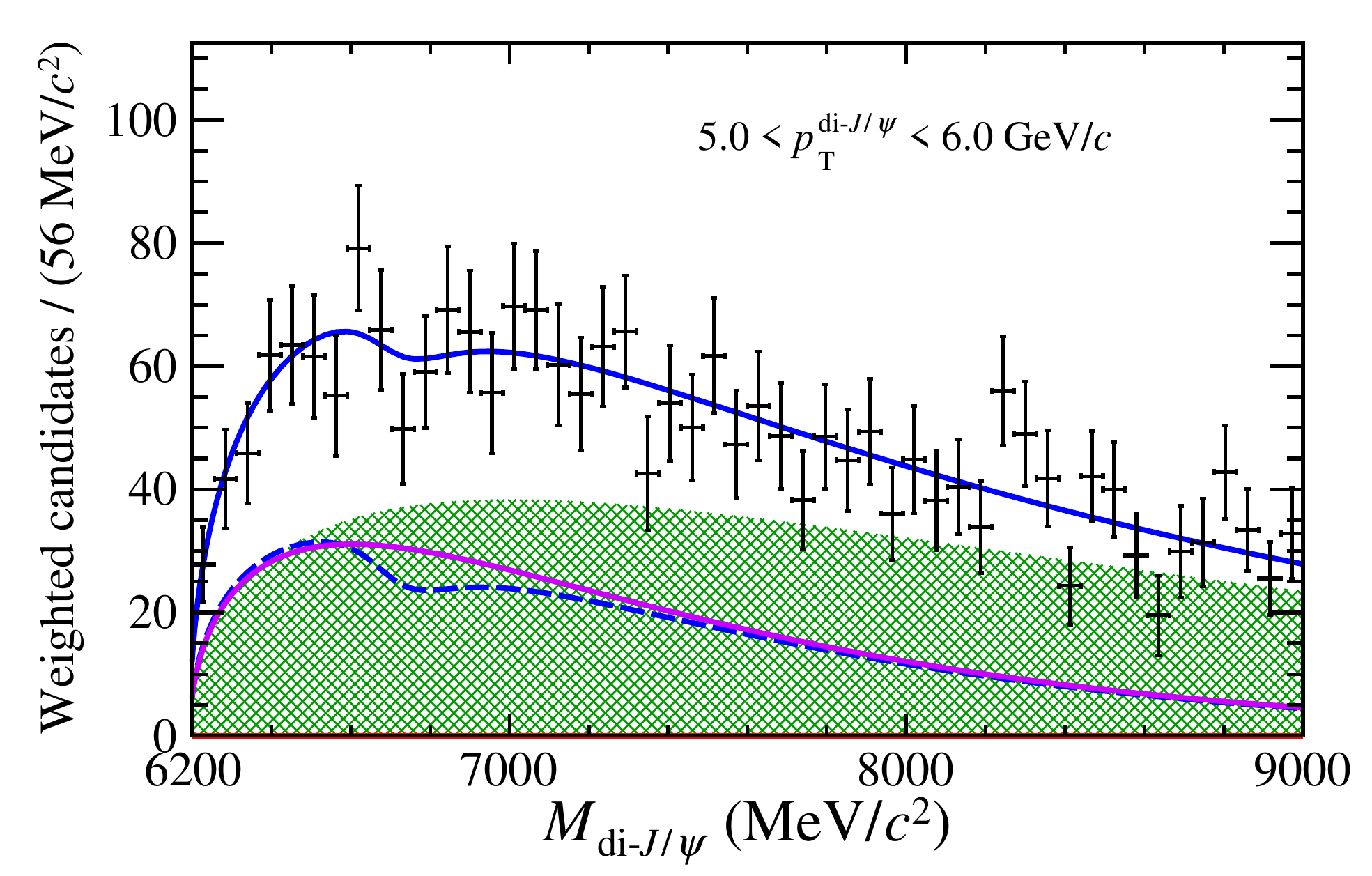}}\\
   \sidesubfloat[]{\includegraphics[width=0.45\linewidth]{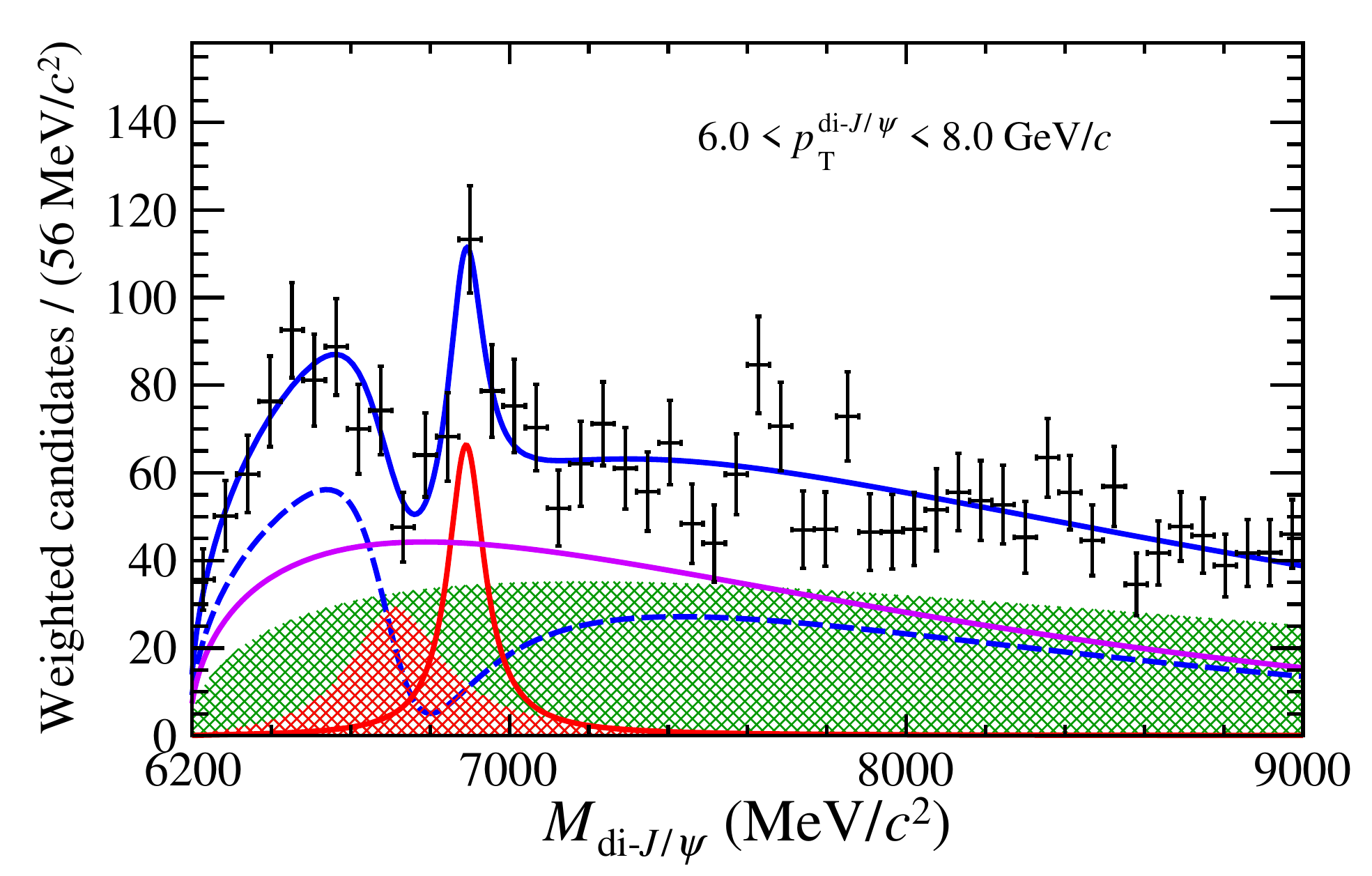}}
   \sidesubfloat[]{\includegraphics[width=0.45\linewidth]{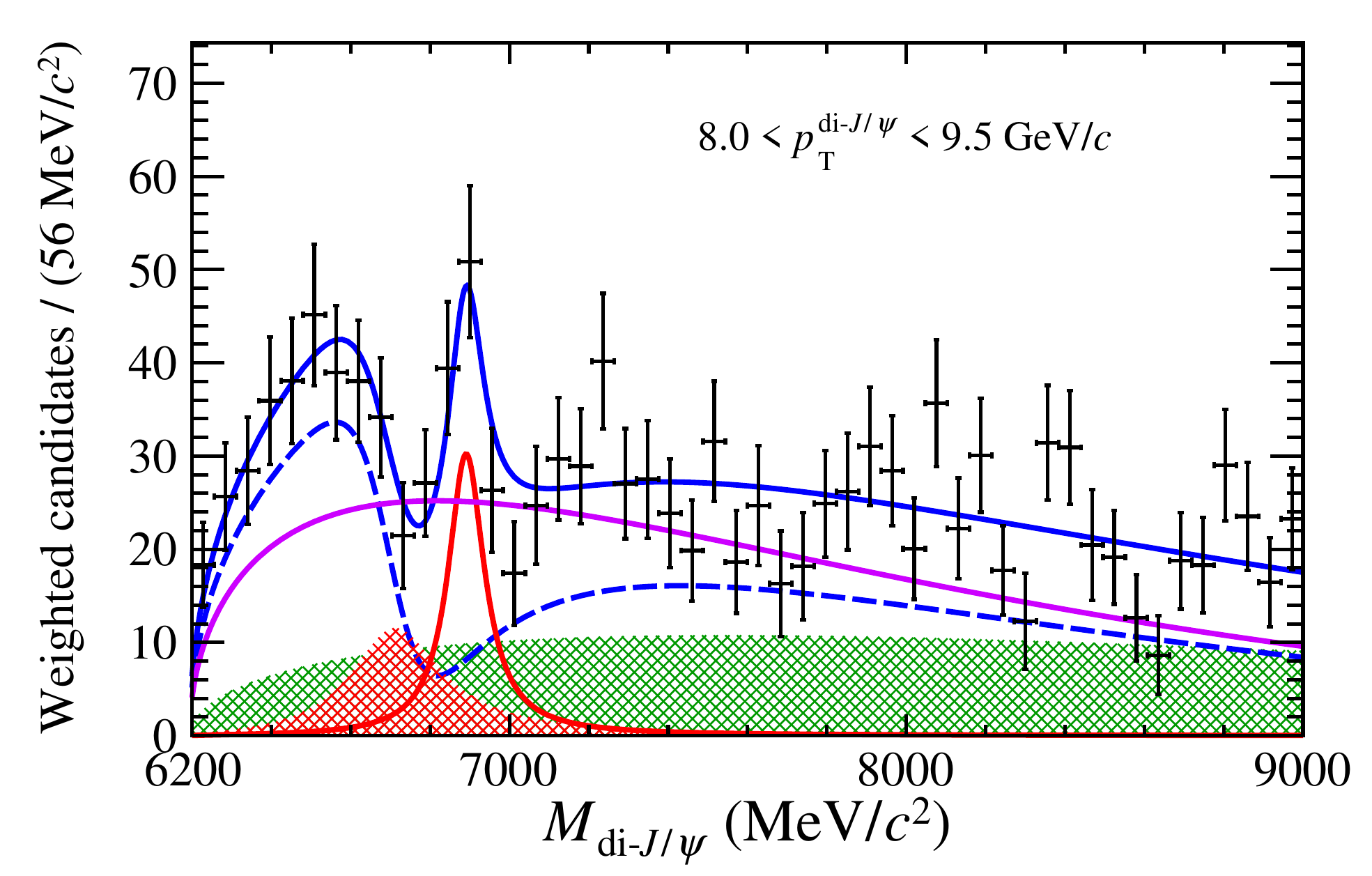}}\\
   \sidesubfloat[]{\includegraphics[width=0.45\linewidth]{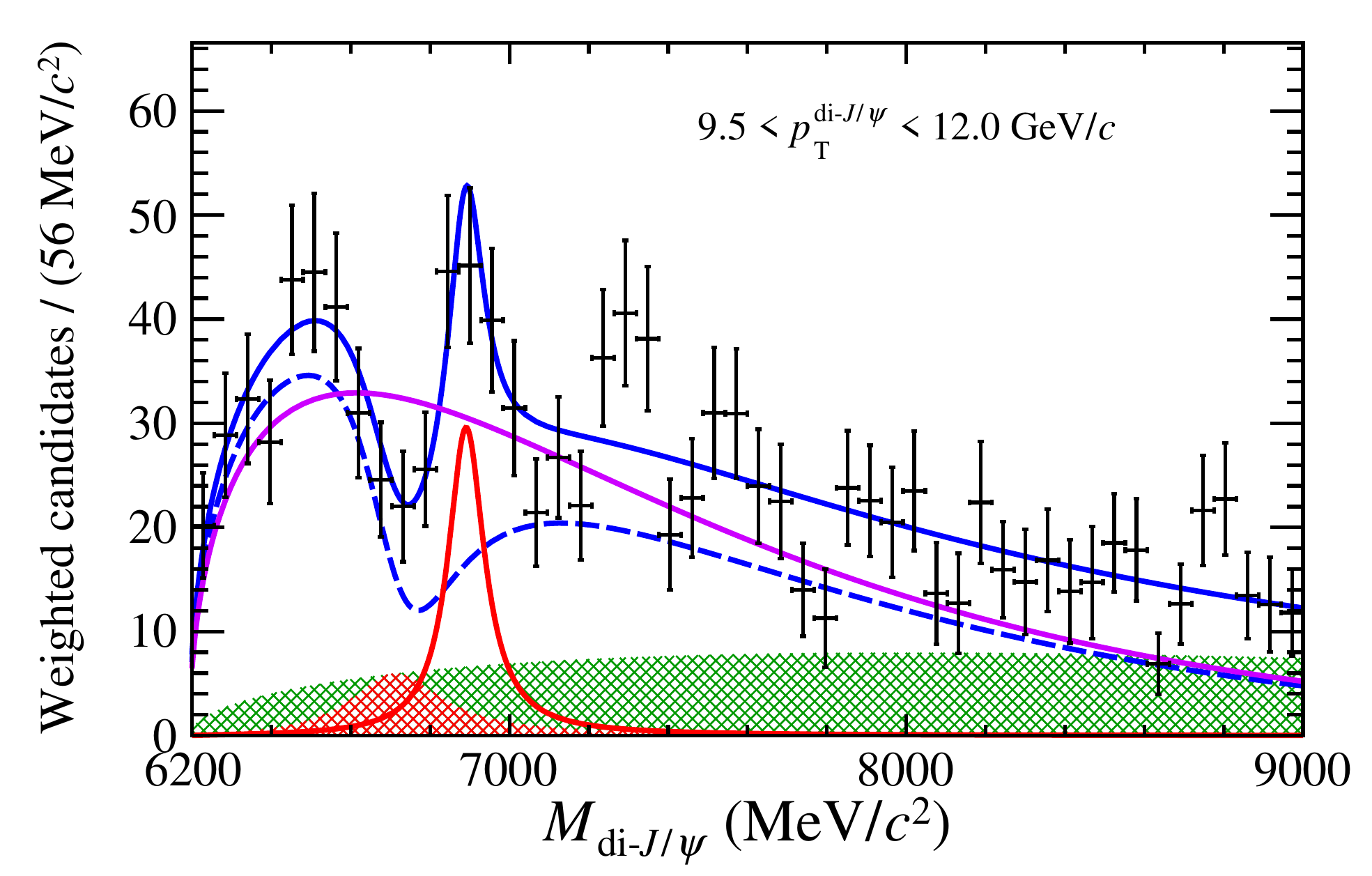}}
   \sidesubfloat[]{\includegraphics[width=0.45\linewidth]{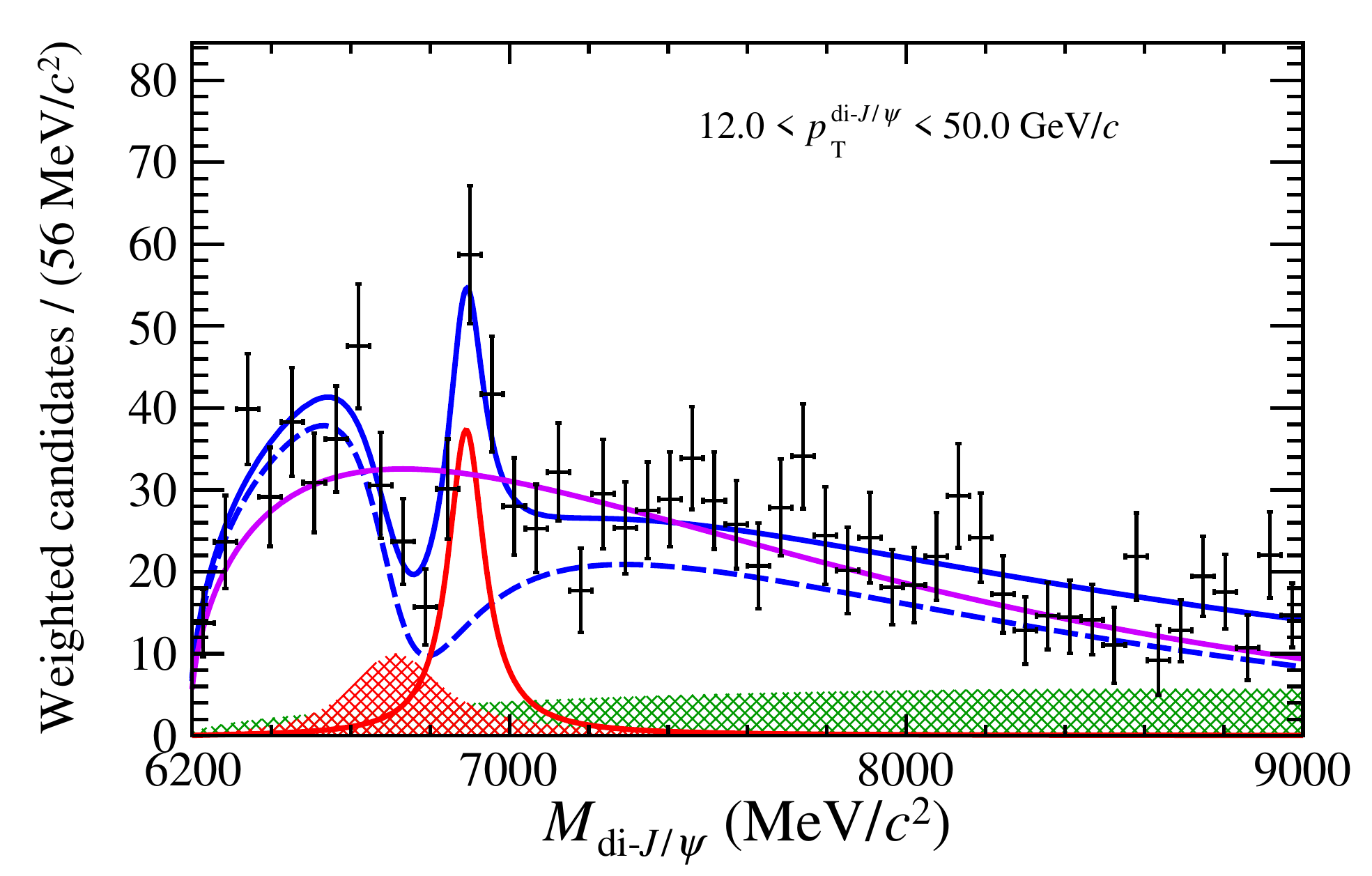}}\\
\vspace*{-0.1cm}
\end{center}
   \caption{Invariant mass spectra of weighted di-$/\jpsi$ candidates in bins of $\dipt$ and overlaid projections of the $\dipt$-binned fit assuming that the threshold structure interferes with the SPS continuum.
   }\label{fig:fitXbinned_itf}
\end{figure}

Projections of the fit to the $\jpsi$-pair invariant mass spectra in bins of $\dipt$ assuming the interference between the threshold structure and the SPS continuum are shown in Fig.~\ref{fig:fitXbinned_itf}.

\boldmath
\section{Dependence of experimental efficiency on $\diM$}
\begin{figure}[!htbp]
\begin{center}
   \includegraphics[width=0.8\linewidth]{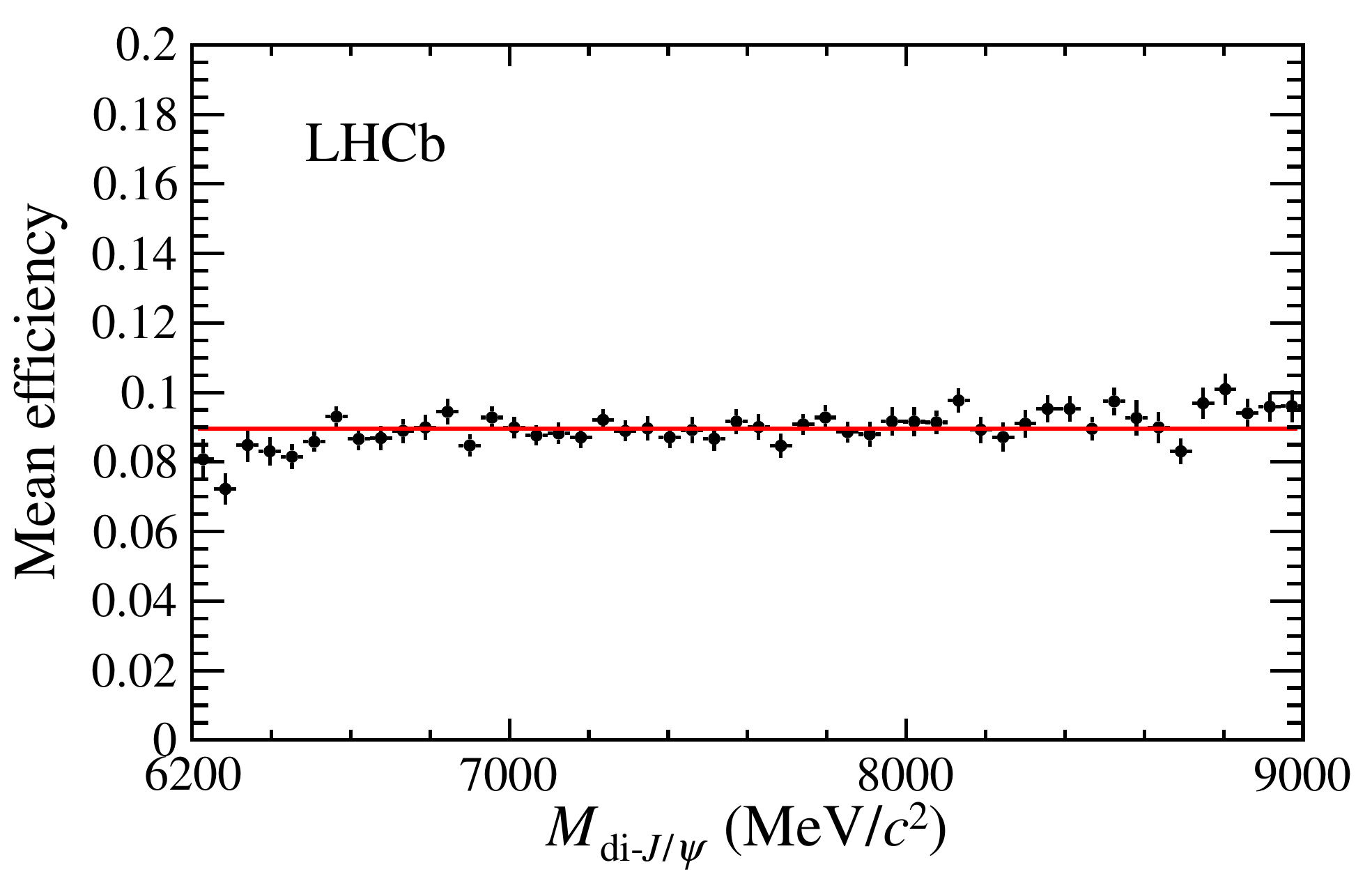}
\vspace*{-0.1cm}
\end{center}
   \caption{Dependence of the experimental efficiency on $\diM$. The error bars include the statistical uncertainties only.
   }\label{fig:eff}
\end{figure}

\unboldmath
The variation of the experimental efficiency with respect to $\diM$ is shown in Fig.~\ref{fig:eff}, which is marginal across the whole $\diM$ range. The efficiency is estimated in the same way as described in Ref.~\cite{LHCb-PAPER-2016-057}.

\clearpage

%% file: LHCb_Authorship_28-Apr-2020.tex
% LHCb collaboration author list
% Data extracted on September 21st, 2020 at 1:10pm for reference date 28-Apr-2020
\centerline
{\large\bf LHCb collaboration}
\begin
{flushleft}
\small
R.~Aaij$^{31}$,
C.~Abell{\'a}n~Beteta$^{49}$,
T.~Ackernley$^{59}$,
B.~Adeva$^{45}$,
M.~Adinolfi$^{53}$,
H.~Afsharnia$^{9}$,
C.A.~Aidala$^{83}$,
S.~Aiola$^{25}$,
Z.~Ajaltouni$^{9}$,
S.~Akar$^{64}$
% ...

{\footnotesize \it

% ...
$ ^{9}$Universit{\'e} Clermont Auvergne, CNRS/IN2P3, LPC, Clermont-Ferrand, France\\
% ...
$ ^{25}$INFN Sezione di Milano, Milano, Italy\\
% ...
$ ^{31}$Nikhef National Institute for Subatomic Physics, Amsterdam, Netherlands\\
% ...
$ ^{45}$Instituto Galego de F{\'\i}sica de Altas Enerx{\'\i}as (IGFAE), Universidade de Santiago de Compostela, Santiago de Compostela, Spain\\
% ...
$ ^{49}$Physik-Institut, Universit{\"a}t Z{\"u}rich, Z{\"u}rich, Switzerland\\
% ...
$ ^{53}$H.H. Wills Physics Laboratory, University of Bristol, Bristol, United Kingdom\\
% ...
$ ^{59}$Oliver Lodge Laboratory, University of Liverpool, Liverpool, United Kingdom\\
% ...
$ ^{64}$University of Cincinnati, Cincinnati, OH, United States\\
% ...
$ ^{67}$Syracuse University, Syracuse, NY, United States\\
% ...
$ ^{83}$University of Michigan, Ann Arbor, United States, associated to $^{67}$\\
\bigskip
% ...
\medskip
}
\end{flushleft}

%% file: main.bbl
\ifx\mcitethebibliography\mciteundefinedmacro
\PackageError{LHCb.bst}{mciteplus.sty has not been loaded}
{This bibstyle requires the use of the mciteplus package.}\fi
\providecommand{\href}[2]{#2}
\begin{mcitethebibliography}{10}
\mciteSetBstSublistMode{n}
\mciteSetBstMaxWidthForm{subitem}{\alph{mcitesubitemcount})}
\mciteSetBstSublistLabelBeginEnd{\mcitemaxwidthsubitemform\space}
{\relax}{\relax}

\bibitem{GellMann:1964nj}
M.~Gell-Mann, \ifthenelse{\boolean{articletitles}}{\emph{{A schematic model of
  baryons and mesons}},
  }{}\href{https://doi.org/10.1016/S0031-9163(64)92001-3}{Phys.\ Lett.\
  \textbf{8} (1964) 214}\relax
\mciteBstWouldAddEndPuncttrue
\mciteSetBstMidEndSepPunct{\mcitedefaultmidpunct}
{\mcitedefaultendpunct}{\mcitedefaultseppunct}\relax
\EndOfBibitem
\bibitem{Zweig:352337}
G.~Zweig, \ifthenelse{\boolean{articletitles}}{\emph{{An SU$_3$ model for
  strong interaction symmetry and its breaking; Version 1}}}{}
  \href{http://cds.cern.ch/record/352337}{CERN-TH-401}, CERN, Geneva,
  1964\relax
\mciteBstWouldAddEndPuncttrue
\mciteSetBstMidEndSepPunct{\mcitedefaultmidpunct}
{\mcitedefaultendpunct}{\mcitedefaultseppunct}\relax
\EndOfBibitem
\bibitem{Brambilla:2014jmp}
N.~Brambilla, S.~Eidelman, P.~Foka {\em et~al.},
  \ifthenelse{\boolean{articletitles}}{\emph{{QCD and strongly coupled gauge
  theories: challenges and perspectives}},
  }{}\href{https://doi.org/10.1140/epjc/s10052-014-2981-5}{Eur.\ Phys.\ J.\
  \textbf{C74} (2014) 2981},
  \href{http://arxiv.org/abs/1404.3723}{{\normalfont\ttfamily
  arXiv:1404.3723}}\relax
\mciteBstWouldAddEndPuncttrue
\mciteSetBstMidEndSepPunct{\mcitedefaultmidpunct}
{\mcitedefaultendpunct}{\mcitedefaultseppunct}\relax
\EndOfBibitem
\bibitem{X3872}
Belle Collaboration, S.~K. Choi {\em et~al.},
  \ifthenelse{\boolean{articletitles}}{\emph{{Observation of a narrow
  charmonium-like state in exclusive $\decay{B^{\pm}}{K^{\pm} \pi^{+}\pi^{-}
  J/\psi}$ decays}},
  }{}\href{https://doi.org/10.1103/PhysRevLett.91.262001}{Phys.\ Rev.\ Lett.\
  \textbf{91} (2003) 262001},
  \href{http://arxiv.org/abs/hep-ex/0309032}{{\normalfont\ttfamily
  arXiv:hep-ex/0309032}}\relax
\mciteBstWouldAddEndPuncttrue
\mciteSetBstMidEndSepPunct{\mcitedefaultmidpunct}
{\mcitedefaultendpunct}{\mcitedefaultseppunct}\relax
\EndOfBibitem
\bibitem{Olsen:2017bmm}
S.~L. Olsen, T.~Skwarnicki, and D.~Zieminska,
  \ifthenelse{\boolean{articletitles}}{\emph{{Nonstandard heavy mesons and
  baryons: Experimental evidence}},
  }{}\href{https://doi.org/10.1103/RevModPhys.90.015003}{Rev.\ Mod.\ Phys.\
  \textbf{90} (2018) 015003},
  \href{http://arxiv.org/abs/1708.04012}{{\normalfont\ttfamily
  arXiv:1708.04012}}\relax
\mciteBstWouldAddEndPuncttrue
\mciteSetBstMidEndSepPunct{\mcitedefaultmidpunct}
{\mcitedefaultendpunct}{\mcitedefaultseppunct}\relax
\EndOfBibitem
\bibitem{LHCb-PAPER-2015-029}
LHCb Collaboration, R.~Aaij {\em et~al.},
  \ifthenelse{\boolean{articletitles}}{\emph{{Observation of $\jpsi\proton$
  resonances consistent with pentaquark states in
  \mbox{\decay{\Lb}{\jpsi\proton\Km}} decays}},
  }{}\href{https://doi.org/10.1103/PhysRevLett.115.072001}{Phys.\ Rev.\ Lett.\
  \textbf{115} (2015) 072001},
  \href{http://arxiv.org/abs/1507.03414}{{\normalfont\ttfamily
  arXiv:1507.03414}}\relax
\mciteBstWouldAddEndPuncttrue
\mciteSetBstMidEndSepPunct{\mcitedefaultmidpunct}
{\mcitedefaultendpunct}{\mcitedefaultseppunct}\relax
\EndOfBibitem
\bibitem{LHCb-PAPER-2016-009}
LHCb Collaboration, R.~Aaij {\em et~al.},
  \ifthenelse{\boolean{articletitles}}{\emph{{Model-independent evidence for
  $\jpsi\proton$ contributions to \mbox{\decay{\Lb}{\jpsi\proton\Km}} decays}},
  }{}\href{https://doi.org/10.1103/PhysRevLett.117.082002}{Phys.\ Rev.\ Lett.\
  \textbf{117} (2016) 082002},
  \href{http://arxiv.org/abs/1604.05708}{{\normalfont\ttfamily
  arXiv:1604.05708}}\relax
\mciteBstWouldAddEndPuncttrue
\mciteSetBstMidEndSepPunct{\mcitedefaultmidpunct}
{\mcitedefaultendpunct}{\mcitedefaultseppunct}\relax
\EndOfBibitem
\bibitem{LHCb-PAPER-2016-015}
LHCb Collaboration, R.~Aaij {\em et~al.},
  \ifthenelse{\boolean{articletitles}}{\emph{{Evidence for exotic hadron
  contributions to \mbox{\decay{\Lb}{\jpsi\proton\pim}} decays}},
  }{}\href{https://doi.org/10.1103/PhysRevLett.117.082003}{Phys.\ Rev.\ Lett.\
  \textbf{117} (2016) 082003},
  \href{http://arxiv.org/abs/1606.06999}{{\normalfont\ttfamily
  arXiv:1606.06999}}\relax
\mciteBstWouldAddEndPuncttrue
\mciteSetBstMidEndSepPunct{\mcitedefaultmidpunct}
{\mcitedefaultendpunct}{\mcitedefaultseppunct}\relax
\EndOfBibitem
\bibitem{LHCb-PAPER-2019-014}
LHCb Collaboration, R.~Aaij {\em et~al.},
  \ifthenelse{\boolean{articletitles}}{\emph{{Observation of a narrow
  $P_c(4312)^+$ state, and of two-peak structure of the $P_c(4450)^+$}},
  }{}\href{https://doi.org/10.1103/PhysRevLett.122.222001}{Phys.\ Rev.\ Lett.\
  \textbf{122} (2019) 222001},
  \href{http://arxiv.org/abs/1904.03947}{{\normalfont\ttfamily
  arXiv:1904.03947}}\relax
\mciteBstWouldAddEndPuncttrue
\mciteSetBstMidEndSepPunct{\mcitedefaultmidpunct}
{\mcitedefaultendpunct}{\mcitedefaultseppunct}\relax
\EndOfBibitem
\bibitem{Iwasaki:1976cn}
Y.~Iwasaki, \ifthenelse{\boolean{articletitles}}{\emph{{Is a state
  $c\bar{c}c\bar{c}$ found at 6.0 \gev?}},
  }{}\href{https://doi.org/10.1103/PhysRevLett.36.1266}{Phys.\ Rev.\ Lett.\
  \textbf{36} (1976) 1266}\relax
\mciteBstWouldAddEndPuncttrue
\mciteSetBstMidEndSepPunct{\mcitedefaultmidpunct}
{\mcitedefaultendpunct}{\mcitedefaultseppunct}\relax
\EndOfBibitem
\bibitem{Chao:1980dv}
K.-T. Chao, \ifthenelse{\boolean{articletitles}}{\emph{{The (cc) - ($\bar{cc}$)
  (diquark-antidiquark) states in $e^+ e^-$ annihilation}},
  }{}\href{https://doi.org/10.1007/BF01431564}{Z.\ Phys.\  \textbf{C7} (1981)
  317}\relax
\mciteBstWouldAddEndPuncttrue
\mciteSetBstMidEndSepPunct{\mcitedefaultmidpunct}
{\mcitedefaultendpunct}{\mcitedefaultseppunct}\relax
\EndOfBibitem
\bibitem{PhysRevD.25.2370}
J.-P. Ader, J.-M. Richard, and P.~Taxil,
  \ifthenelse{\boolean{articletitles}}{\emph{Do narrow heavy multiquark states
  exist?}, }{}\href{https://doi.org/10.1103/PhysRevD.25.2370}{Phys.\ Rev.\
  \textbf{D25} (1982) 2370}\relax
\mciteBstWouldAddEndPuncttrue
\mciteSetBstMidEndSepPunct{\mcitedefaultmidpunct}
{\mcitedefaultendpunct}{\mcitedefaultseppunct}\relax
\EndOfBibitem
\bibitem{PhysRevD.29.426}
B.-A. Li and K.-F. Liu, \ifthenelse{\boolean{articletitles}}{\emph{$\jpsi$ pair
  production in hadronic collisions},
  }{}\href{https://doi.org/10.1103/PhysRevD.29.426}{Phys.\ Rev.\  \textbf{D29}
  (1984) 426}\relax
\mciteBstWouldAddEndPuncttrue
\mciteSetBstMidEndSepPunct{\mcitedefaultmidpunct}
{\mcitedefaultendpunct}{\mcitedefaultseppunct}\relax
\EndOfBibitem
\bibitem{Badalian:1985es}
A.~M. Badalian, B.~L. Ioffe, and A.~V. Smilga,
  \ifthenelse{\boolean{articletitles}}{\emph{{Four quark states in heavy quark
  systems}}, }{}\href{https://doi.org/10.1016/0550-3213(87)90248-3}{Nucl.\
  Phys.\  \textbf{281} (1987) B85}\relax
\mciteBstWouldAddEndPuncttrue
\mciteSetBstMidEndSepPunct{\mcitedefaultmidpunct}
{\mcitedefaultendpunct}{\mcitedefaultseppunct}\relax
\EndOfBibitem
\bibitem{Berezhnoy:2011xn}
A.~V. Berezhnoy, A.~V. Luchinsky, and A.~A. Novoselov,
  \ifthenelse{\boolean{articletitles}}{\emph{{Heavy tetraquarks production at
  the LHC}}, }{}\href{https://doi.org/10.1103/PhysRevD.86.034004}{Phys.\ Rev.\
  \textbf{D86} (2012) 034004},
  \href{http://arxiv.org/abs/1111.1867}{{\normalfont\ttfamily
  arXiv:1111.1867}}\relax
\mciteBstWouldAddEndPuncttrue
\mciteSetBstMidEndSepPunct{\mcitedefaultmidpunct}
{\mcitedefaultendpunct}{\mcitedefaultseppunct}\relax
\EndOfBibitem
\bibitem{Wu:2016vtq}
J.~Wu, Y.-R. Liu, K.~Chen {\em et~al.},
  \ifthenelse{\boolean{articletitles}}{\emph{{Heavy-flavored tetraquark states
  with the $QQ\bar{Q}\bar{Q}$ configuration}},
  }{}\href{https://doi.org/10.1103/PhysRevD.97.094015}{Phys.\ Rev.\
  \textbf{D97} (2018) 094015},
  \href{http://arxiv.org/abs/1605.01134}{{\normalfont\ttfamily
  arXiv:1605.01134}}\relax
\mciteBstWouldAddEndPuncttrue
\mciteSetBstMidEndSepPunct{\mcitedefaultmidpunct}
{\mcitedefaultendpunct}{\mcitedefaultseppunct}\relax
\EndOfBibitem
\bibitem{Karliner:2016zzc}
M.~Karliner, S.~Nussinov, and J.~L. Rosner,
  \ifthenelse{\boolean{articletitles}}{\emph{{$Q Q \bar Q \bar Q$ states:
  masses, production, and decays}},
  }{}\href{https://doi.org/10.1103/PhysRevD.95.034011}{Phys.\ Rev.\
  \textbf{D95} (2017) 034011},
  \href{http://arxiv.org/abs/1611.00348}{{\normalfont\ttfamily
  arXiv:1611.00348}}\relax
\mciteBstWouldAddEndPuncttrue
\mciteSetBstMidEndSepPunct{\mcitedefaultmidpunct}
{\mcitedefaultendpunct}{\mcitedefaultseppunct}\relax
\EndOfBibitem
\bibitem{Barnea:2006sd}
N.~Barnea, J.~Vijande, and A.~Valcarce,
  \ifthenelse{\boolean{articletitles}}{\emph{{Four-quark spectroscopy within
  the hyperspherical formalism}},
  }{}\href{https://doi.org/10.1103/PhysRevD.73.054004}{Phys.\ Rev.\
  \textbf{D73} (2006) 054004},
  \href{http://arxiv.org/abs/hep-ph/0604010}{{\normalfont\ttfamily
  arXiv:hep-ph/0604010}}\relax
\mciteBstWouldAddEndPuncttrue
\mciteSetBstMidEndSepPunct{\mcitedefaultmidpunct}
{\mcitedefaultendpunct}{\mcitedefaultseppunct}\relax
\EndOfBibitem
\bibitem{Debastiani:2017msn}
V.~R. Debastiani and F.~S. Navarra,
  \ifthenelse{\boolean{articletitles}}{\emph{{A non-relativistic model for the
  $[cc][\bar{c}\bar{c}]$ tetraquark}},
  }{}\href{https://doi.org/10.1088/1674-1137/43/1/013105}{Chin.\ Phys.\
  \textbf{C43} (2019) 013105},
  \href{http://arxiv.org/abs/1706.07553}{{\normalfont\ttfamily
  arXiv:1706.07553}}\relax
\mciteBstWouldAddEndPuncttrue
\mciteSetBstMidEndSepPunct{\mcitedefaultmidpunct}
{\mcitedefaultendpunct}{\mcitedefaultseppunct}\relax
\EndOfBibitem
\bibitem{Liu:2019zuc}
M.-S. Liu, Q.-F. Lü, X.-H. Zhong {\em et~al.},
  \ifthenelse{\boolean{articletitles}}{\emph{{All-heavy tetraquarks}},
  }{}\href{https://doi.org/10.1103/PhysRevD.100.016006}{Phys.\ Rev.\
  \textbf{D100} (2019) 016006},
  \href{http://arxiv.org/abs/1901.02564}{{\normalfont\ttfamily
  arXiv:1901.02564}}\relax
\mciteBstWouldAddEndPuncttrue
\mciteSetBstMidEndSepPunct{\mcitedefaultmidpunct}
{\mcitedefaultendpunct}{\mcitedefaultseppunct}\relax
\EndOfBibitem
\bibitem{Chen:2016jxd}
W.~Chen, H.-X. Chen, X.~Liu {\em et~al.},
  \ifthenelse{\boolean{articletitles}}{\emph{{Hunting for exotic doubly
  hidden-charm/bottom tetraquark states}},
  }{}\href{https://doi.org/10.1016/j.physletb.2017.08.034}{Phys.\ Lett.\
  \textbf{B773} (2017) 247},
  \href{http://arxiv.org/abs/1605.01647}{{\normalfont\ttfamily
  arXiv:1605.01647}}\relax
\mciteBstWouldAddEndPuncttrue
\mciteSetBstMidEndSepPunct{\mcitedefaultmidpunct}
{\mcitedefaultendpunct}{\mcitedefaultseppunct}\relax
\EndOfBibitem
\bibitem{Wang:2019rdo}
G.-J. Wang, L.~Meng, and S.-L. Zhu,
  \ifthenelse{\boolean{articletitles}}{\emph{{Spectrum of the fully-heavy
  tetraquark state $QQ\bar Q' \bar Q'$}},
  }{}\href{https://doi.org/10.1103/PhysRevD.100.096013}{Phys.\ Rev.\
  \textbf{D100} (2019) 096013},
  \href{http://arxiv.org/abs/1907.05177}{{\normalfont\ttfamily
  arXiv:1907.05177}}\relax
\mciteBstWouldAddEndPuncttrue
\mciteSetBstMidEndSepPunct{\mcitedefaultmidpunct}
{\mcitedefaultendpunct}{\mcitedefaultseppunct}\relax
\EndOfBibitem
\bibitem{Bedolla:2019zwg}
M.~A. Bedolla, J.~Ferretti, C.~D. Roberts {\em et~al.},
  \ifthenelse{\boolean{articletitles}}{\emph{{Spectrum of fully-heavy
  tetraquarks from a diquark+antidiquark perspective}},
  }{}\href{http://arxiv.org/abs/1911.00960}{{\normalfont\ttfamily
  arXiv:1911.00960}}\relax
\mciteBstWouldAddEndPuncttrue
\mciteSetBstMidEndSepPunct{\mcitedefaultmidpunct}
{\mcitedefaultendpunct}{\mcitedefaultseppunct}\relax
\EndOfBibitem
\bibitem{Lloyd:2003yc}
R.~J. Lloyd and J.~P. Vary,
  \ifthenelse{\boolean{articletitles}}{\emph{{All-charm tetraquarks}},
  }{}\href{https://doi.org/10.1103/PhysRevD.70.014009}{Phys.\ Rev.\
  \textbf{D70} (2004) 014009},
  \href{http://arxiv.org/abs/hep-ph/0311179}{{\normalfont\ttfamily
  arXiv:hep-ph/0311179}}\relax
\mciteBstWouldAddEndPuncttrue
\mciteSetBstMidEndSepPunct{\mcitedefaultmidpunct}
{\mcitedefaultendpunct}{\mcitedefaultseppunct}\relax
\EndOfBibitem
\bibitem{Chen:2020lgj}
X.~Chen, \ifthenelse{\boolean{articletitles}}{\emph{{Fully-charm tetraquarks:
  $cc\bar{c}\bar{c}$}},
  }{}\href{http://arxiv.org/abs/2001.06755}{{\normalfont\ttfamily
  arXiv:2001.06755}}\relax
\mciteBstWouldAddEndPuncttrue
\mciteSetBstMidEndSepPunct{\mcitedefaultmidpunct}
{\mcitedefaultendpunct}{\mcitedefaultseppunct}\relax
\EndOfBibitem
\bibitem{Wang:2018poa}
Z.-G. Wang and Z.-Y. Di, \ifthenelse{\boolean{articletitles}}{\emph{{Analysis
  of the vector and axialvector $QQ\bar{Q}\bar{Q}$ tetraquark states with QCD
  sum rules}}, }{}\href{https://doi.org/10.5506/APhysPolB.50.1335}{Acta Phys.\
  Polon.\  \textbf{B50} (2019) 1335},
  \href{http://arxiv.org/abs/1807.08520}{{\normalfont\ttfamily
  arXiv:1807.08520}}\relax
\mciteBstWouldAddEndPuncttrue
\mciteSetBstMidEndSepPunct{\mcitedefaultmidpunct}
{\mcitedefaultendpunct}{\mcitedefaultseppunct}\relax
\EndOfBibitem
\bibitem{Anwar:2017toa}
M.~N. Anwar, J.~Ferretti, F.-K. Guo {\em et~al.},
  \ifthenelse{\boolean{articletitles}}{\emph{{Spectroscopy and decays of the
  fully-heavy tetraquarks}},
  }{}\href{https://doi.org/10.1140/epjc/s10052-018-6073-9}{Eur.\ Phys.\ J.\
  \textbf{C78} (2018) 647},
  \href{http://arxiv.org/abs/1710.02540}{{\normalfont\ttfamily
  arXiv:1710.02540}}\relax
\mciteBstWouldAddEndPuncttrue
\mciteSetBstMidEndSepPunct{\mcitedefaultmidpunct}
{\mcitedefaultendpunct}{\mcitedefaultseppunct}\relax
\EndOfBibitem
\bibitem{Liu:2019zoy}
Y.-R. Liu, H.-X. Chen, W.~Chen {\em et~al.},
  \ifthenelse{\boolean{articletitles}}{\emph{{Pentaquark and tetraquark
  states}}, }{}\href{https://doi.org/10.1016/j.ppnp.2019.04.003}{Prog.\ Part.\
  Nucl.\ Phys.\  \textbf{107} (2019) 237},
  \href{http://arxiv.org/abs/1903.11976}{{\normalfont\ttfamily
  arXiv:1903.11976}}\relax
\mciteBstWouldAddEndPuncttrue
\mciteSetBstMidEndSepPunct{\mcitedefaultmidpunct}
{\mcitedefaultendpunct}{\mcitedefaultseppunct}\relax
\EndOfBibitem
\bibitem{Esposito:2018cwh}
A.~Esposito and A.~D. Polosa, \ifthenelse{\boolean{articletitles}}{\emph{{A
  $bb\bar b\bar b$ di-bottomonium at the LHC?}},
  }{}\href{https://doi.org/10.1140/epjc/s10052-018-6269-z}{Eur.\ Phys.\ J.\
  \textbf{C78} (2018) 782},
  \href{http://arxiv.org/abs/1807.06040}{{\normalfont\ttfamily
  arXiv:1807.06040}}\relax
\mciteBstWouldAddEndPuncttrue
\mciteSetBstMidEndSepPunct{\mcitedefaultmidpunct}
{\mcitedefaultendpunct}{\mcitedefaultseppunct}\relax
\EndOfBibitem
\bibitem{Becchi:2020mjz}
C.~Becchi, A.~Giachino, L.~Maiani {\em et~al.},
  \ifthenelse{\boolean{articletitles}}{\emph{{Search for $bb\bar{b}\bar{b}$
  tetraquark decays in 4 muons, $B^{+} B^{-}$, $B^0 \bar{B}^0$ and $B_s^0
  \bar{B}_s^0$ channels at LHC}},
  }{}\href{https://doi.org/10.1016/j.physletb.2020.135495}{Phys.\ Lett.\
  \textbf{B806} (2020) 135495},
  \href{http://arxiv.org/abs/2002.11077}{{\normalfont\ttfamily
  arXiv:2002.11077}}\relax
\mciteBstWouldAddEndPuncttrue
\mciteSetBstMidEndSepPunct{\mcitedefaultmidpunct}
{\mcitedefaultendpunct}{\mcitedefaultseppunct}\relax
\EndOfBibitem
\bibitem{Bai:2016int}
Y.~Bai, S.~Lu, and J.~Osborne,
  \ifthenelse{\boolean{articletitles}}{\emph{{Beauty-full tetraquarks}},
  }{}\href{https://doi.org/10.1016/j.physletb.2019.134930}{Phys.\ Lett.\
  \textbf{B798} (2019) 134930},
  \href{http://arxiv.org/abs/1612.00012}{{\normalfont\ttfamily
  arXiv:1612.00012}}\relax
\mciteBstWouldAddEndPuncttrue
\mciteSetBstMidEndSepPunct{\mcitedefaultmidpunct}
{\mcitedefaultendpunct}{\mcitedefaultseppunct}\relax
\EndOfBibitem
\bibitem{Richard:2017vry}
J.-M. Richard, A.~Valcarce, and J.~Vijande,
  \ifthenelse{\boolean{articletitles}}{\emph{{String dynamics and metastability
  of all-heavy tetraquarks}},
  }{}\href{https://doi.org/10.1103/PhysRevD.95.054019}{Phys.\ Rev.\
  \textbf{D95} (2017) 054019},
  \href{http://arxiv.org/abs/1703.00783}{{\normalfont\ttfamily
  arXiv:1703.00783}}\relax
\mciteBstWouldAddEndPuncttrue
\mciteSetBstMidEndSepPunct{\mcitedefaultmidpunct}
{\mcitedefaultendpunct}{\mcitedefaultseppunct}\relax
\EndOfBibitem
\bibitem{Vega-Morales:2017pmm}
Y.~Chen and R.~Vega-Morales, \ifthenelse{\boolean{articletitles}}{\emph{{Golden
  probe of the di$-\Upsilonres$ threshold}},
  }{}\href{http://arxiv.org/abs/1710.02738}{{\normalfont\ttfamily
  arXiv:1710.02738}}\relax
\mciteBstWouldAddEndPuncttrue
\mciteSetBstMidEndSepPunct{\mcitedefaultmidpunct}
{\mcitedefaultendpunct}{\mcitedefaultseppunct}\relax
\EndOfBibitem
\bibitem{Chen:2019vrj}
X.~Chen, \ifthenelse{\boolean{articletitles}}{\emph{{Fully-heavy tetraquarks:
  $bb\bar{c}\bar{c}$ and $bc\bar{b}\bar{c}$}},
  }{}\href{https://doi.org/10.1103/PhysRevD.100.094009}{Phys.\ Rev.\
  \textbf{D100} (2019) 094009},
  \href{http://arxiv.org/abs/1908.08811}{{\normalfont\ttfamily
  arXiv:1908.08811}}\relax
\mciteBstWouldAddEndPuncttrue
\mciteSetBstMidEndSepPunct{\mcitedefaultmidpunct}
{\mcitedefaultendpunct}{\mcitedefaultseppunct}\relax
\EndOfBibitem
\bibitem{Berezhnoy:2012tu}
A.~V. Berezhnoy, A.~K. Likhoded, and A.~A. Novoselov,
  \ifthenelse{\boolean{articletitles}}{\emph{{$\Upsilonres$-meson pair
  production at LHC}},
  }{}\href{https://doi.org/10.1103/PhysRevD.87.054023}{Phys.\ Rev.\
  \textbf{D87} (2013) 054023},
  \href{http://arxiv.org/abs/1210.5754}{{\normalfont\ttfamily
  arXiv:1210.5754}}\relax
\mciteBstWouldAddEndPuncttrue
\mciteSetBstMidEndSepPunct{\mcitedefaultmidpunct}
{\mcitedefaultendpunct}{\mcitedefaultseppunct}\relax
\EndOfBibitem
\bibitem{Karliner:2014gca}
M.~Karliner and J.~L. Rosner,
  \ifthenelse{\boolean{articletitles}}{\emph{{Baryons with two heavy quarks:
  Masses, production, decays, and detection}},
  }{}\href{https://doi.org/10.1103/PhysRevD.90.094007}{Phys.\ Rev.\
  \textbf{D90} (2014) 094007},
  \href{http://arxiv.org/abs/1408.5877}{{\normalfont\ttfamily
  arXiv:1408.5877}}\relax
\mciteBstWouldAddEndPuncttrue
\mciteSetBstMidEndSepPunct{\mcitedefaultmidpunct}
{\mcitedefaultendpunct}{\mcitedefaultseppunct}\relax
\EndOfBibitem
\bibitem{LHCb-PAPER-2017-018}
LHCb Collaboration, R.~Aaij {\em et~al.},
  \ifthenelse{\boolean{articletitles}}{\emph{{Observation of the doubly charmed
  baryon \Xiccpp}},
  }{}\href{https://doi.org/10.1103/PhysRevLett.119.112001}{Phys.\ Rev.\ Lett.\
  \textbf{119} (2017) 112001},
  \href{http://arxiv.org/abs/1707.01621}{{\normalfont\ttfamily
  arXiv:1707.01621}}\relax
\mciteBstWouldAddEndPuncttrue
\mciteSetBstMidEndSepPunct{\mcitedefaultmidpunct}
{\mcitedefaultendpunct}{\mcitedefaultseppunct}\relax
\EndOfBibitem
\bibitem{LHCb-PAPER-2019-039}
LHCb Collaboration, R.~Aaij {\em et~al.},
  \ifthenelse{\boolean{articletitles}}{\emph{{Isospin amplitudes in
  \mbox{\decay{\Lb}{\jpsi\Lz (\Sigmares^0)}} and
  \mbox{\decay{\Xibz}{\jpsi\Xires^0(\Lz)}} decays}},
  }{}\href{https://doi.org/10.1103/PhysRevLett.124.111802}{Phys.\ Rev.\ Lett.\
  \textbf{124} (2020) 111802},
  \href{http://arxiv.org/abs/1912.02110}{{\normalfont\ttfamily
  arXiv:1912.02110}}\relax
\mciteBstWouldAddEndPuncttrue
\mciteSetBstMidEndSepPunct{\mcitedefaultmidpunct}
{\mcitedefaultendpunct}{\mcitedefaultseppunct}\relax
\EndOfBibitem
\bibitem{LHCb-PAPER-2018-027}
LHCb Collaboration, R.~Aaij {\em et~al.},
  \ifthenelse{\boolean{articletitles}}{\emph{{Search for beautiful tetraquarks
  in the $\Upsilonres$(1S)$\mumu$ invariant-mass spectrum}},
  }{}\href{https://doi.org/10.1007/JHEP10(2018)086}{J.\ High Energy Phys.\
  \textbf{10} (2018) 086},
  \href{http://arxiv.org/abs/1806.09707}{{\normalfont\ttfamily
  arXiv:1806.09707}}\relax
\mciteBstWouldAddEndPuncttrue
\mciteSetBstMidEndSepPunct{\mcitedefaultmidpunct}
{\mcitedefaultendpunct}{\mcitedefaultseppunct}\relax
\EndOfBibitem
\bibitem{Sirunyan:2020txn}
CMS Collaboration, A.~M. Sirunyan {\em et~al.},
  \ifthenelse{\boolean{articletitles}}{\emph{{Measurement of the
  $\Upsilonres$(1S) pair production cross section and search for resonances
  decaying to $\Upsilonres$(1S)$\mu^+\mu^-$ in proton-proton collisions at
  $\sqrt{s}=$ 13\tev}},
  }{}\href{http://arxiv.org/abs/2002.06393}{{\normalfont\ttfamily
  arXiv:2002.06393}}\relax
\mciteBstWouldAddEndPuncttrue
\mciteSetBstMidEndSepPunct{\mcitedefaultmidpunct}
{\mcitedefaultendpunct}{\mcitedefaultseppunct}\relax
\EndOfBibitem
\bibitem{Calucci:1997ii}
G.~Calucci and D.~Treleani,
  \ifthenelse{\boolean{articletitles}}{\emph{{Minijets and the two-body parton
  correlation}}, }{}\href{https://doi.org/10.1103/PhysRevD.57.503}{Phys.\ Rev.\
   \textbf{D57} (1998) 503},
  \href{http://arxiv.org/abs/hep-ph/9707389}{{\normalfont\ttfamily
  arXiv:hep-ph/9707389}}\relax
\mciteBstWouldAddEndPuncttrue
\mciteSetBstMidEndSepPunct{\mcitedefaultmidpunct}
{\mcitedefaultendpunct}{\mcitedefaultseppunct}\relax
\EndOfBibitem
\bibitem{Calucci:1999yz}
G.~Calucci and D.~Treleani, \ifthenelse{\boolean{articletitles}}{\emph{{Proton
  structure in transverse space and the effective cross section}},
  }{}\href{https://doi.org/10.1103/PhysRevD.60.054023}{Phys.\ Rev.\
  \textbf{D60} (1999) 054023},
  \href{http://arxiv.org/abs/hep-ph/9902479}{{\normalfont\ttfamily
  arXiv:hep-ph/9902479}}\relax
\mciteBstWouldAddEndPuncttrue
\mciteSetBstMidEndSepPunct{\mcitedefaultmidpunct}
{\mcitedefaultendpunct}{\mcitedefaultseppunct}\relax
\EndOfBibitem
\bibitem{DelFabbro:2000ds}
A.~Del~Fabbro and D.~Treleani,
  \ifthenelse{\boolean{articletitles}}{\emph{{Scale factor in double parton
  collisions and parton densities in transverse space}},
  }{}\href{https://doi.org/10.1103/PhysRevD.63.057901}{Phys.\ Rev.\
  \textbf{D63} (2001) 057901},
  \href{http://arxiv.org/abs/hep-ph/0005273}{{\normalfont\ttfamily
  arXiv:hep-ph/0005273}}\relax
\mciteBstWouldAddEndPuncttrue
\mciteSetBstMidEndSepPunct{\mcitedefaultmidpunct}
{\mcitedefaultendpunct}{\mcitedefaultseppunct}\relax
\EndOfBibitem
\bibitem{Sun:2014gca}
L.-P. Sun, H.~Han, and K.-T. Chao,
  \ifthenelse{\boolean{articletitles}}{\emph{{Impact of $\jpsi$ pair production
  at the LHC and predictions in nonrelativistic QCD}},
  }{}\href{https://doi.org/10.1103/PhysRevD.94.074033}{Phys.\ Rev.\
  \textbf{D94} (2016) 074033},
  \href{http://arxiv.org/abs/1404.4042}{{\normalfont\ttfamily
  arXiv:1404.4042}}\relax
\mciteBstWouldAddEndPuncttrue
\mciteSetBstMidEndSepPunct{\mcitedefaultmidpunct}
{\mcitedefaultendpunct}{\mcitedefaultseppunct}\relax
\EndOfBibitem
\bibitem{Likhoded:2016zmk}
A.~K. Likhoded, A.~V. Luchinsky, and S.~V. Poslavsky,
  \ifthenelse{\boolean{articletitles}}{\emph{{Production of $\jpsi + \chi_c$
  and $\jpsi + \jpsi$ with real gluon emission at LHC}},
  }{}\href{https://doi.org/10.1103/PhysRevD.94.054017}{Phys.\ Rev.\
  \textbf{D94} (2016) 054017},
  \href{http://arxiv.org/abs/1606.06767}{{\normalfont\ttfamily
  arXiv:1606.06767}}\relax
\mciteBstWouldAddEndPuncttrue
\mciteSetBstMidEndSepPunct{\mcitedefaultmidpunct}
{\mcitedefaultendpunct}{\mcitedefaultseppunct}\relax
\EndOfBibitem
\bibitem{Shao:2012iz}
H.-S. Shao, \ifthenelse{\boolean{articletitles}}{\emph{{{\sc{HELAC-Onia}}: An
  automatic matrix element generator for heavy quarkonium physics}},
  }{}\href{https://doi.org/10.1016/j.cpc.2013.05.023}{Comput.\ Phys.\ Commun.\
  \textbf{184} (2013) 2562},
  \href{http://arxiv.org/abs/1212.5293}{{\normalfont\ttfamily
  arXiv:1212.5293}}\relax
\mciteBstWouldAddEndPuncttrue
\mciteSetBstMidEndSepPunct{\mcitedefaultmidpunct}
{\mcitedefaultendpunct}{\mcitedefaultseppunct}\relax
\EndOfBibitem
\bibitem{Shao:2015vga}
H.-S. Shao, \ifthenelse{\boolean{articletitles}}{\emph{{{\sc{HELAC-Onia~2.0}}:
  An~upgraded matrix-element and event generator for heavy quarkonium
  physics}}, }{}\href{https://doi.org/10.1016/j.cpc.2015.09.011}{Comput.\
  Phys.\ Commun.\  \textbf{198} (2016) 238},
  \href{http://arxiv.org/abs/1507.03435}{{\normalfont\ttfamily
  arXiv:1507.03435}}\relax
\mciteBstWouldAddEndPuncttrue
\mciteSetBstMidEndSepPunct{\mcitedefaultmidpunct}
{\mcitedefaultendpunct}{\mcitedefaultseppunct}\relax
\EndOfBibitem
\bibitem{Baranov:2011zz}
S.~P. Baranov, \ifthenelse{\boolean{articletitles}}{\emph{{Pair production of
  \jpsi~mesons in the $k_{t}$-factorization approach}},
  }{}\href{https://doi.org/10.1103/PhysRevD.84.054012}{Phys.\ Rev.\
  \textbf{D84} (2011) 054012}\relax
\mciteBstWouldAddEndPuncttrue
\mciteSetBstMidEndSepPunct{\mcitedefaultmidpunct}
{\mcitedefaultendpunct}{\mcitedefaultseppunct}\relax
\EndOfBibitem
\bibitem{Lansberg:2013qka}
J.-P. Lansberg and H.-S. Shao,
  \ifthenelse{\boolean{articletitles}}{\emph{{Production of $\jpsi +
  \Peta_{\cquark}$~versus $\jpsi + \jpsi$~at the LHC: Importance of real
  $\upalpha^{5}_{\mathrm{s}}$~corrections}},
  }{}\href{https://doi.org/10.1103/PhysRevLett.111.122001}{Phys.\ Rev.\ Lett.\
  \textbf{111} (2013) 122001},
  \href{http://arxiv.org/abs/1308.0474}{{\normalfont\ttfamily
  arXiv:1308.0474}}\relax
\mciteBstWouldAddEndPuncttrue
\mciteSetBstMidEndSepPunct{\mcitedefaultmidpunct}
{\mcitedefaultendpunct}{\mcitedefaultseppunct}\relax
\EndOfBibitem
\bibitem{Lansberg:2014swa}
J.-P. Lansberg and H.-S. Shao,
  \ifthenelse{\boolean{articletitles}}{\emph{{\jpsi-pair production at large
  momenta: indications for double parton scatterings and large
  $\upalpha_{\mathrm{s}}^5$~contributions}},
  }{}\href{https://doi.org/10.1016/j.physletb.2015.10.083}{Phys.\ Lett.\
  \textbf{B751} (2015) 479},
  \href{http://arxiv.org/abs/1410.8822}{{\normalfont\ttfamily
  arXiv:1410.8822}}\relax
\mciteBstWouldAddEndPuncttrue
\mciteSetBstMidEndSepPunct{\mcitedefaultmidpunct}
{\mcitedefaultendpunct}{\mcitedefaultseppunct}\relax
\EndOfBibitem
\bibitem{Lansberg:2015lva}
J.-P. Lansberg and H.-S. Shao,
  \ifthenelse{\boolean{articletitles}}{\emph{{Double-quarkonium production at a
  fixed-target experiment at the LHC (AFTER@LHC)}},
  }{}\href{https://doi.org/10.1016/j.nuclphysb.2015.09.005}{Nucl.\ Phys.\
  \textbf{B900} (2015) 273},
  \href{http://arxiv.org/abs/1504.06531}{{\normalfont\ttfamily
  arXiv:1504.06531}}\relax
\mciteBstWouldAddEndPuncttrue
\mciteSetBstMidEndSepPunct{\mcitedefaultmidpunct}
{\mcitedefaultendpunct}{\mcitedefaultseppunct}\relax
\EndOfBibitem
\bibitem{W2jets}
CMS Collaboration, S.~Chatrchyan {\em et~al.},
  \ifthenelse{\boolean{articletitles}}{\emph{{Study of double parton scattering
  using W + 2-jet events in proton-proton collisions at $\sqrt{s} = 7$ TeV}},
  }{}\href{https://doi.org/10.1007/JHEP03(2014)032}{J.\ High Energy Phys.\
  \textbf{03} (2014) 032},
  \href{http://arxiv.org/abs/1312.5729}{{\normalfont\ttfamily
  arXiv:1312.5729}}\relax
\mciteBstWouldAddEndPuncttrue
\mciteSetBstMidEndSepPunct{\mcitedefaultmidpunct}
{\mcitedefaultendpunct}{\mcitedefaultseppunct}\relax
\EndOfBibitem
\bibitem{CMSDJ}
CMS Collaboration, V.~Khachatryan {\em et~al.},
  \ifthenelse{\boolean{articletitles}}{\emph{{Measurement of prompt $\jpsi$
  pair production in pp collisions at $ \sqrt{s} = $7\tev}},
  }{}\href{https://doi.org/10.1007/JHEP09(2014)094}{J.\ High Energy Phys.\
  \textbf{09} (2014) 094},
  \href{http://arxiv.org/abs/1406.0484}{{\normalfont\ttfamily
  arXiv:1406.0484}}\relax
\mciteBstWouldAddEndPuncttrue
\mciteSetBstMidEndSepPunct{\mcitedefaultmidpunct}
{\mcitedefaultendpunct}{\mcitedefaultseppunct}\relax
\EndOfBibitem
\bibitem{DUpsilon}
CMS Collaboration, V.~Khachatryan {\em et~al.},
  \ifthenelse{\boolean{articletitles}}{\emph{{Observation of $\Upsilonres({\rm
  1S})$ pair production in proton-proton collisions at $\sqrt{s} =$ 8 TeV}},
  }{}\href{http://arxiv.org/abs/1610.07095}{{\normalfont\ttfamily
  arXiv:1610.07095}}\relax
\mciteBstWouldAddEndPuncttrue
\mciteSetBstMidEndSepPunct{\mcitedefaultmidpunct}
{\mcitedefaultendpunct}{\mcitedefaultseppunct}\relax
\EndOfBibitem
\bibitem{JpsiW}
ATLAS Collaboration, G.~Aad {\em et~al.},
  \ifthenelse{\boolean{articletitles}}{\emph{{Measurement of the production of
  a $W$ boson in association with a charm quark in $pp$ collisions at $\sqrt{s}
  = 7$ TeV with the ATLAS detector}},
  }{}\href{https://doi.org/10.1007/JHEP05(2014)068}{J.\ High Energy Phys.\
  \textbf{05} (2014) 068},
  \href{http://arxiv.org/abs/1402.6263}{{\normalfont\ttfamily
  arXiv:1402.6263}}\relax
\mciteBstWouldAddEndPuncttrue
\mciteSetBstMidEndSepPunct{\mcitedefaultmidpunct}
{\mcitedefaultendpunct}{\mcitedefaultseppunct}\relax
\EndOfBibitem
\bibitem{JpsiZ}
ATLAS Collaboration, G.~Aad {\em et~al.},
  \ifthenelse{\boolean{articletitles}}{\emph{{Observation and measurements of
  the production of prompt and non-prompt $\jpsi$ mesons in association with a
  $Z$ boson in $pp$ collisions at $\sqrt{s}$ = 8 TeV with the ATLAS detector}},
  }{}\href{https://doi.org/10.1140/epjc/s10052-015-3406-9}{Eur.\ Phys.\ J.\
  \textbf{C75} (2015) 229},
  \href{http://arxiv.org/abs/1412.6428}{{\normalfont\ttfamily
  arXiv:1412.6428}}\relax
\mciteBstWouldAddEndPuncttrue
\mciteSetBstMidEndSepPunct{\mcitedefaultmidpunct}
{\mcitedefaultendpunct}{\mcitedefaultseppunct}\relax
\EndOfBibitem
\bibitem{Aaboud:2016dea}
ATLAS Collaboration, M.~Aaboud {\em et~al.},
  \ifthenelse{\boolean{articletitles}}{\emph{{Study of hard double-parton
  scattering in four-jet events in pp collisions at $ \sqrt{s}=7 $ TeV with the
  ATLAS experiment}}, }{}\href{https://doi.org/10.1007/JHEP11(2016)110}{J.\
  High Energy Phys.\  \textbf{11} (2016) 110},
  \href{http://arxiv.org/abs/1608.01857}{{\normalfont\ttfamily
  arXiv:1608.01857}}\relax
\mciteBstWouldAddEndPuncttrue
\mciteSetBstMidEndSepPunct{\mcitedefaultmidpunct}
{\mcitedefaultendpunct}{\mcitedefaultseppunct}\relax
\EndOfBibitem
\bibitem{ATLASDJ}
ATLAS Collaboration, M.~Aaboud {\em et~al.},
  \ifthenelse{\boolean{articletitles}}{\emph{{Measurement of the prompt \jpsi
  pair production cross-section in pp collisions at $\sqrt{s} = $8\tev with the
  ATLAS detector}},
  }{}\href{https://doi.org/10.1140/epjc/s10052-017-4644-9}{Eur.\ Phys.\ J.\
  \textbf{C77} (2017) 76},
  \href{http://arxiv.org/abs/1612.02950}{{\normalfont\ttfamily
  arXiv:1612.02950}}\relax
\mciteBstWouldAddEndPuncttrue
\mciteSetBstMidEndSepPunct{\mcitedefaultmidpunct}
{\mcitedefaultendpunct}{\mcitedefaultseppunct}\relax
\EndOfBibitem
\bibitem{LHCb-PAPER-2012-003}
LHCb Collaboration, R.~Aaij {\em et~al.},
  \ifthenelse{\boolean{articletitles}}{\emph{{Observation of double charm
  production involving open charm in \proton\proton collisions at
  \mbox{$\sqs=$7\tev}}}, }{}\href{https://doi.org/10.1007/JHEP06(2012)141}{J.\
  High Energy Phys.\  \textbf{06} (2012) 141}, Addendum
  \href{https://doi.org/10.1007/JHEP03(2014)108}{ibid.\   \textbf{03} (2014)
  108}, \href{http://arxiv.org/abs/1205.0975}{{\normalfont\ttfamily
  arXiv:1205.0975}}\relax
\mciteBstWouldAddEndPuncttrue
\mciteSetBstMidEndSepPunct{\mcitedefaultmidpunct}
{\mcitedefaultendpunct}{\mcitedefaultseppunct}\relax
\EndOfBibitem
\bibitem{LHCb-PAPER-2013-062}
LHCb Collaboration, R.~Aaij {\em et~al.},
  \ifthenelse{\boolean{articletitles}}{\emph{{Observation of associated
  production of a \Z boson with a \D meson in the forward region}},
  }{}\href{https://doi.org/10.1007/JHEP04(2014)091}{J.\ High Energy Phys.\
  \textbf{04} (2014) 091},
  \href{http://arxiv.org/abs/1401.3245}{{\normalfont\ttfamily
  arXiv:1401.3245}}\relax
\mciteBstWouldAddEndPuncttrue
\mciteSetBstMidEndSepPunct{\mcitedefaultmidpunct}
{\mcitedefaultendpunct}{\mcitedefaultseppunct}\relax
\EndOfBibitem
\bibitem{LHCb-PAPER-2015-046}
LHCb Collaboration, R.~Aaij {\em et~al.},
  \ifthenelse{\boolean{articletitles}}{\emph{{Production of associated
  \Upsilonres and open charm hadrons in \proton\proton collisions at $\sqs = $7
  and 8\tev via double parton scattering}},
  }{}\href{https://doi.org/10.1007/JHEP07(2016)052}{J.\ High Energy Phys.\
  \textbf{07} (2016) 052},
  \href{http://arxiv.org/abs/1510.05949}{{\normalfont\ttfamily
  arXiv:1510.05949}}\relax
\mciteBstWouldAddEndPuncttrue
\mciteSetBstMidEndSepPunct{\mcitedefaultmidpunct}
{\mcitedefaultendpunct}{\mcitedefaultseppunct}\relax
\EndOfBibitem
\bibitem{4jets1}
Axial Field Spectrometer Collaboration, T.~Akesson {\em et~al.},
  \ifthenelse{\boolean{articletitles}}{\emph{{Double parton scattering in $p p$
  collisions at $\sqrt{s}=63\gev$}},
  }{}\href{https://doi.org/10.1007/BF01566757}{Z.\ Phys.\  \textbf{C34} (1987)
  163}\relax
\mciteBstWouldAddEndPuncttrue
\mciteSetBstMidEndSepPunct{\mcitedefaultmidpunct}
{\mcitedefaultendpunct}{\mcitedefaultseppunct}\relax
\EndOfBibitem
\bibitem{Abe:1993rv}
CDF Collaboration, F.~Abe {\em et~al.},
  \ifthenelse{\boolean{articletitles}}{\emph{{Study of four jet events and
  evidence for double parton interactions in $\proton\antiproton$ collisions at
  $\sqs=1.8$ TeV}}, }{}\href{https://doi.org/10.1103/PhysRevD.47.4857}{Phys.\
  Rev.\  \textbf{D47} (1993) 4857}\relax
\mciteBstWouldAddEndPuncttrue
\mciteSetBstMidEndSepPunct{\mcitedefaultmidpunct}
{\mcitedefaultendpunct}{\mcitedefaultseppunct}\relax
\EndOfBibitem
\bibitem{Gamma3jets1}
CDF Collaboration, F.~Abe {\em et~al.},
  \ifthenelse{\boolean{articletitles}}{\emph{{Measurement of double parton
  scattering in $\bar{p}p$ collisions at $\sqrt{s} = 1.8$ TeV}},
  }{}\href{https://doi.org/10.1103/PhysRevLett.79.584}{Phys.\ Rev.\ Lett.\
  \textbf{79} (1997) 584}\relax
\mciteBstWouldAddEndPuncttrue
\mciteSetBstMidEndSepPunct{\mcitedefaultmidpunct}
{\mcitedefaultendpunct}{\mcitedefaultseppunct}\relax
\EndOfBibitem
\bibitem{Gamma3jets2}
D0 Collaboration, V.~M. Abazov {\em et~al.},
  \ifthenelse{\boolean{articletitles}}{\emph{{Double parton interactions in
  $\gamma$+3 jet events in $p\bar{p}$ collisions at $\sqrt{s}=1.96$ TeV}},
  }{}\href{https://doi.org/10.1103/PhysRevD.81.052012}{Phys.\ Rev.\
  \textbf{D81} (2010) 052012},
  \href{http://arxiv.org/abs/0912.5104}{{\normalfont\ttfamily
  arXiv:0912.5104}}\relax
\mciteBstWouldAddEndPuncttrue
\mciteSetBstMidEndSepPunct{\mcitedefaultmidpunct}
{\mcitedefaultendpunct}{\mcitedefaultseppunct}\relax
\EndOfBibitem
\bibitem{Abazov:2014fha}
D0 Collaboration, V.~M. Abazov {\em et~al.},
  \ifthenelse{\boolean{articletitles}}{\emph{{Double parton interactions in
  $\gamma+3$ jet and $\gamma+b/c \ jet+2$ jet events in $p \bar p$ collisions
  at $\sqrt s=1.96$ TeV}},
  }{}\href{https://doi.org/10.1103/PhysRevD.89.072006}{Phys.\ Rev.\
  \textbf{D89} (2014) 072006},
  \href{http://arxiv.org/abs/1402.1550}{{\normalfont\ttfamily
  arXiv:1402.1550}}\relax
\mciteBstWouldAddEndPuncttrue
\mciteSetBstMidEndSepPunct{\mcitedefaultmidpunct}
{\mcitedefaultendpunct}{\mcitedefaultseppunct}\relax
\EndOfBibitem
\bibitem{Abazov:2015nnn}
D0 Collaboration, V.~M. Abazov {\em et~al.},
  \ifthenelse{\boolean{articletitles}}{\emph{{Study of double parton
  interactions in diphoton + dijet events in $p\bar{p}$ collisions at $\sqrt{s}
  = 1.96$ TeV}}, }{}\href{https://doi.org/10.1103/PhysRevD.93.052008}{Phys.\
  Rev.\  \textbf{D93} (2016) 052008},
  \href{http://arxiv.org/abs/1512.05291}{{\normalfont\ttfamily
  arXiv:1512.05291}}\relax
\mciteBstWouldAddEndPuncttrue
\mciteSetBstMidEndSepPunct{\mcitedefaultmidpunct}
{\mcitedefaultendpunct}{\mcitedefaultseppunct}\relax
\EndOfBibitem
\bibitem{4jets2}
UA2 Collaboration, J.~Alitti {\em et~al.},
  \ifthenelse{\boolean{articletitles}}{\emph{{A study of multi-jet events at
  the CERN $\bar{p}p$ collider and a search for double parton scattering}},
  }{}\href{https://doi.org/10.1016/0370-2693(91)90937-L}{Phys.\ Lett.\
  \textbf{B268} (1991) 145}\relax
\mciteBstWouldAddEndPuncttrue
\mciteSetBstMidEndSepPunct{\mcitedefaultmidpunct}
{\mcitedefaultendpunct}{\mcitedefaultseppunct}\relax
\EndOfBibitem
\bibitem{LHCb-PAPER-2011-013}
LHCb Collaboration, R.~Aaij {\em et~al.},
  \ifthenelse{\boolean{articletitles}}{\emph{{Observation of \jpsi-pair
  production in \proton\proton collisions at \mbox{$\sqs=$7\tev}}},
  }{}\href{https://doi.org/10.1016/j.physletb.2011.12.015}{Phys.\ Lett.\
  \textbf{B707} (2012) 52},
  \href{http://arxiv.org/abs/1109.0963}{{\normalfont\ttfamily
  arXiv:1109.0963}}\relax
\mciteBstWouldAddEndPuncttrue
\mciteSetBstMidEndSepPunct{\mcitedefaultmidpunct}
{\mcitedefaultendpunct}{\mcitedefaultseppunct}\relax
\EndOfBibitem
\bibitem{LHCb-PAPER-2016-057}
LHCb Collaboration, R.~Aaij {\em et~al.},
  \ifthenelse{\boolean{articletitles}}{\emph{{Measurement of the \jpsi pair
  production cross-section in \proton\proton collisions at
  \mbox{$\sqs=$13\tev}}}, }{}\href{https://doi.org/10.1007/JHEP06(2017)047}{J.\
  High Energy Phys.\  \textbf{06} (2017) 047}, Erratum
  \href{https://doi.org/10.1007/JHEP10(2017)068}{ibid.\   \textbf{10} (2017)
  068}, \href{http://arxiv.org/abs/1612.07451}{{\normalfont\ttfamily
  arXiv:1612.07451}}\relax
\mciteBstWouldAddEndPuncttrue
\mciteSetBstMidEndSepPunct{\mcitedefaultmidpunct}
{\mcitedefaultendpunct}{\mcitedefaultseppunct}\relax
\EndOfBibitem
\bibitem{LHCb-DP-2008-001}
LHCb collaboration, A.~A. Alves~Jr.\ {\em et~al.},
  \ifthenelse{\boolean{articletitles}}{\emph{{The \lhcb detector at the LHC}},
  }{}\href{https://doi.org/10.1088/1748-0221/3/08/S08005}{JINST \textbf{3}
  (2008) S08005}\relax
\mciteBstWouldAddEndPuncttrue
\mciteSetBstMidEndSepPunct{\mcitedefaultmidpunct}
{\mcitedefaultendpunct}{\mcitedefaultseppunct}\relax
\EndOfBibitem
\bibitem{LHCb-DP-2014-002}
LHCb collaboration, R.~Aaij {\em et~al.},
  \ifthenelse{\boolean{articletitles}}{\emph{{LHCb detector performance}},
  }{}\href{https://doi.org/10.1142/S0217751X15300227}{Int.\ J.\ Mod.\ Phys.\
  \textbf{A30} (2015) 1530022},
  \href{http://arxiv.org/abs/1412.6352}{{\normalfont\ttfamily
  arXiv:1412.6352}}\relax
\mciteBstWouldAddEndPuncttrue
\mciteSetBstMidEndSepPunct{\mcitedefaultmidpunct}
{\mcitedefaultendpunct}{\mcitedefaultseppunct}\relax
\EndOfBibitem
\bibitem{LHCb-PAPER-2011-035}
LHCb Collaboration, R.~Aaij {\em et~al.},
  \ifthenelse{\boolean{articletitles}}{\emph{{Measurement of \bquark-hadron
  masses}}, }{}\href{https://doi.org/10.1016/j.physletb.2012.01.058}{Phys.\
  Lett.\  \textbf{B708} (2012) 241},
  \href{http://arxiv.org/abs/1112.4896}{{\normalfont\ttfamily
  arXiv:1112.4896}}\relax
\mciteBstWouldAddEndPuncttrue
\mciteSetBstMidEndSepPunct{\mcitedefaultmidpunct}
{\mcitedefaultendpunct}{\mcitedefaultseppunct}\relax
\EndOfBibitem
\bibitem{Sjostrand:2007gs}
T.~Sj\"{o}strand, S.~Mrenna, and P.~Skands,
  \ifthenelse{\boolean{articletitles}}{\emph{{A brief introduction to PYTHIA
  8.1}}, }{}\href{https://doi.org/10.1016/j.cpc.2008.01.036}{Comput.\ Phys.\
  Commun.\  \textbf{178} (2008) 852},
  \href{http://arxiv.org/abs/0710.3820}{{\normalfont\ttfamily
  arXiv:0710.3820}}\relax
\mciteBstWouldAddEndPuncttrue
\mciteSetBstMidEndSepPunct{\mcitedefaultmidpunct}
{\mcitedefaultendpunct}{\mcitedefaultseppunct}\relax
\EndOfBibitem
\bibitem{LHCb-PROC-2010-056}
I.~Belyaev {\em et~al.}, \ifthenelse{\boolean{articletitles}}{\emph{{Handling
  of the generation of primary events in Gauss, the LHCb simulation
  framework}}, }{}\href{https://doi.org/10.1088/1742-6596/331/3/032047}{J.\
  Phys.\ Conf.\ Ser.\  \textbf{331} (2011) 032047}\relax
\mciteBstWouldAddEndPuncttrue
\mciteSetBstMidEndSepPunct{\mcitedefaultmidpunct}
{\mcitedefaultendpunct}{\mcitedefaultseppunct}\relax
\EndOfBibitem
\bibitem{Lange:2001uf}
D.~J. Lange, \ifthenelse{\boolean{articletitles}}{\emph{{The EvtGen particle
  decay simulation package}},
  }{}\href{https://doi.org/10.1016/S0168-9002(01)00089-4}{Nucl.\ Instrum.\
  Meth.\  \textbf{A462} (2001) 152}\relax
\mciteBstWouldAddEndPuncttrue
\mciteSetBstMidEndSepPunct{\mcitedefaultmidpunct}
{\mcitedefaultendpunct}{\mcitedefaultseppunct}\relax
\EndOfBibitem
\bibitem{Golonka:2005pn}
P.~Golonka and Z.~Was, \ifthenelse{\boolean{articletitles}}{\emph{{PHOTOS Monte
  Carlo: A precision tool for QED corrections in $Z$ and $W$ decays}},
  }{}\href{https://doi.org/10.1140/epjc/s2005-02396-4}{Eur.\ Phys.\ J.\
  \textbf{C45} (2006) 97},
  \href{http://arxiv.org/abs/hep-ph/0506026}{{\normalfont\ttfamily
  arXiv:hep-ph/0506026}}\relax
\mciteBstWouldAddEndPuncttrue
\mciteSetBstMidEndSepPunct{\mcitedefaultmidpunct}
{\mcitedefaultendpunct}{\mcitedefaultseppunct}\relax
\EndOfBibitem
\bibitem{Allison:2006ve}
Geant4 collaboration, J.~Allison, K.~Amako, J.~Apostolakis {\em et~al.},
  \ifthenelse{\boolean{articletitles}}{\emph{{Geant4 developments and
  applications}}, }{}\href{https://doi.org/10.1109/TNS.2006.869826}{IEEE
  Trans.\ Nucl.\ Sci.\  \textbf{53} (2006) 270}\relax
\mciteBstWouldAddEndPuncttrue
\mciteSetBstMidEndSepPunct{\mcitedefaultmidpunct}
{\mcitedefaultendpunct}{\mcitedefaultseppunct}\relax
\EndOfBibitem
\bibitem{LHCb-PROC-2011-006}
M.~Clemencic {\em et~al.}, \ifthenelse{\boolean{articletitles}}{\emph{{The
  \lhcb simulation application, Gauss: Design, evolution and experience}},
  }{}\href{https://doi.org/10.1088/1742-6596/331/3/032023}{J.\ Phys.\ Conf.\
  Ser.\  \textbf{331} (2011) 032023}\relax
\mciteBstWouldAddEndPuncttrue
\mciteSetBstMidEndSepPunct{\mcitedefaultmidpunct}
{\mcitedefaultendpunct}{\mcitedefaultseppunct}\relax
\EndOfBibitem
\bibitem{Hulsbergen:2005pu}
W.~D. Hulsbergen, \ifthenelse{\boolean{articletitles}}{\emph{{Decay chain
  fitting with a Kalman filter}},
  }{}\href{https://doi.org/10.1016/j.nima.2005.06.078}{Nucl.\ Instrum.\ Meth.\
  \textbf{A552} (2005) 566},
  \href{http://arxiv.org/abs/physics/0503191}{{\normalfont\ttfamily
  arXiv:physics/0503191}}\relax
\mciteBstWouldAddEndPuncttrue
\mciteSetBstMidEndSepPunct{\mcitedefaultmidpunct}
{\mcitedefaultendpunct}{\mcitedefaultseppunct}\relax
\EndOfBibitem
\bibitem{Skwarnicki:1986xj}
T.~Skwarnicki, {\em {A study of the radiative cascade transitions between the
  Upsilon-prime and Upsilon resonances}}, PhD thesis, Institute of Nuclear
  Physics, Krakow, 1986,
  {\href{http://inspirehep.net/record/230779/}{DESY-F31-86-02}}\relax
\mciteBstWouldAddEndPuncttrue
\mciteSetBstMidEndSepPunct{\mcitedefaultmidpunct}
{\mcitedefaultendpunct}{\mcitedefaultseppunct}\relax
\EndOfBibitem
\bibitem{LHCb-PAPER-2011-003}
LHCb Collaboration, R.~Aaij {\em et~al.},
  \ifthenelse{\boolean{articletitles}}{\emph{{Measurement of \jpsi production
  in \proton\proton collisions at \mbox{$\sqs=$7\tev}}},
  }{}\href{https://doi.org/10.1140/epjc/s10052-011-1645-y}{Eur.\ Phys.\ J.\
  \textbf{C71} (2011) 1645},
  \href{http://arxiv.org/abs/1103.0423}{{\normalfont\ttfamily
  arXiv:1103.0423}}\relax
\mciteBstWouldAddEndPuncttrue
\mciteSetBstMidEndSepPunct{\mcitedefaultmidpunct}
{\mcitedefaultendpunct}{\mcitedefaultseppunct}\relax
\EndOfBibitem
\bibitem{LHCb-PAPER-2013-016}
LHCb Collaboration, R.~Aaij {\em et~al.},
  \ifthenelse{\boolean{articletitles}}{\emph{{Production of \jpsi and
  \Upsilonres mesons in \proton\proton collisions at \mbox{$\sqs=$8\tev}}},
  }{}\href{https://doi.org/10.1007/JHEP06(2013)064}{J.\ High Energy Phys.\
  \textbf{06} (2013) 064},
  \href{http://arxiv.org/abs/1304.6977}{{\normalfont\ttfamily
  arXiv:1304.6977}}\relax
\mciteBstWouldAddEndPuncttrue
\mciteSetBstMidEndSepPunct{\mcitedefaultmidpunct}
{\mcitedefaultendpunct}{\mcitedefaultseppunct}\relax
\EndOfBibitem
\bibitem{LHCb-PAPER-2015-037}
LHCb Collaboration, R.~Aaij {\em et~al.},
  \ifthenelse{\boolean{articletitles}}{\emph{{Measurement of forward \jpsi
  production cross-sections in \proton\proton collisions at
  \mbox{$\sqs=$13\tev}}}, }{}\href{https://doi.org/10.1007/JHEP10(2015)172}{J.\
  High Energy Phys.\  \textbf{10} (2015) 172}, Erratum
  \href{https://doi.org/10.1007/JHEP05(2017)063}{ibid.\   \textbf{05} (2017)
  063}, \href{http://arxiv.org/abs/1509.00771}{{\normalfont\ttfamily
  arXiv:1509.00771}}\relax
\mciteBstWouldAddEndPuncttrue
\mciteSetBstMidEndSepPunct{\mcitedefaultmidpunct}
{\mcitedefaultendpunct}{\mcitedefaultseppunct}\relax
\EndOfBibitem
\bibitem{PDG2019}
Particle Data Group, M.~Tanabashi {\em et~al.},
  \ifthenelse{\boolean{articletitles}}{\emph{{\href{http://pdg.lbl.gov/}{Review
  of particle physics}}},
  }{}\href{https://doi.org/10.1103/PhysRevD.98.030001}{Phys.\ Rev.\
  \textbf{D98} (2018) 030001}, and {\href{http://pdglive.lbl.gov/}{2019
  update}}\relax
\mciteBstWouldAddEndPuncttrue
\mciteSetBstMidEndSepPunct{\mcitedefaultmidpunct}
{\mcitedefaultendpunct}{\mcitedefaultseppunct}\relax
\EndOfBibitem
\bibitem{Pivk:2004ty}
M.~Pivk and F.~R. Le~Diberder,
  \ifthenelse{\boolean{articletitles}}{\emph{{sPlot: A statistical tool to
  unfold data distributions}},
  }{}\href{https://doi.org/10.1016/j.nima.2005.08.106}{Nucl.\ Instrum.\ Meth.\
  \textbf{A555} (2005) 356},
  \href{http://arxiv.org/abs/physics/0402083}{{\normalfont\ttfamily
  arXiv:physics/0402083}}\relax
\mciteBstWouldAddEndPuncttrue
\mciteSetBstMidEndSepPunct{\mcitedefaultmidpunct}
{\mcitedefaultendpunct}{\mcitedefaultseppunct}\relax
\EndOfBibitem
\bibitem{Xie:2009rka}
Y.~Xie, \ifthenelse{\boolean{articletitles}}{\emph{{sFit: a method for
  background subtraction in maximum likelihood fit}},
  }{}\href{http://arxiv.org/abs/0905.0724}{{\normalfont\ttfamily
  arXiv:0905.0724}}\relax
\mciteBstWouldAddEndPuncttrue
\mciteSetBstMidEndSepPunct{\mcitedefaultmidpunct}
{\mcitedefaultendpunct}{\mcitedefaultseppunct}\relax
\EndOfBibitem
\bibitem{Braaten:2020iye}
E.~Braaten, L.-P. He, K.~Ingles {\em et~al.},
  \ifthenelse{\boolean{articletitles}}{\emph{{Charm-meson triangle singularity
  in ${e^+e^-}$ annihilation into ${ D^{*0} \bar{D}^0 + \gamma }$}},
  }{}\href{https://doi.org/10.1103/PhysRevD.101.096020}{Phys.\ Rev.\
  \textbf{D101} (2020) 096020},
  \href{http://arxiv.org/abs/2004.12841}{{\normalfont\ttfamily
  arXiv:2004.12841}}\relax
\mciteBstWouldAddEndPuncttrue
\mciteSetBstMidEndSepPunct{\mcitedefaultmidpunct}
{\mcitedefaultendpunct}{\mcitedefaultseppunct}\relax
\EndOfBibitem
\bibitem{Liu:2015fea}
X.-H. Liu, Q.~Wang, and Q.~Zhao,
  \ifthenelse{\boolean{articletitles}}{\emph{{Understanding the newly observed
  heavy pentaquark candidates}},
  }{}\href{https://doi.org/10.1016/j.physletb.2016.03.089}{Phys.\ Lett.\
  \textbf{B757} (2016) 231},
  \href{http://arxiv.org/abs/1507.05359}{{\normalfont\ttfamily
  arXiv:1507.05359}}\relax
\mciteBstWouldAddEndPuncttrue
\mciteSetBstMidEndSepPunct{\mcitedefaultmidpunct}
{\mcitedefaultendpunct}{\mcitedefaultseppunct}\relax
\EndOfBibitem
\bibitem{Xie:2016lvs}
J.-J. Xie, L.-S. Geng, and E.~Oset,
  \ifthenelse{\boolean{articletitles}}{\emph{{$f_2$(1810) as a triangle
  singularity}}, }{}\href{https://doi.org/10.1103/PhysRevD.95.034004}{Phys.\
  Rev.\  \textbf{D95} (2017) 034004},
  \href{http://arxiv.org/abs/1610.09592}{{\normalfont\ttfamily
  arXiv:1610.09592}}\relax
\mciteBstWouldAddEndPuncttrue
\mciteSetBstMidEndSepPunct{\mcitedefaultmidpunct}
{\mcitedefaultendpunct}{\mcitedefaultseppunct}\relax
\EndOfBibitem
\bibitem{Guo:2016bjq}
F.-K. Guo, C.~Hanhart, Y.~S. Kalashnikova {\em et~al.},
  \ifthenelse{\boolean{articletitles}}{\emph{{Interplay of quark and meson
  degrees of freedom in near-threshold states: A practical parametrization for
  line shapes}}, }{}\href{https://doi.org/10.1103/PhysRevD.93.074031}{Phys.\
  Rev.\  \textbf{D93} (2016) 074031},
  \href{http://arxiv.org/abs/1602.00940}{{\normalfont\ttfamily
  arXiv:1602.00940}}\relax
\mciteBstWouldAddEndPuncttrue
\mciteSetBstMidEndSepPunct{\mcitedefaultmidpunct}
{\mcitedefaultendpunct}{\mcitedefaultseppunct}\relax
\EndOfBibitem
\bibitem{Langenbruch:2019nwe}
C.~Langenbruch, \ifthenelse{\boolean{articletitles}}{\emph{{Parameter
  uncertainties in weighted unbinned maximum likelihood fits}},
  }{}\href{http://arxiv.org/abs/1911.01303}{{\normalfont\ttfamily
  arXiv:1911.01303}}\relax
\mciteBstWouldAddEndPuncttrue
\mciteSetBstMidEndSepPunct{\mcitedefaultmidpunct}
{\mcitedefaultendpunct}{\mcitedefaultseppunct}\relax
\EndOfBibitem
\bibitem{Cowan:2010js}
G.~Cowan, K.~Cranmer, E.~Gross {\em et~al.},
  \ifthenelse{\boolean{articletitles}}{\emph{{Asymptotic formulae for
  likelihood-based tests of new physics}},
  }{}\href{https://doi.org/10.1140/epjc/s10052-011-1554-0}{Eur.\ Phys.\ J.\
  \textbf{C71} (2011) 1554}, Erratum
  \href{https://doi.org/10.1140/epjc/s10052-013-2501-z}{ibid.\   \textbf{C73}
  (2013) 2501}, \href{http://arxiv.org/abs/1007.1727}{{\normalfont\ttfamily
  arXiv:1007.1727}}\relax
\mciteBstWouldAddEndPuncttrue
\mciteSetBstMidEndSepPunct{\mcitedefaultmidpunct}
{\mcitedefaultendpunct}{\mcitedefaultseppunct}\relax
\EndOfBibitem
\end{mcitethebibliography}
